\documentclass[aps,prd,final,groupedaddress,showpacs,showkeys,floatfix,preprintnumbers,nofootinbib,letterpaper,notitlepage]{revtex4-1}

\usepackage{graphicx}
\usepackage{float}

\usepackage[utf8]{inputenc}                      
\usepackage{latexsym}                            
\usepackage{amsfonts}                            
\usepackage{amssymb}                             
\usepackage{amsmath}                             
\usepackage[mathscr]{eucal}                      
\usepackage{dcolumn}                             
\usepackage{subfigure}
\usepackage[english]{babel}
\usepackage{theorem}                            
\usepackage{hyperref}
\hypersetup{unicode=true}
\usepackage{slashed}
\usepackage{color}

\newcommand{\beq}{\begin{equation}}
\newcommand{\eeq}{\end{equation}}
\newcommand{\bgath}{\begin{gathered}}
\newcommand{\egath}{\end{gathered}}
\newcommand{\bsplit}{\begin{split}}
\newcommand{\esplit}{\end{split}}
\newcommand{\bdm}{\begin{displaymath}}
\newcommand{\edm}{\end{displaymath}}
\newcommand{\beqn}{\begin{eqnarray}}
\newcommand{\eeqn}{\end{eqnarray}}
\newcommand{\bea}[1]{\beq\begin{array}{#1}}
\newcommand{\eea}{\end{array}\eeq}
\newcommand{\bi}{\begin{itemize}}
\newcommand{\ei}{\end{itemize}}
\newcommand{\ben}{\begin{enumerate}}
\newcommand{\een}{\end{enumerate}}
\newcommand{\ba}[1]{\begin{array}{#1}}
\newcommand{\ea}{\end{array}}

\newcommand{\bc}{\begin{center}}
\newcommand{\ec}{\end{center}}
\newcommand{\bfr}{\begin{flushright}}
\newcommand{\efr}{\end{flushright}}
\newcommand{\bfl}{\begin{flushleft}}
\newcommand{\efl}{\end{flushleft}}

\newcommand{\bv}{\begin{verbatim}}
\newcommand{\ev}{\end{verbatim}}

\newcommand{\Tr}{\mathrm{Tr}}

\newcommand{\la}{\langle}
\newcommand{\ra}{\rangle}

\begin{document}

\color{black}

\preprint{MIT-CTP 4456}
\preprint{WUB/13-13}
\preprint{BNL-104912-2014-JA}
\title[]{Nucleon electromagnetic form factors from lattice QCD using a nearly physical pion mass}

\author{J.~R.~Green}
\altaffiliation[Current address: ]{\emph{Institut f\"ur Kernphysik, Johannes Gutenberg-Universit\"at Mainz, D-55099 Mainz, Germany}}
\author{J.~W.~Negele}
\author{A.~V.~Pochinsky} 
  \affiliation{Center for Theoretical Physics, Massachusetts Institute of Technology, 
               Cambridge, Massachusetts 02139, USA}
\author{S.~N.~Syritsyn} 
\altaffiliation[Current address: ]{\emph{RIKEN BNL Research Center, Brookhaven National Laboratory, Upton, NY 11973, USA}}
  \affiliation{Nuclear Science Division, Lawrence Berkeley National Laboratory, 
               Berkeley, California 94720, USA}

\author{M.~Engelhardt}
  \affiliation{Department of Physics, New Mexico State University, Las Cruces, NM 88003-8001, USA}

\author{S.~Krieg}
  \affiliation{Bergische Universit\"at Wuppertal, D-42119 Wuppertal, Germany}
  \affiliation{IAS, J\"ulich Supercomputing Centre, Forschungszentrum J\"ulich, D-52425 J\"ulich, Germany}
   
\date{\today}

\begin{abstract}

We present lattice QCD calculations of nucleon electromagnetic form
factors using pion masses $m_\pi = 149$, 202, and 254 MeV and an
action with clover-improved Wilson quarks coupled to smeared gauge
fields, as used by the Budapest-Marseille-Wuppertal
collaboration. Particular attention is given to removal of the effects
of excited state contamination by calculation at three source-sink
separations and use of the summation and generalized
pencil-of-function methods. The combination of calculation at the
nearly physical mass $m_\pi = 149$ MeV in a large spatial volume
($m_\pi L_s = 4.2$) and removal of excited state effects yields
agreement with experiment for the electric and magnetic form factors
$G_E(Q^2)$ and $G_M(Q^2)$ up to $Q^2 = 0.5\text{ GeV}^2$.

\end{abstract}



\maketitle

\section{\label{sect:Introduction}Introduction}

 Electromagnetic form  factors are of great interest theoretically and experimentally because they specify fundamental aspects of the structure of nucleons. 
At low momentum transfer, they reveal the spatial distribution of charge and current. 
In a  non-relativistic system, the electric and magnetic form factors $G_E(Q^2)$ and $G_M(Q^2)$ defined below would be the Fourier transforms of the distribution of charge and magnetization in the nucleon and the rms  charge and magnetization radii would be given by their slopes at zero momentum transfer,
$\langle r^2\rangle_{E,M} = -6 G_{E,M}^\prime(0)/G_{E,M}(0)$.
Relativistically, these Sachs form factors may be regarded as three dimensional Fourier transforms of charge and current densities suitably defined in the Breit frame. In addition, 
Burkardt~\cite{Burkardt:2000za,Burkardt:2002hr} has shown that
the Dirac and Pauli form factors $F_1(Q^2)$ and $F_2(Q^2)$ also correspond to two dimensional Fourier transforms of transverse charge and current densities defined in the infinite momentum frame, complementing our knowledge of quark distributions in the infinite momentum frame from deep inelastic scattering.
 At sufficiently high momentum transfer $Q^2$,  asymptotic scaling sets in and elastic form factors follow simple counting rules based on the minimum number of gluon exchanges required to divide the momentum transfer  equally among all the  quarks in the hadron.  
 In the nucleon, at least two gluon exchanges are required so that the electric form factor  falls off as  $Q^{-4} $.   The scale determining the onset of asymptotic scaling is of great interest in non-perturbative QCD.
 
  Because of their fundamental physical content, electromagnetic form factors have  continued to be studied extensively  experimentally throughout the world as technology has improved, but even now, significant questions remain. The most  accurately measured form factor is the dominant $F_1(Q^2)$  form factor for the proton.  However, its slope at very low $Q^2$   is still uncertain.
One problem, which has generated considerable theoretical and experimental interest, is that there is a 7$\sigma$ discrepancy between the 2010 CODATA value~\cite{Mohr:2012tt}  for the rms charge radius measured using electron-proton elastic scattering and spectroscopy, and the smaller value recently measured using the Lamb shift in muonic hydrogen~\cite{Pohl:2010zza}. 
Another  problem is that phenomenological fits to experimental electron scattering form 
factors~\cite{Friedrich:2003iz,Arrington:2007ux} have been inconsistent with analyses based on dispersion theory~\cite{Hohler:1976ax,Mergell:1995bf,Belushkin:2006qa,Lorenz:2012tm}. Interestingly, the charge radius determined using dispersion theory agrees with  the Lamb shift result.
 Measurements of  $F_2(Q^2)$ using spin polarization~\cite{Milbrath:1997de,Pospischil:2001pp,Gayou:2001qd,Gayou:2001qt,Punjabi:2005wq}
differ significantly  from traditional measurements based on Rosenbluth separation. Athough  two-photon exchange processes contribute much more strongly to the backward cross section used in Rosenbluth separation than to polarization transfer~\cite{Arrington:2007ux},  there are not yet precise theoretical calculations of two photon exchange that resolve the discrepancy. To measure the two photon exchange contribution directly, experiments using     $e^+\text{--}\,p$  scattering, for which the relative contribution of the two-photon term changes sign, have been performed by the CLAS experiment at Jeffserson Lab Hall B~\cite{Bennett:2012zza,Moteabbed:2013isu}, at the VEPP-3 Storage ring  in Novosibirsk~\cite{Gramolin:2011tr,Nikolenko_2014},  and by the OLYMPUS experiment  at the DORIS storage ring at DESY~\cite{Milner:2012zz,Milner20141}, although none of the three has published final results.
Finally, neutron form factors are less accurately determined
 than proton form factors because of uncertainty in nuclear wave functions for deuterium
  or $^3\mathrm{He}$.   Hence, for all these reasons, definitive lattice calculations can play an important role in resolving significant experimental uncertainties.

The Dirac and Pauli form factors, $F_1^q(Q^2)$ and $F_2^q(Q^2)$,  parameterize matrix elements of the
vector current between proton states:
\begin{equation}
  \label{eq:formfac_def}
  \langle \vec p\,',\lambda '|V^\mu_q|\vec p,\lambda\rangle = \bar u(\vec p\,',\lambda ')\left[ \gamma^\mu F_1^q(Q^2) + \frac{i\sigma^{\mu\nu}(p'-p)_\nu}{2 m_N} F_2^q(Q^2)\right] u(p,\lambda),
\end{equation}
where $Q^2=-(p'-p)^2$ and $V^\mu_q=\bar q\gamma^\mu q$. In comparing
with experiment, we also consider  form factors of the electromagnetic
current $V^\mu_\text{em}=\frac{2}{3}\bar u\gamma^\mu u -
\frac{1}{3}\bar d\gamma^\mu d$ in a proton and in a neutron,
$F_{1,2}^{p,n}(Q^2)$. Isovector and isoscalar form factors are defined by
\begin{gather}
F_{1,2}^v(Q^2) = F_{1,2}^p(Q^2) - F_{1,2}^n(Q^2) = F_{1,2}^u(Q^2) - F_{1,2}^d(Q^2) \equiv F_{1,2}^{u-d}(Q^2)\\
F_{1,2}^s(Q^2) = F_{1,2}^p(Q^2) + F_{1,2}^n(Q^2) = \frac{1}{3}\left(F_{1,2}^u(Q^2) + F_{1,2}^d(Q^2)\right) \equiv \frac{1}{3}F_{1,2}^{u+d}(Q^2).
\end{gather}

 The electric and magnetic Sachs form factors  $G_E(Q^2)$ and $G_M(Q^2)$ are defined by:
\begin{align}
G_E(Q^2) & = F_1(Q^2) - \frac{Q^2}{(2 m_N)^2} F_2(Q^2) \\
G_M(Q^2) & = F_1(Q^2) +  F_2(Q^2)\, .
\end{align}

Electromagnetic form factors have previously been calculated in
lattice QCD using a variety of actions, but so far using pion masses
substantially higher than the physical pion mass. Early calculations
have been described in review
articles~\cite{Arrington:2006zm,Hagler:2009ni}, including the
pioneering calculations of nucleon electric~\cite{Martinelli:1988rr}
and magnetic~\cite{Draper:1989pi} form factors using quenched
fermions, as well as later quenched
calculations~\cite{Leinweber:1990dv,Wilcox:1991cq,Gockeler:2003ay,Dong:1997xr,Tang:2003jh,Boinepalli:2006xd,Alexandrou:2006ru}. Calculations
with $N_f = 2$ flavors have been performed using
Wilson~\cite{Alexandrou:2006ru}, clover-improved
Wilson~\cite{Gockeler:2007hj,Collins:2011mk}, domain
wall~\cite{Lin:2008uz}, and twisted
mass~\cite{Alexandrou:2009xk,Alexandrou:2011db} actions.  $N_f = 2+1$
calculations have used clover-improved
Wilson~\cite{Shanahan:2014uka,Shanahan:2014cga} and domain
wall~\cite{Syritsyn:2009mx,Yamazaki:2009zq,Lin:2014saa} actions, and a
mixed action with domain wall valence quarks and Asqtad sea
quarks~\cite{Hagler:2007xi,Bratt:2010jn}.  Finally, calculations with
$N_f = 2+1+1$ flavors have been performed using twisted mass
action~\cite{Alexandrou:2013joa} and a mixed action with
clover-improved Wilson valence quarks and HISQ sea
quarks~\cite{Bhattacharya:2013ehc}.

This present work advances the calculation of electromagnetic form factors using lattice QCD in two crucial ways. One essential advance is calculation at the nearly physical pion mass of 149 MeV.  Previous calculations referenced above  clearly show that for large pion masses, the form factors   $F_1(Q^2)$ and $F_2(Q^2)$
at low $Q^2$ lie significantly above the physical values and monotonically decrease toward them as the pion mass is decreased. This behavior is clear physically, because the size of the pion cloud increases strongly as the pion mass decreases so that the rms radius and consequently the slope of the form factor at $Q^2 = 0$  increase strongly. Quantitatively, the dramatic increase in the  isovector Dirac radius as the pion mass decreases arises from the $\log(m_\pi)$ term in chiral perturbation theory.
The second crucial advance is  the removal of contamination due to excited states. Having already seen~\cite{Green:2012ud}  the importance of the removal of excited state contaminants in obtaining agreement with experiment for  the radii $(r_{1,2}^2)^v$, it is clearly important to do the same for the full $Q^2$  dependence and we do this using two methods described below. The removal of excited state contaminants in form factors has also been addressed recently~\cite{Capitani:2012ef,Jager:2013kha}  for form factors calculated with $N_f=2$ Wilson-clover fermions at $m_\pi \geq 195$ MeV. We find that the combination of calculation at the nearly physical mass of 149 MeV and removal of contamination due to excited states produces excellent agreement with experiment.

The outline of the paper is as follows. Section~\ref{sect:lqcd_method} presents the lattice methodology, beginning with the description of the clover-improved Wilson action from the Budapest-Marseille-Wuppertal (BMW) collaboration and the ensembles of configurations that are used. Three methods of calculating the relevant matrix elements of the electromagnetic current are then described,
 the standard ratio method, the summation method, and the generalized pencil-of-function (GPoF) method,  from which form factors are  extracted by an overdetermined analysis  to minimize the statistical uncertainty.
 In section~\ref{sec:isovector}, we present our results for isovector  observables.  Dirac form factors  
$F_1^v(Q^2)$  and Pauli form factors  $F_2^v(Q^2)$  are calculated for  ensembles  with a range of pion masses and results  using the ratio, summation, and GPoF methods are compared.
 For use in calculating rms radii, dipole fits to these form factors are performed for several ranges of $Q^2$  and compared to establish insensitivity to the  $Q^2$ range  for sufficiently low $Q^2$. In one of the highlights of this work, Sachs  form factors, $G_E(Q^2)$ and $G_M(Q^2)$,  are calculated at the lowest pion mass, 149 MeV,  and shown to produce excellent agreement with phenomenological fits to electron scattering data.  The Dirac radius, $(r_1^2)^v$, Pauli radius,   $(r_2^2)^v$ and anomalous magnetic moment $\kappa^v$ are calculated for ensembles with a range  of pion masses and chirally extrapolated to the physical pion mass.
 Section~\ref{sec:isoscalar} presents analogous results for isoscalar observables. Finally, we show the proton Sachs form factors in section~\ref{sec:proton} and present our conclusions in section~\ref{sec:conclusions}.

We include three appendices. Appendix~\ref{app:chpt} gives details on chiral
extrapolation formulae and phenomenological inputs for isovector observables.
Appendix~\ref{app:exc_states} includes additional plots comparing, for
observables where this was omitted in the main text, results computed on
each ensemble using the ratio, summation, and GPoF methods; this is
intended to be useful for others performing similar lattice
QCD calculations. Finally, Appendix~\ref{app:table} has tables listing form
factors for four ensembles.

\section{\label{sect:lqcd_method}Lattice methodology}

\subsection{Lattice action and gauge ensembles}
We perform lattice QCD calculations using a tree-level Symanzik-improved
gauge action and 2+1 flavors of tree-level improved Wilson-clover
quarks, which couple to the gauge links via two levels of HEX
smearing as motivated by Ref.~\cite{Capitani:2006ni}.
For a detailed description of the action and smearing procedure we refer
the reader to~\cite{Durr:2010aw}. The $s$ quarks are tuned to have a mass
close to physical, and the light quark mass (with $m_u=m_d$) is
varied, yielding pion masses between 149 and 356~MeV. The algorithms
used to generate the gauge field ensembles are described in~\cite{Durr:2010aw}.

\begin{table}
  \caption{\label{tab:gauge_ens}Gauge configuration ensembles and
    measurement counts for form factor calculations. The coarse ensembles
    have gauge coupling $\beta=3.31$ and bare strange quark mass
    $am_s=-0.04$, while the fine ensemble has $\beta=3.5$ and $am_s=-0.006$.}
  \begin{tabular}{cl|ccccD{.}{.}{2}|rr}
    \hline\hline
    $m_\pi\text{ [MeV]}$ &  $m_N\text{ [GeV]}$ &  
      $a\text{ [fm]}$ &  $am_{ud}$ & $L_s^3\times L_t$ &  $m_\pi L_s$ &
      \multicolumn{1}{c}{$m_\pi L_t$} &
      $N_\text{conf}$ &  $N_\text{meas}$ \\
    \hline
    149(1) &  0.929(19) &  
      0.116 & $-0.09900$ & $48^3\times48$ & 4.21 & 4.21 &
      646 &  7752 \\
    202(1) & 1.003(22) &
      0.116 & $-0.09756$ & $32^3\times48$ & 3.81 & 5.71 &
      457 & 5484 \\
    253(1) & 1.030(23)  &
      0.116 & $-0.09530$ & $32^3\times96$ & 4.78 & 14.34 &
      202 & 2424 \\
    254(1) & 1.051(13)  & 
      0.116 & $-0.09530$ & $32^3\times48$ & 4.79 & 7.18 &
      420 & 5040 \\
    254(1) & 1.041(15) &
      0.116 & $-0.09530$ & $32^3\times24$ & 4.79 & 3.59 &
      2074 & 12444 \\
    254(1) & 1.072(7)  &
      0.116 & $-0.09530$ & $24^3\times48$ & 3.60 & 7.19 &
      1019 &  24456 \\
    252(2) & 1.072(7) &
      0.116 & $-0.09530$ & $24^3\times24$ & 3.56 & 3.56 &
      3999 & 23994 \\
    303(2) & 1.043(51)  & 
      0.116 & $-0.09300$ & $24^3\times48$ & 4.28 & 8.56 &
      128 & 768 \\
    317(2) & 1.153(20) &
      0.093 & $-0.04630$ & $32^3\times64$ & 4.76 & 9.52 &
      103 & 824 \\
    356(2) & 1.175(18) & 
      0.116 & $-0.09000$ & $24^3\times48$ & 5.04 & 10.08 &
      127 & 762 \\
    351(2) & 1.163(13) & 
      0.116 & $-0.09000$ & $24^3\times24$ & 4.97 & 4.97 &
      420 & 2520 \\
    \hline\hline
  \end{tabular}
\end{table} 

\begin{figure}
  \centering\includegraphics[width=0.7\textwidth]{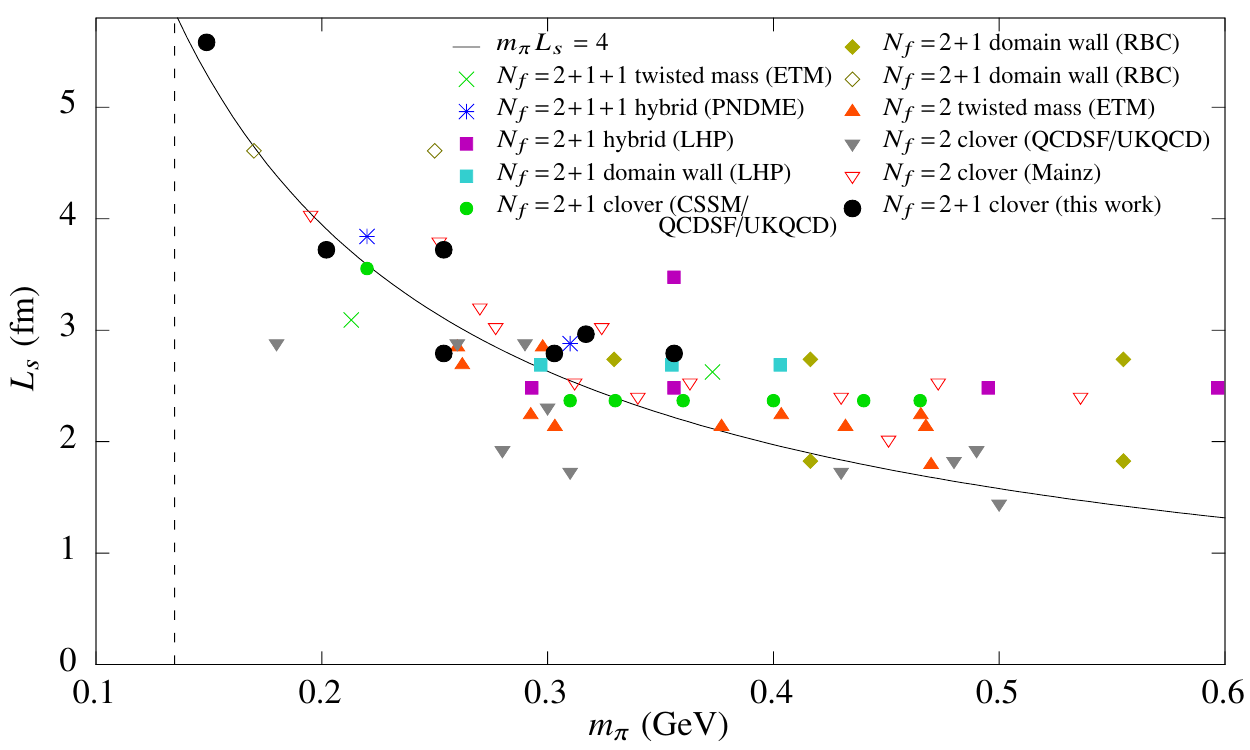}
  \caption{\label{fig:mpi_vs_L}Summary of pion masses and volumes in
    existing lattice calculations of nucleon structure. Open symbols
    are used to indicate cases where results were described by the
    authors as ``preliminary''.}
\end{figure}

In Tab.~\ref{tab:gauge_ens} we list the gauge ensembles analyzed in this paper.
In Fig.~\ref{fig:mpi_vs_L} we show $(m_\pi,L)$ values in comparison to other 
lattice calculations of nucleon structure~\cite{Yamazaki:2009zq, Syritsyn:2009mx, Bratt:2010jn, Collins:2011mk, Alexandrou:2011db, Alexandrou:2013joa, Bhattacharya:2013ehc, Lin:2014saa, Jager:2013kha,Shanahan:2014uka,Shanahan:2014cga}.
We check the volume dependence of our results at $m_\pi=254\text{ MeV}$ 
by varying the physical volume from $(3.7\text{ fm})^3$ to $(2.8\text{ fm})^3$.
We perform all calculations with $a=0.116\text{ fm}$ except 
one with $m_\pi=317\text{ MeV}$ and $a=0.093\text{ fm}$ to check for 
discretization effects.
In addition, at $m_\pi\approx250$ and $\approx350\text{ MeV}$ we vary the time extent of the
lattices between $2.8\text{ fm}$ and $11.1\text{ fm}$ to check whether thermal
states~\cite{Beane:2009kya} have any effect on the nucleon structure observables that we
calculate.

\subsection{\label{sec:ff_extract}Computation of matrix elements}
In order to measure nucleon matrix elements in lattice QCD, we compute
nucleon two-point and three-point functions,
\begin{gather}
  C_\text{2pt}(\vec p, t) = \sum_{\vec x} e^{-i\vec p\cdot\vec x} \Tr [\Gamma_\text{pol} \la N(\vec x, t) \bar N(\vec 0, 0) \ra ]\\
  C_\text{3pt}^{V^\mu_q}(\vec p,\vec p\,',\tau,T) = \sum_{\vec x,\vec y} e^{-i\vec p\,'\cdot\vec x}e^{i(\vec p\,'-\vec p)\cdot y} \Tr[\Gamma_\text{pol} \la N(\vec x,T) V^\mu_q(\vec y,\tau) \bar N(\vec 0, 0) \ra ],
\end{gather}
where $N=\epsilon^{abc}(\tilde
u^T_aC\gamma_5\frac{1+\gamma_4}{2}\tilde d_b)\tilde u_c$ is a proton
interpolating operator constructed using smeared quark fields,
$V^\mu_q=\bar q\gamma^\mu q$ is the site-local vector current, and
$\Gamma_\text{pol}=\frac{1+\gamma_4}{2}\frac{1-i\gamma_3\gamma_5}{2}$
is a spin and parity projection matrix. For smearing, we use
approximately-Gaussian Wuppertal smearing~\cite{Gusken:1989qx} with
the same double-HEX-smeared links as used for the fermion action.
We compute $C_\text{3pt}$ with
both $\vec p\,'=\vec 0$ and $\vec p\,'=\frac{2\pi}{L_s}(-1,0,0)$, and
for quark flavors $q\in\{u,d\}$. The three-point correlators have
contributions from both connected and disconnected quark contractions,
but we compute only the connected part. Omitting the disconnected part
(where the vector current is attached to a quark loop) introduces an
uncontrolled systematic error except when taking the $u-d$ (isovector)
flavor combination, where the disconnected contributions cancel
out. The magnitude of disconnected contributions is discussed in the
conclusions (Sec.~\ref{sec:conclusions}).

On a lattice with finite time extent $L_t$, the transfer matrix
formalism yields
\begin{gather}
  C_\text{2pt}(\vec p, t) = \sum_{n,m} e^{-E_m L_t}e^{-(E_n-E_m)t} \sum_{\alpha,\beta} (\Gamma_\text{pol})_{\alpha\beta} \sum_{\vec x} e^{-i\vec p\cdot\vec x}\la m|N_\beta(\vec x)|n\ra \la n|\bar N_\alpha(\vec 0)|m\ra \\
\begin{aligned}
  C_\text{3pt}^{V^\mu_q}(\vec p,\vec p\,',\tau, T) &= \sum_{n,n',m} e^{-E_m L_t}e^{-(E_n-E_m)\tau}e^{-(E_{n'}-E_m)(T-\tau)} \sum_{\alpha\beta}(\Gamma_\text{pol})_{\alpha\beta} \\ &\qquad\times\sum_{\vec x,\vec y}e^{-i\vec p\,'\cdot\vec x}e^{i(\vec p\,'-\vec p)\cdot y}\la m|N_\beta(\vec x)|n'\ra \la n'|V^\mu_q(\vec y)|n\ra\la n|\bar N_\alpha(\vec 0)|m\ra.
\end{aligned}
\end{gather}
Thermal contamination is eliminated in the large $L_t$
(zero-temperature) limit, in which state $m$ is the vacuum, and states
$n$ and $n'$ are restricted to having the quantum numbers of a proton
with momentum $\vec p$ and $\vec p\,'$, respectively. Unwanted
contributions from excited states can be eliminated by then taking
$\tau$ and $T-\tau$ to be large.

In order to compute $C_\text{3pt}$, we use sequential propagators
through the sink~\cite{Martinelli:1988rr}. This has
the advantage of allowing for any operator
to be measured at any time using a fixed set of quark propagators, but
new backward propagators must be computed for each source-sink
separation $T$. Increasing $T$ suppresses excited-state contamination,
but it also increases the noise; the signal-to-noise ratio is expected
to decay asymptotically as
$e^{-(m_N-\frac{3}{2}m_\pi)T}$~\cite{Lepage:1989hd}. Past calculations
have often used a single source-sink separation, which only allows for
a limited ability to identify and remove excited state
contamination. In particular, when computing forward matrix elements,
there is no way of distinguishing contributions from excited states
with $n'=n$ from the ground state contribution, when using
$C_\text{3pt}$ with a single $T$. Therefore, in this work, we perform
measurements using three source-sink separations on all ensembles:
$T/a\in\{8,10,12\}$ for the coarse lattices and $T/a\in\{10,13,16\}$
for the fine lattice.

\subsubsection{Ratio method}
We label proton states as $|\vec p,\lambda\ra$ and use the relativistic
normalization, $\la \vec p\,',\lambda'|\vec p,\lambda\ra = 2E L_s^3
\delta_{\vec p\,',\vec p} \delta_{\lambda',\lambda}$. Parameterizing the
overlap of our interpolating operator with the ground-state proton as
$\la \Omega|N_\alpha(\vec x)|\vec p,\lambda\ra = \sqrt{Z(\vec p)}
u_\alpha(\vec p,\lambda) e^{i\vec p\cdot\vec x}$, at zero temperature
we obtain
\begin{gather}
  C_\text{2pt}(\vec p,t) = \frac{Z(\vec p)e^{-E(\vec p)t}}{2E(\vec p)}\Tr[\Gamma_\text{pol}(i\slashed{p}+m_N)] + O(e^{-\Delta E_{10}(\vec p)t}) \\
\begin{aligned}
  C_\text{3pt}^{V^\mu_q}(\vec p,\vec p\,',\tau,T) &= \frac{\sqrt{Z(\vec p)Z(\vec p\,')}e^{-E(\vec p)\tau-E(\vec p\,')(T-\tau)}}{4 E(\vec p\,')E(\vec p)} \sum_{\lambda,\lambda'} \bar u(\vec p,\lambda)\Gamma_\text{pol} u(\vec p\,',\lambda') \la p',\lambda'|V^\mu_q|p,\lambda\ra \\&\qquad + O(e^{-\Delta E_{10}(\vec p)\tau}) + O(e^{-\Delta E_{10}(\vec p\,')(T-\tau)}),
\end{aligned}
\end{gather}
where $\Delta E_{10}(\vec p)$ is the energy gap between the ground and
lowest excited state with momentum $\vec p$. To cancel the overlap
factors and the depedence on Euclidean time, we compute the
\emph{ratios},
\begin{equation}
\begin{aligned}\label{eq:ratio}
  R^\mu_q(\tau,T) &= \frac{C_\text{3pt}^{V^\mu_q}(\vec p,\vec p\,',\tau,T)}{\sqrt{C_\text{2pt}(\vec p,T)C_\text{2pt}(\vec p\,',T)}}\sqrt{\frac{C_\text{2pt}(\vec p,T-\tau)C_\text{2pt}(\vec p\,',\tau)}{C_\text{2pt}(\vec p\,',T-\tau)C_\text{2pt}(\vec p,\tau)}} \\
  &= \frac{\sum_{\lambda,\lambda'} \bar u(\vec p,\lambda)\Gamma_\text{pol} u(\vec p\,',\lambda') \la p',\lambda'|V^\mu_q|p,\lambda\ra}{\sqrt{2E(\vec p)(E(\vec p)+m_N)\cdot 2E(\vec p\,')(E(\vec p\,')+m_N)}} + O(e^{-\Delta E_{10}(\vec p)\tau}) + O(e^{-\Delta E_{10}(\vec p\,')(T-\tau)}).
\end{aligned}
\end{equation}
As a function of $\tau\in[0,T]$ with fixed $T$, the ratios produce a
plateau with ``tails'' at both ends caused by excited states. In
practice, for each fixed $T$, we average over the central two or three
points near $\tau=T/2$, which allows for matrix elements to be
computed with errors that decay asymptotically as $e^{-\Delta
  E_\text{min}T/2}$, where $\Delta E_\text{min}=\min\{\Delta
E_{10}(\vec p),\Delta E_{10}(\vec p\,')\}$.

\subsubsection{Summation method}
Improved asymptotic behavior of excited-state contributions can be
achieved by using the summation method
\cite{Capitani:2010sg,Bulava:2010ej}. Taking the sums of ratios yields
\begin{equation}\label{eq:summation}
  S(T) \equiv \sum_{\tau=\tau_0}^{T-\tau_0} R(\tau,T) = c + TM + O(Te^{-\Delta E_\text{min}T}),
\end{equation}
where $c$ is independent of $T$, and $M$ contains the desired
ground-state matrix element. (We choose $\tau_0=1$ and thus omit the
first and last points of each plateau.) Thus finite differences,
$(\delta T)^{-1}(S(T+\delta T)-S(T))$, yield the ground-state matrix
element with excited-state contamination that asymptotically decays as
$Te^{-\Delta E_\text{min}T}$. In particular, transitions between the
ground and lowest excited state, which were the dominant excited-state
contribution for the ratio method at large time separations, are
highly suppressed, now decaying as $e^{-\Delta E_\text{min}T}$.

With our three source-sink separations,
we can compute this finite difference at two values of $T$, however
the result at the larger value of $T$ tends to have very large
statistical uncertainties. Instead of using a finite difference, we
fit a line $a + bT$ to our three $S(T)$ points, and take the slope $b$
as the extracted matrix element. The result is mostly determined from
the lower two source-sink separations, as their sums have smaller
errors, but choosing this fit over a finite difference allows the
larger source-sink separation to also have some influence.

\subsubsection{Generalized pencil-of-function method}
By using $n$ interpolating operators, the variational method
\cite{Luscher:1990ck,Blossier:2009kd} allows for asymptotically
removing the unwanted contributions from the first $n-1$ excited
states. We are able to make use of the variational method via the
generalized pencil-of-function (GPoF) method \cite{Aubin:2010jc},
which is based on the recognition that if $N(t)$ and $\bar N(t)$ are
our interpolating operators for annihilating and creating the nucleon,
then the time-displaced operators
\begin{align}
  N^\delta(t) &\equiv e^{H\delta}N(t)e^{-H\delta} = N(t+\delta)\\
\bar N^\delta(t) &\equiv e^{-H\delta}\bar N(t)e^{H\delta} = \bar N(t-\delta)
\end{align}
are linearly independent interpolating operators for the nucleon. This
enables us to construct a matrix of two-point functions,
\begin{equation}
  \mathbf{C}_\text{2pt}(t) =
  \begin{pmatrix}
    \langle N(t)\bar N(0)\rangle & \langle N^\delta(t)\bar N(0)\rangle \\
    \langle N(t)\bar N^\delta(0)\rangle & \langle N^\delta(t)\bar N^\delta(0)\rangle
  \end{pmatrix} =
  \begin{pmatrix}
    C_\text{2pt}(t) & C_\text{2pt}(t+\delta) \\
    C_\text{2pt}(t+\delta) & C_\text{2pt}(t+2\delta)
  \end{pmatrix},
\end{equation}
using our ordinary two-point function $C_\text{2pt}(t)$. By solving
the generalized eigenvalue problem,
\begin{equation}\label{eqn:gevp}
  \mathbf{C}_\text{2pt}(t) \mathbf{v}(t_0,t) = \lambda(t_0,t) \mathbf{C}_\text{2pt}(t_0) \mathbf{v}(t_0,t)
\end{equation}
we can find eigenvectors $\mathbf{v}(t_0,t)$ that asymptotically give linear
combinations of $N$ and $N^\delta$ which have zero overlap with the
first excited state. Then using also the matrix of three-point
functions,
\begin{equation}
  \mathbf{C}_\text{3pt}(\tau,T) =
  \begin{pmatrix}
    \langle N(T) \mathcal{O}(\tau) \bar N(0)\rangle &
    \langle N^\delta(T) \mathcal{O}(\tau) \bar N(0)\rangle \\
    \langle N(T) \mathcal{O}(\tau) \bar N^\delta(0)\rangle &
    \langle N^\delta(T) \mathcal{O}(\tau) \bar N^\delta(0)\rangle
  \end{pmatrix} =
  \begin{pmatrix}
    C_\text{3pt}(\tau,T) & C_\text{3pt}(\tau,T+\delta) \\
    C_\text{3pt}(\tau+\delta,T+\delta) & C_\text{3pt}(\tau+\delta,T+2\delta)
  \end{pmatrix},
\end{equation}
we compute two-point and three-point functions using a particular
linear combination:
\begin{equation}\label{eqn:gpof_c2_c3}
  C^\text{GPoF}_\text{2pt}(t) = \mathbf{v}^\dagger \mathbf{C}_\text{2pt}(t) \mathbf{v} \qquad 
  C^\text{GPoF}_\text{3pt}(\tau,T) = \mathbf{v}^\dagger \mathbf{C}_\text{3pt}(\tau,T) \mathbf{v},
\end{equation}
and then proceed with the usual ratio-plateau analysis. Note that this
requires computing three-point functions at three equally spaced
source-sink separations, which is precisely what we have, and thus we
can only compute $C^\text{GPoF}_\text{3pt}(\tau,T)$ at our shortest
source-sink separation $T$.

Consider, for example, a $2\times2$ GPoF analysis applied to a system with 
exactly two states, $E_0$ and $E_1$.
It is trivial to show that the eigenstates $\lambda(t_0,t)$ in Eq.~(\ref{eqn:gevp})
are equal to $e^{-E_{0,1}(t-t_0)}$ and the eigenvectors are
${\mathbf v}^T_{0,1}=(-e^{-E_{1,0}\delta}, 1)$.
Substituting the ground state eigenvector $\mathbf{v}_0$ into Eq.~(\ref{eqn:gpof_c2_c3}), 
we obtain
\begin{equation}
\begin{aligned}
  C^\text{GPoF}_\text{2pt}(t) 
    &= C_\text{2pt}(t+2\delta) - 2e^{-E_1 \delta} C_\text{2pt}(t+\delta) 
      + e^{-2 E_1 \delta} C_\text{2pt}(t)\,,\\
  C^\text{GPoF}_\text{3pt}(\tau,T) 
    &= C_\text{3pt}(\tau+\delta,T+2\delta) 
      - e^{-E_1 \delta} \big(C_\text{3pt}(\tau,T+\delta) +
        C_\text{3pt}(\tau+\delta,T+\delta)\big)
      + e^{-2 E_1 \delta} C_\text{3pt}(\tau,T)\,,\\
\end{aligned}
\end{equation}
indicating that, if  computed using the GPoF method, the ground state matrix 
elements and their uncertainties will be mostly determined by the values of 
correlators with the largest separation $T$.

\begin{figure}
  \centering
  \includegraphics[width=0.495\columnwidth]{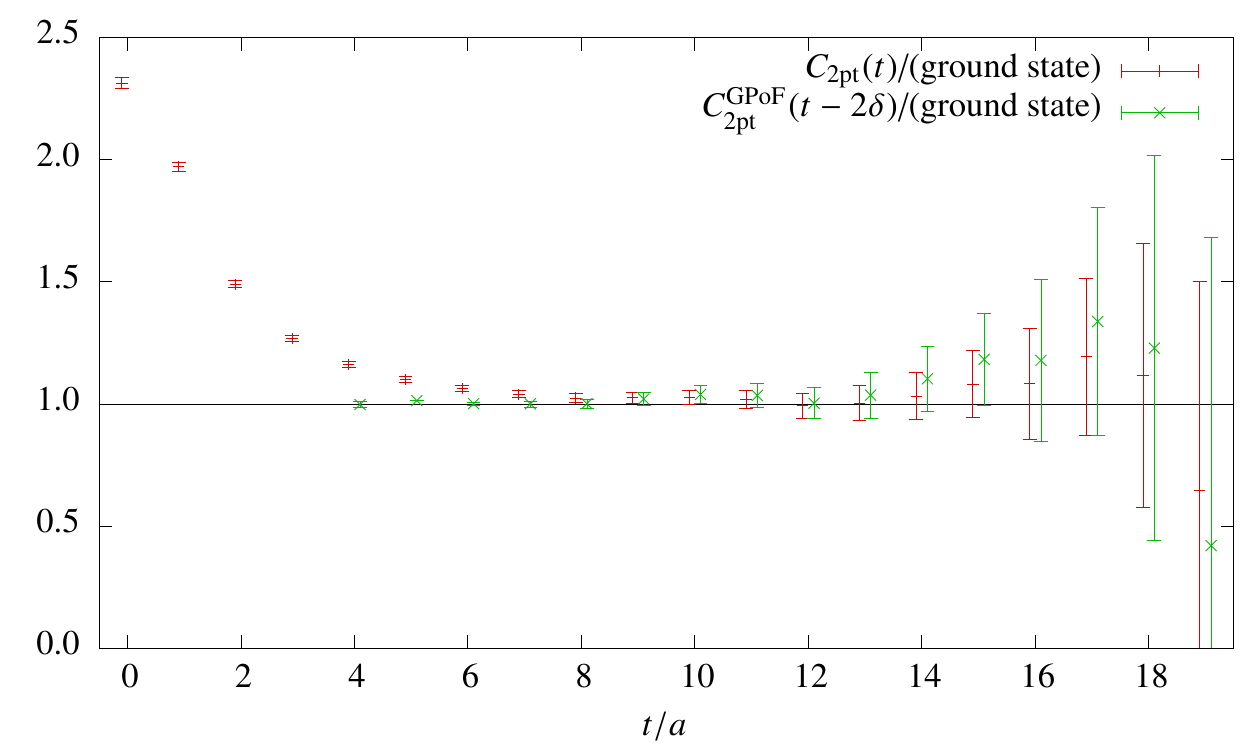}
  \includegraphics[width=0.495\columnwidth]{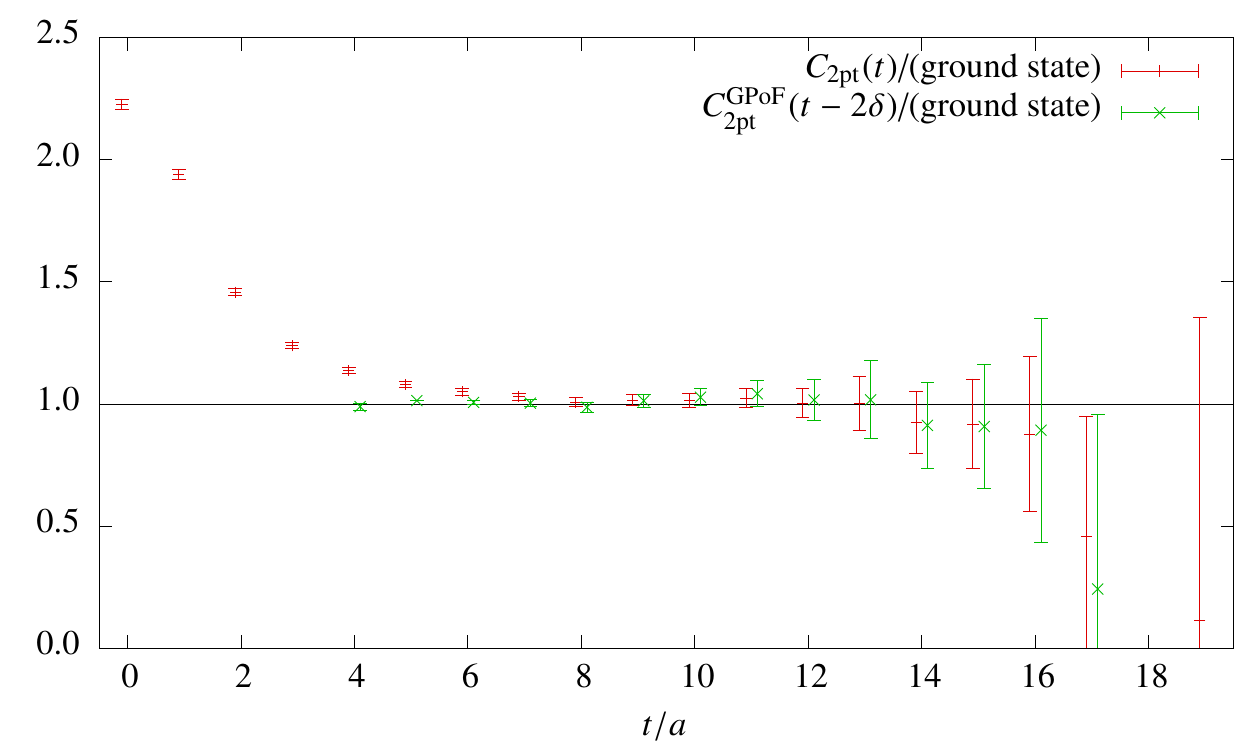}
  \caption{Two-point correlators $C_\text{2pt}(t,\vec p)$ and
    $C_\text{2pt}^\text{GPoF}(t,\vec p)$ for the 149~MeV ensemble,
    divided by their ground-state contributions, as determined from a
    two-state fit to the former with $t/a\in [3,14]$ and a one-state
    fit to the latter with $t/a\in [2,12]$. The left plot has $\vec
    p=\vec 0$, and the right plot is averaged over $a\vec p$
    equivalent to $\frac{2\pi}{48}(2,1,1)$, which is the largest used
    for computing matrix elements on this ensemble. The GPoF
    correlators are shifted to show that at large times their dominant
    contribution is from the time-displaced nucleon interpolating
    operator.}
  \label{fig:proton_c2_gpof}
\end{figure}

In practice, for each class of lattice momenta $\vec p$ equivalent
under the group of lattice rotations and reflections, we average the
two-point correlators $C_\text{2pt}(t,\vec p)$ and then use the GPoF
method and solve the generalized eigenvalue problem. This produces a
different linear combination of the original and the time-displaced
nucleon operator for each class of equivalent lattice momenta. It has
been shown~\cite{Blossier:2009kd} that by appropriately increasing
$t_0$ and $t$ as $\tau$ and $T-\tau$ are increased, the contributions
from the lowest-lying excited state can be completely removed
asymptotically; however, in this work, we find the eigenvector using
the fixed values $t_0/a=1$ and $t/a=2$. As shown in
Fig.~\ref{fig:proton_c2_gpof}, this is sufficient to remove the effect
of excited-state contamination in $C_\text{2pt}^\text{GPoF}$ at the
present level of statistics.

\subsubsection{Comparison of methods for computing matrix elements}
Given our level of statistical error and that we have only three
source-sink separations, there are trade-offs between the three
methods for computing matrix elements:
\begin{itemize}
\item Although the ratio method has the worst asymptotic behavior, we
  are able to compute one result for each source-sink separation,
  which gives an indication of the approach to the ground-state matrix
  element.
\item The summation method asymptotically suppresses excited-state
  contributions without requiring knowledge about any particular
  state. In particular, this method is most effective at suppressing
  the contributions from transition matrix-elements between the ground
  state and an excited state.
\item If excited-state contributions to two-point and three-point
  functions come mostly from a single state, then (given sufficient
  statistics) the GPoF method is effective at removing them. In
  particular, this removal will include contributions to the two-point
  function and both ground-to-excited and excited-to-excited matrix
  elements in the three-point function.
\end{itemize}
The case of contamination from transition matrix elements is, in
particular, one where the GPoF method could in practice be not very
successful at removing the effect of excited states. Consider
an excited state with a small amplitude relative
to the ground state. That is, $r\equiv\sqrt{Z'/Z}$ is small, where $Z$ is
defined as above and $Z'$ is defined analogously for the excited
state. Then the contribution from this state to the
two-point function would be suppressed as $r^2$, such that it could
disappear into statistical noise. However, its contribution to
three-point functions via transitions to the ground state would only
be suppressed by the factor $r$. Since the GPoF method relies on the
two-point function for optimizing its effective interpolating
operator, it could fail to remove such excited-state contributions.

For a concise presentation, we select a single method for our primary
results, namely the summation method, as it is effective at
suppressing contributions from all excited states, is fairly simple,
and has been used successfully in computing the nucleon axial
charge~\cite{Capitani:2012gj,Jager:2013kha}. The GPoF method has not
seen widespread use, and our set of results using it should be
considered an exploratory study. We will see (in the main text and in
Appendix~\ref{app:exc_states}) that, with the present level of
statistics, results using the ratio (with the largest source-sink
separation), summation, and GPoF methods are consistent with one
another and therefore this choice does not have a significant effect
on the results.

The two main methods not considered here are multi-state fitting and
broader application of the variational method with different
interpolating operators (beyond just the time-displacements used by
GPoF). In recent years, the former has been
applied to nucleon matrix elements in
Refs.~\cite{Green:2011fg,Bhattacharya:2013ehc,Bali:2013nla,Hippel:2014kta},
typically with the assumption that only two states contribute in the
range of probed time separations. The latter has been used extensively
in spectroscopy calculations together with a large number of
interpolating operators; see, e.g., Ref.~\cite{Edwards:2012fx} for an
application to excited baryons. It has also seen some use for nucleon
matrix elements, such as the calculations in
Refs.~\cite{Owen:2012ts,Owen:2013pfa}. These used sequential
propagators from a fixed current rather than a fixed sink as was used
in this work; this reduces the cost of including several interpolating
operators, with the drawback of requiring additional propagators for
each current insertion.

\subsection{Extraction of form factors}
The renormalized $O(a)$-improved vector current is given
by~\cite{Luscher:1996iy}
\begin{equation}
(V^\mu_q)_R = Z_V(1 + b\,am_q)(V^\mu_q + c\,a\partial^\nu T^{\mu\nu}_q),
\end{equation}
where $T^{\mu\nu}_q=i\bar q\sigma^{\mu\nu}q$. For tree-level
improvement, as used in the lattice action, $b=1$ and $c=0$. We keep
the latter and use only the site-local vector current, but rather than
controlling the quark-mass dependence via two parameters $(Z_V,b)$, we
instead compute a separate $Z_V$ renormalization factor on each
ensemble.

\begin{figure}
  \centering
  \includegraphics[width=0.7\textwidth]{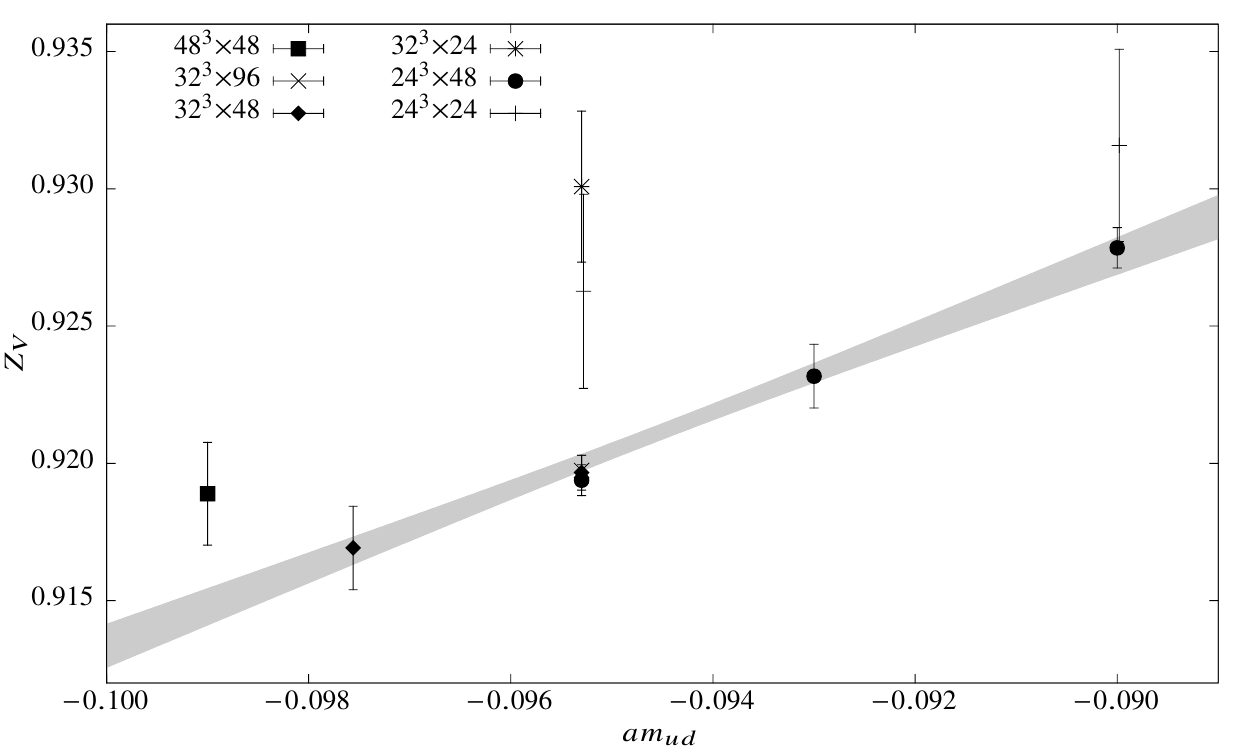}
  \caption{Vector current renormalization factor $Z_V$ versus bare quark
   mass, for the coarse ensembles. The band is from a fit assuming a
   linear relationship.}
  \label{fig:ZV_vs_mq}
\end{figure}

We do this by measuring the time-component of the vector current at
$\vec p\,'-\vec p=0$, which is (up to renormalization and lattice
artifacts) the quark number, a conserved charge. Specifically, we take
the $u-d$ flavor combination and, on each ensemble, impose that it
equals 1 for the proton in order to obtain $Z_V$. This should be
unaffected by excited states in the normal sense; any dependence on
time separations should only occur as a result of lattice artifacts or
thermal effects. We observe no statistically significant dependence on
source-sink separation in our data, and compute $Z_V$ on each ensemble
using the ratio-plateau method with the shortest source-sink
separation. For the coarse ensembles, this is shown in
Fig.~\ref{fig:ZV_vs_mq}. A linear fit has slope $b=1.42(13)$, which is
somewhat larger than the tree-level value.  We note that the ensembles
with small values of $m_\pi L_t$ tend to have values of $Z_V$ that lie
somewhat above the fit, suggesting the presence of some thermal
contamination; this shows up for $Z_V$ in particular because other
sources of uncertainty (including statistical) are smaller than in
other observables.  As the effect is at the percent level, it is
negligible compared to the statistical uncertainty that we later
obtain for electromagnetic form factors.

We do notice another clear apparent thermal effect: the
statistical uncertainty depends strongly on the time extent. Despite
other ensembles having many more measurements, the $32^3\times 96$
ensemble has the smallest uncertainty for $Z_V$. In addition, the three
$L_t=24a$ ensembles have the largest uncertainties for $Z_V$, and the
uncertainties grow more rapidly with the source-sink separation on the
ensembles with shorter time extent (not shown in Fig.~\ref{fig:ZV_vs_mq}).
The more-rapid onset of noise,
arising from the influence of thermal states, has been previously
examined for the case of (multi-)baryon two-point correlators in
Ref.~\cite{Beane:2009kya}.

To compute form factors: for each value of $Q^2$, we parameterize the
corresponding set of matrix elements of the vector current by
$F_1(Q^2)$ and $F_2(Q^2)$, and perform a linear fit to solve the
resulting overdetermined system of equations~\cite{Hagler:2003jd},
after first combining
equivalent matrix elements to improve the condition
number~\cite{Syritsyn:2009mx}. This approach makes use of all
available matrix elements in order to minimize the statistical
uncertainty in the resulting form factors. On our ensembles, the
largest source momentum that we use is $\vec
p=\frac{2\pi}{L_s}(1,1,1)$, except for the $m_\pi=149$~MeV ensemble,
where we use source momenta as large as $\vec
p=\frac{2\pi}{L_s}(1,1,2)$ to compensate for the larger volume.

\section{\label{sec:isovector}Isovector form factors}
Isovector lattice observables are particularly interesting because
they have no disconnected quark contractions and thus may be compared
directly to differences between proton and neutron experimental
results.

\subsection{Form factors}
\begin{figure}
  \centering
  \begin{minipage}{.5\textwidth}
    \centering
    \includegraphics[width=\textwidth]{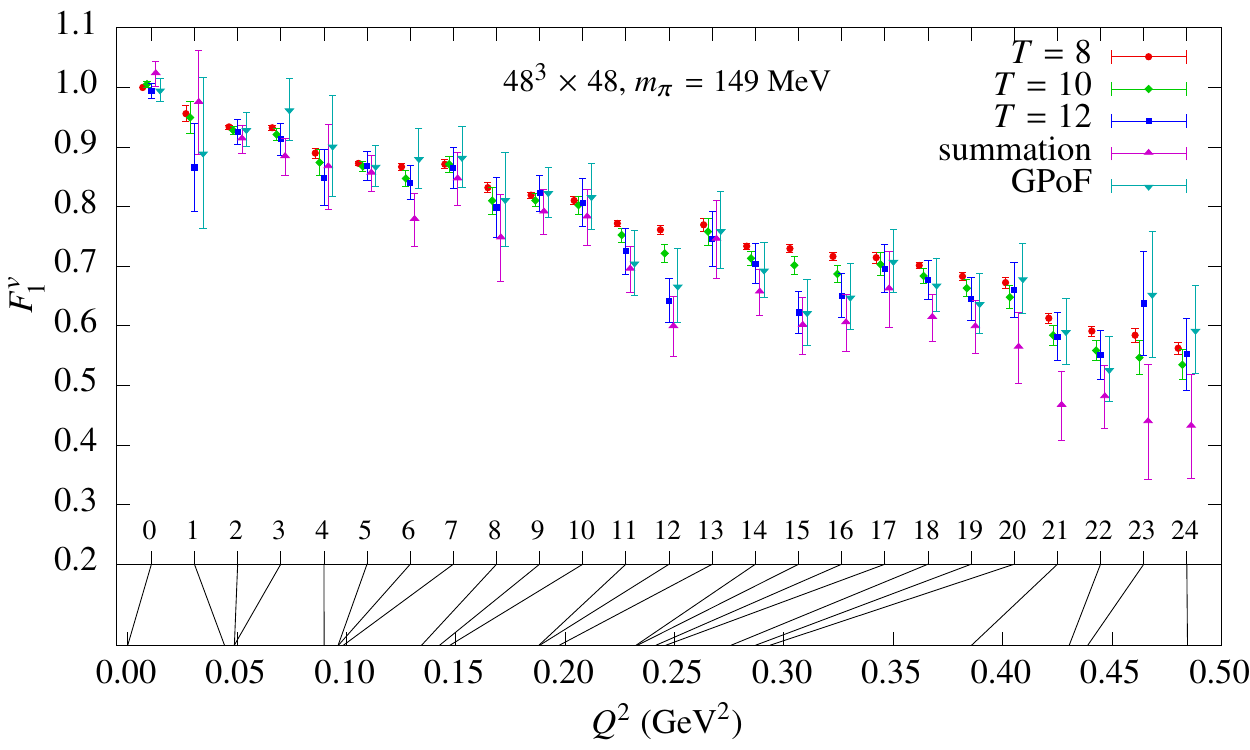}\\
    \includegraphics[width=\textwidth]{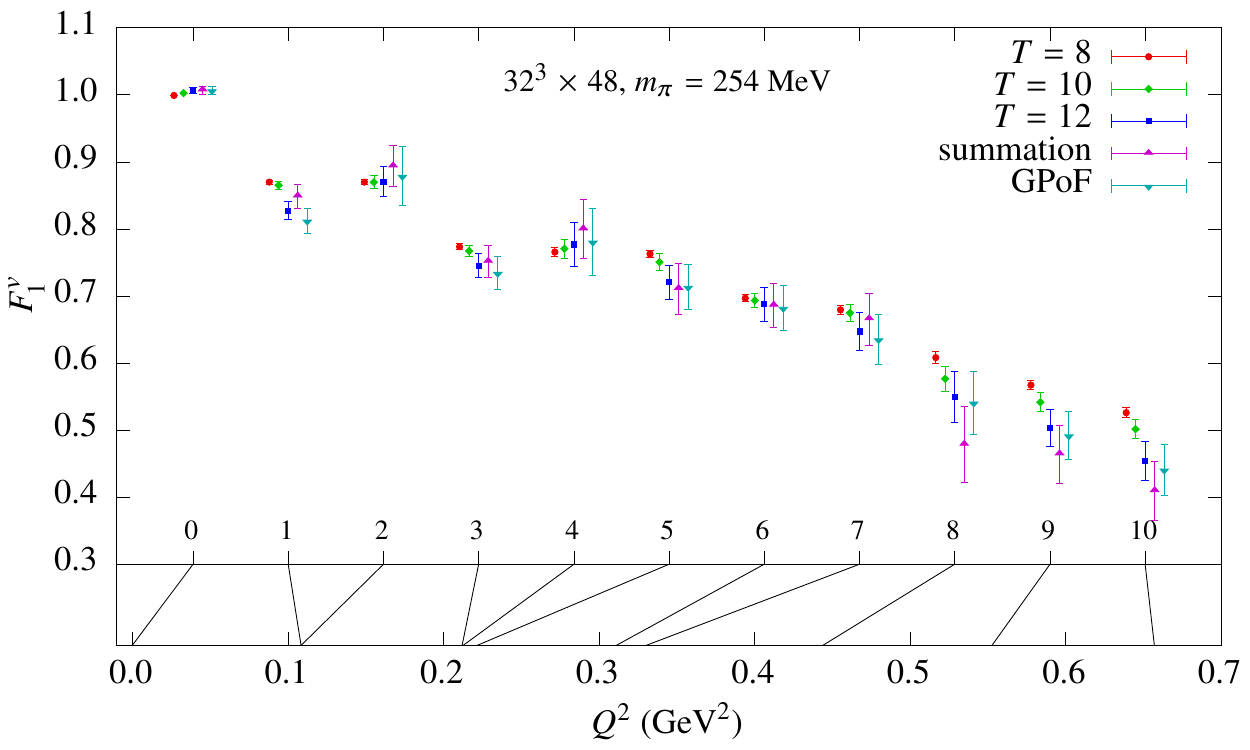}
  \end{minipage}~
  \begin{minipage}{.5\textwidth}
    \centering
    \includegraphics[width=\textwidth]{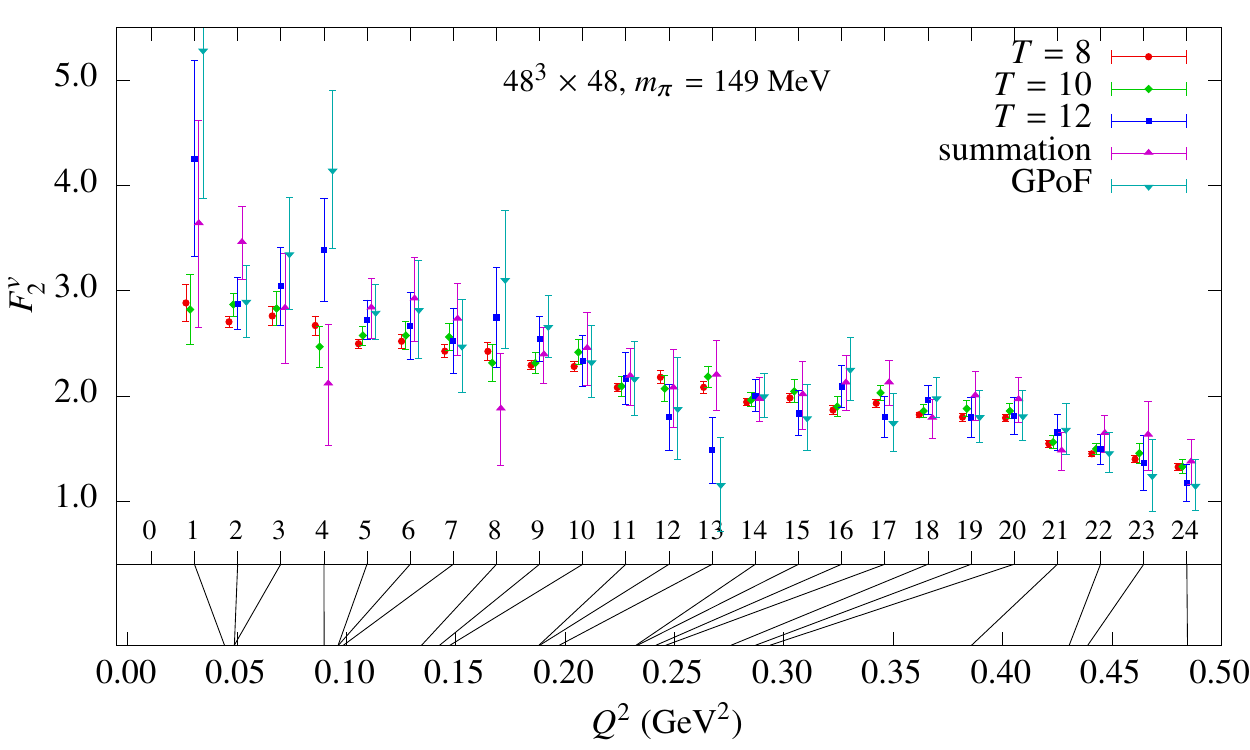}\\
    \includegraphics[width=\textwidth]{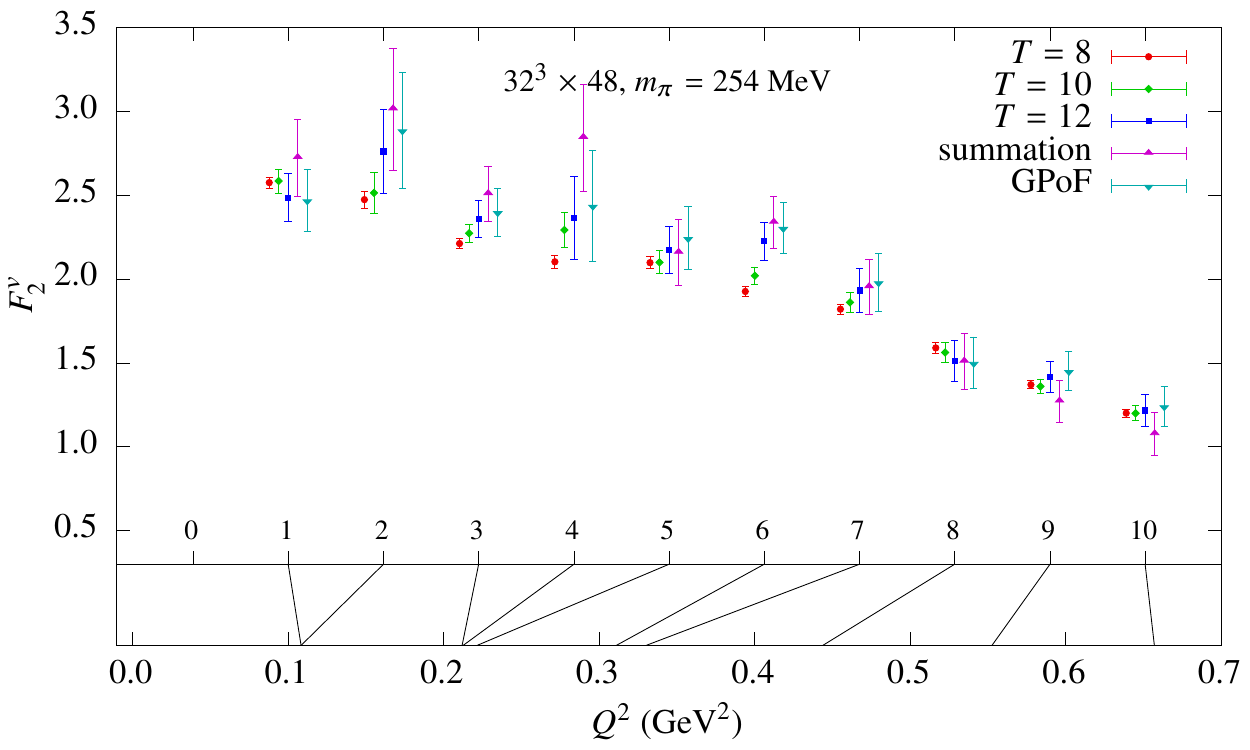}
  \end{minipage}
  \caption{\label{fig:ff_extract_comparison}Comparison
    of different methods to extract the ground state isovector form
    factors $F_1^v(Q^2)$ and $F_2^v(Q^2)$. The upper plots show
    the $m_\pi=149$~MeV ensemble and the lower plots show the
    $m_\pi=254$~MeV, $32^3\times 48$ ensemble.}
\end{figure}

We compute isovector Dirac and Pauli form factors using the different
methods discussed in Sec.~\ref{sec:ff_extract}, and results are shown
for two ensembles in Fig.~\ref{fig:ff_extract_comparison}. A clear
trend when going from the lowest to the middle source-sink separation
is seen for the $m_\pi=149$~MeV ensemble, where $F_1^v$ tends to
decrease and $F_2^v$ tends to increase. The $m_\pi=254$~MeV ensemble
shows similar behavior, although not as strongly. As shown in
Appendix~\ref{app:exc_states}, this trend is even less-clear on
other ensembles. The GPoF and summation results
have similar statistical uncertainties, which are slightly larger than
those of the ratio-plateau method with the largest source-sink
separation. They are in reasonable agreement, except for the
$m_\pi=149$~MeV ensemble, where the summation values for $F_1^v$
consistently lie below the corresponding GPoF values; this suggests
that excited-state effects are not fully under control on this
ensemble.

In general, the GPoF values tend to stay close to the ratio-method
values with the largest source-sink separation, whereas the summation
values tend to appear like an ``extrapolation'' from the trend set by
the lowest two source-sink separations; this is consistent with
expectations from Sec.~\ref{sec:ff_extract}. This tendency can be seen
most clearly when there is a separation between the summation and GPoF
values, such as for momentum \#23 for $F_1^v$ on the $m_\pi=149$~MeV
ensemble.

\subsubsection{Isovector Dirac form factor \texorpdfstring{$F_1^v(Q^2)$}{F₁v(Q²)}}

\begin{figure}
  \centering
  \includegraphics[width=0.49\textwidth]{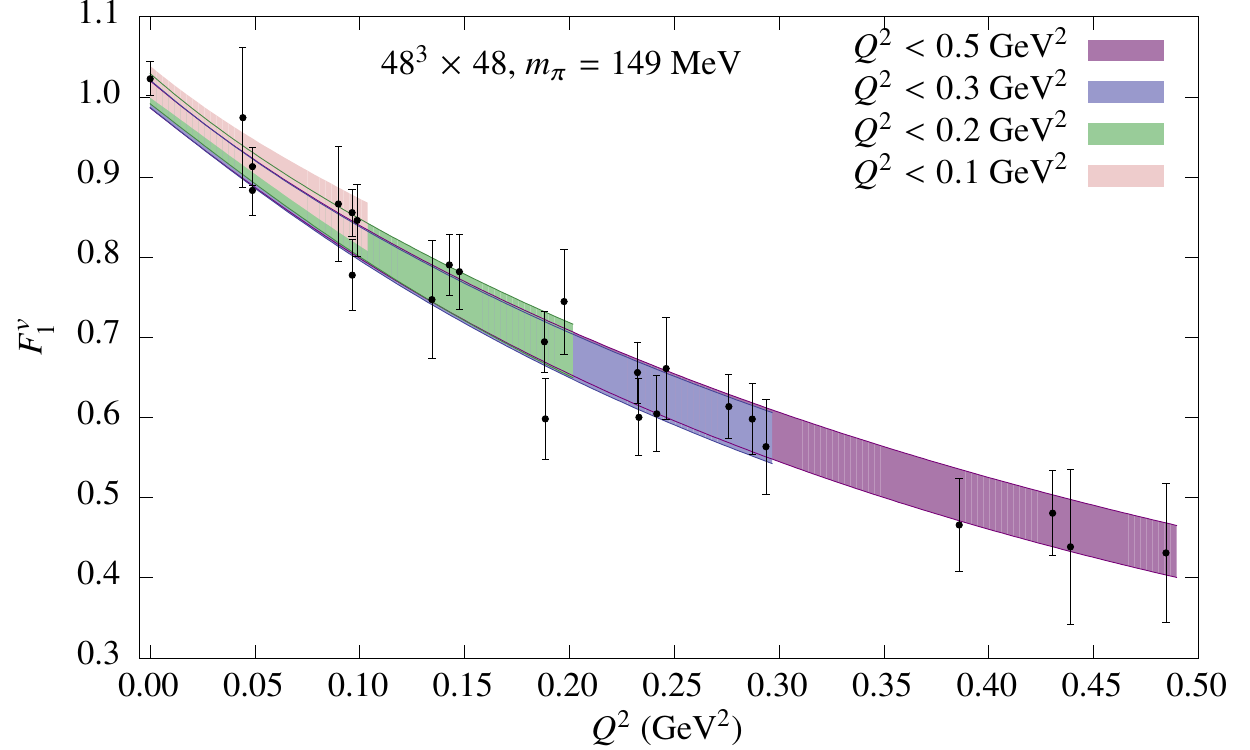}
  \includegraphics[width=0.49\textwidth]{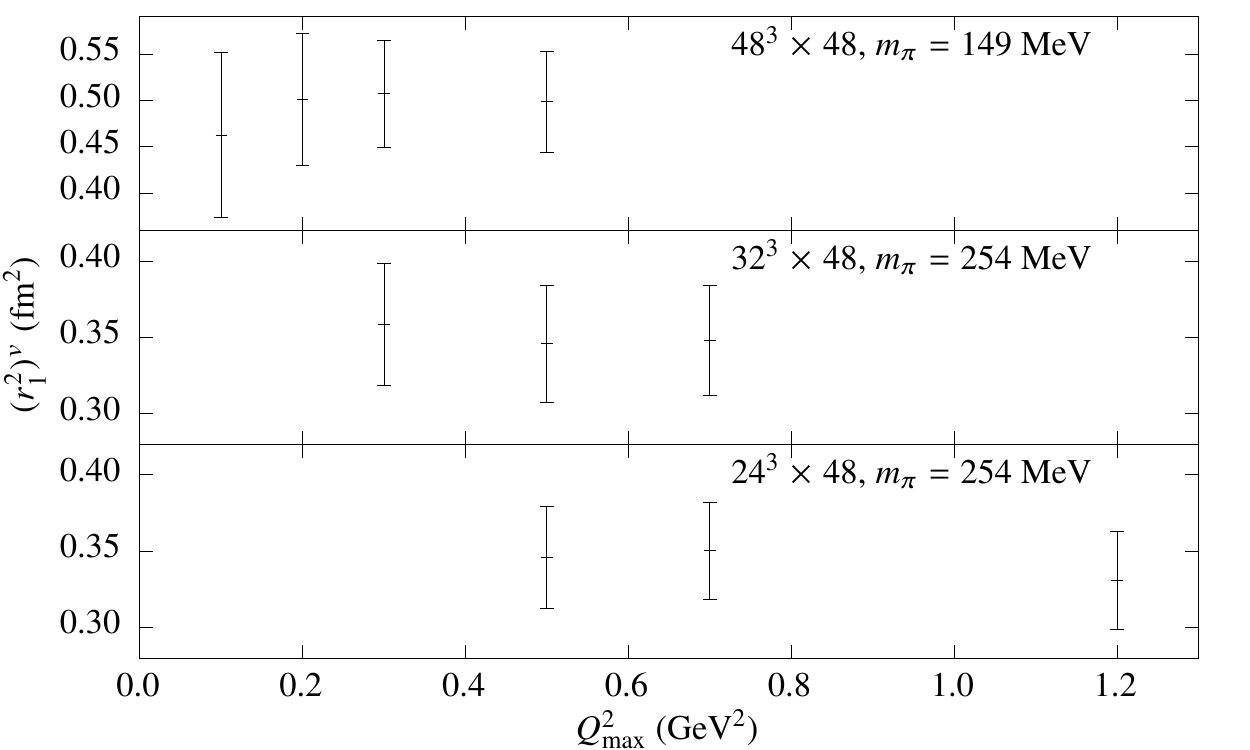}
\\
  \includegraphics[width=0.49\textwidth]{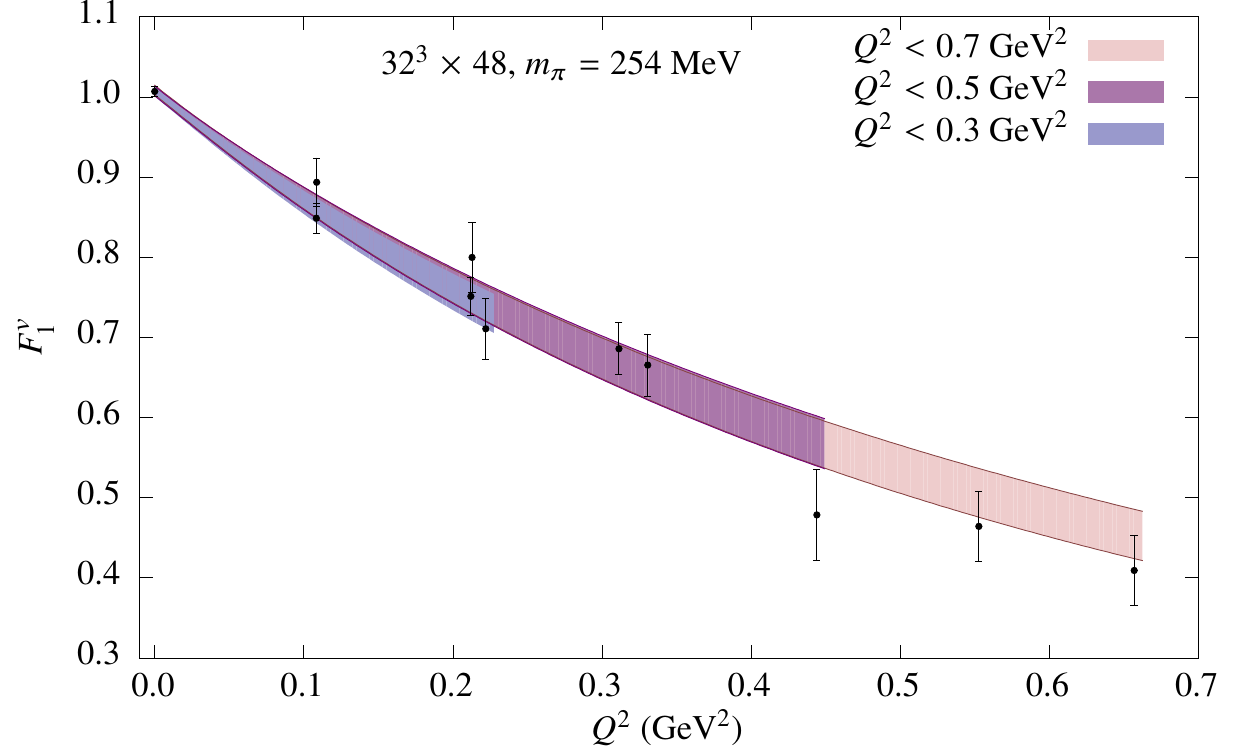}
  \includegraphics[width=0.49\textwidth]{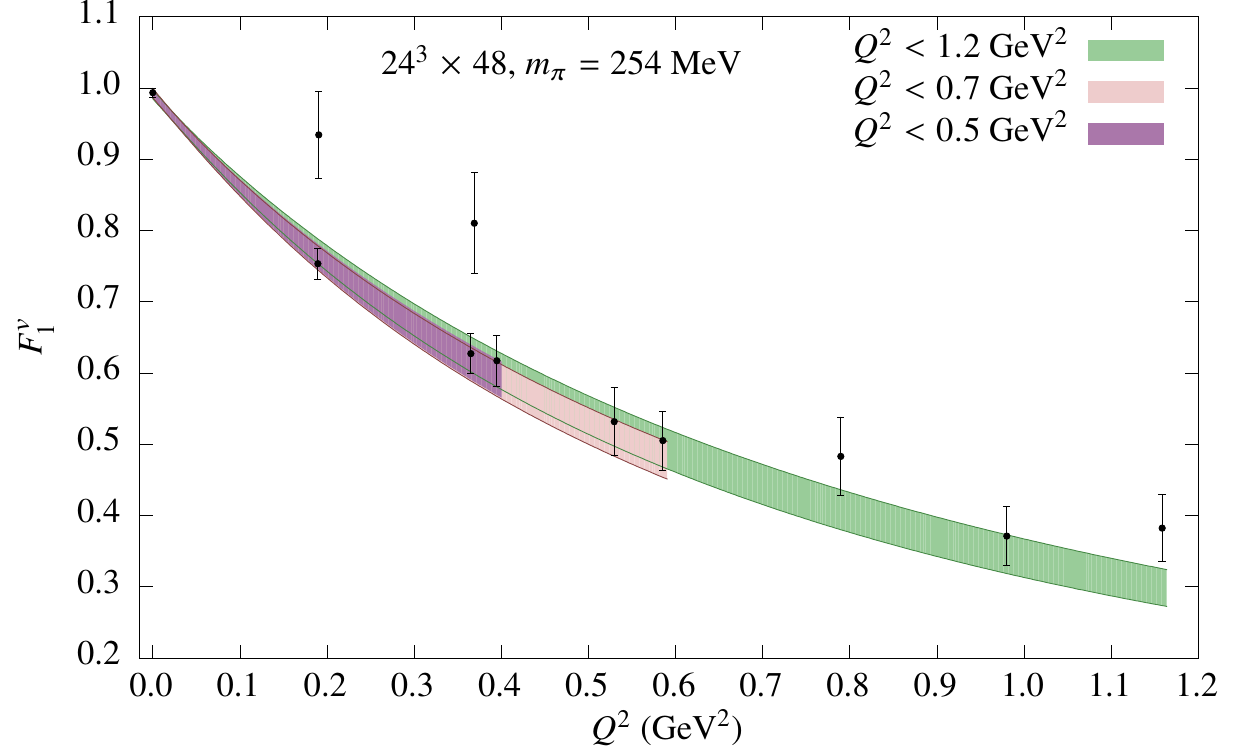}
  \caption{Dipole fits to $F_1^v(Q^2)$ with varying
    $Q^2_\text{max}$. The upper-right plot shows the dependence on
    $Q^2_\text{max}$ of the isovector Dirac radius derived from the
    fits.}
  \label{fig:F1v_dipole}
\end{figure}

We perform two-parameter fits of a dipole form,
\begin{equation}
  \label{eq:dipole}
  F(Q^2) = \frac{F(0)}{\left(1+\frac{Q^2}{m_D^2}\right)^2},  
\end{equation}
in the range $0\leq Q^2<0.5\text{ GeV}^2$, to $F_1^v(Q^2)$ for all of
our ensembles. This produces good fits, except on some ensembles when
using the shortest source-sink separation, where the data have smaller
statistical uncertainties. On the $m_\pi=149$~MeV ensemble, these data
suffer from excited-state contamination, and the fit has
$\chi^2=43(13)$ for 23 degrees of freedom. Because the data at larger
source-sink separations have larger uncertainties, it is unclear
whether this amount of deviation from a dipole form persists when
excited-state effects are reduced. The $24^3\times 48$ and $24^3\times
24$ ensembles at $m_\pi\approx 250$~MeV also suffer from poor fit
quality; this is caused by two momenta that have higher values of
$F_1^v$ than other nearby momenta (visible in
Fig.~\ref{fig:F1v_dipole}; specifically, these are momenta \#2 and \#4
in Fig.~\ref{fig:ff_extract_comparison_2}). This appears to be a
fluctuation, as such a large difference between nearby momenta is not
seen on other ensembles.

To study the dependence on the fit form, we perform dipole fits for
$0\leq Q^2 < Q^2_\text{max}$ with varying $Q^2_\text{max}$, to the
summation data on three ensembles; these are shown in
Fig.~\ref{fig:F1v_dipole}. In all three cases, the fit parameters vary
with $Q^2_\text{max}$ by less than the statistical uncertainty, with
the largest variation occurring on the 149~MeV ensemble, where
$(r_1^2)^v\equiv \frac{12}{m_D^2}$ varies between $0.463(88)\text{
  fm}^2$ and $0.507(58)\text{ fm}^2$, and our choice of
$Q^2_\text{max}=0.5\text{ GeV}^2$ yields $(r_1^2)^v=0.498(55)\text{
  fm}^2$. Therefore we conclude that errors caused by fitting are
smaller than the statistical uncertainty.

\subsubsection{Isovector Pauli form factor \texorpdfstring{$F_2^v(Q^2)$}{F₂v(Q²)}}

For the isovector Pauli form factor, we again perform two-parameter
dipole fits [Eq.~(\ref{eq:dipole})] in the range $0<Q^2<0.5\text{
  GeV}^2$; the main difference is that, because of the kinematic
factor in Eq.~(\ref{eq:formfac_def}), we have no measurement of $F_2$ at
$Q^2=0$. Therefore, understanding behavior near zero momentum transfer
requires an extrapolation below the smallest accessible
$Q^2_\text{min}\sim (\frac{2\pi}{L_s})^2$, and this extrapolation is
more difficult on ensembles with smaller volumes. The quality of fits
is generally reasonable, particularly when not using the shortest
source-sink separation, which has the most precise data. The
most-consistently bad fits are on the $32^3\times 48$, $m_\pi=254$~MeV
ensemble, where $\chi^2$ varies between 11 and 15, depending on how
the matrix elements are computed, for fits with 6 degrees of freedom.

\begin{figure}
  \centering
  \includegraphics[width=0.49\textwidth]{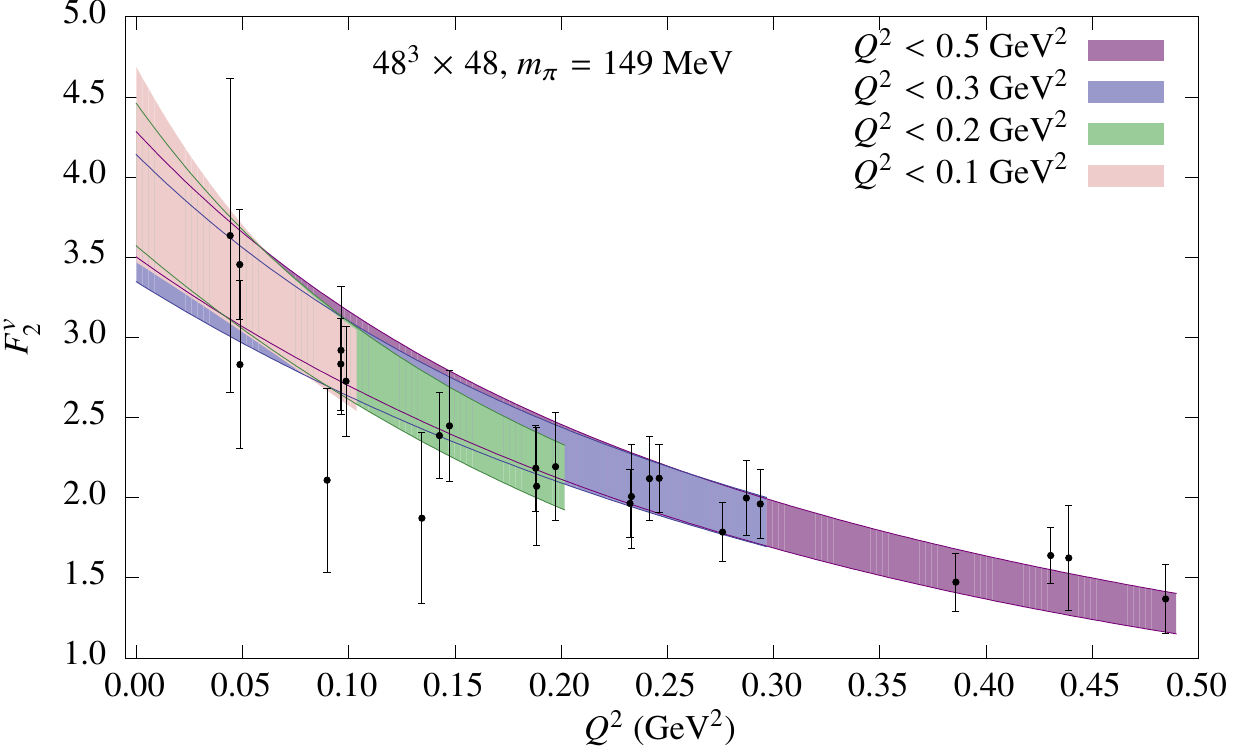}\\
  \includegraphics[width=0.49\textwidth]{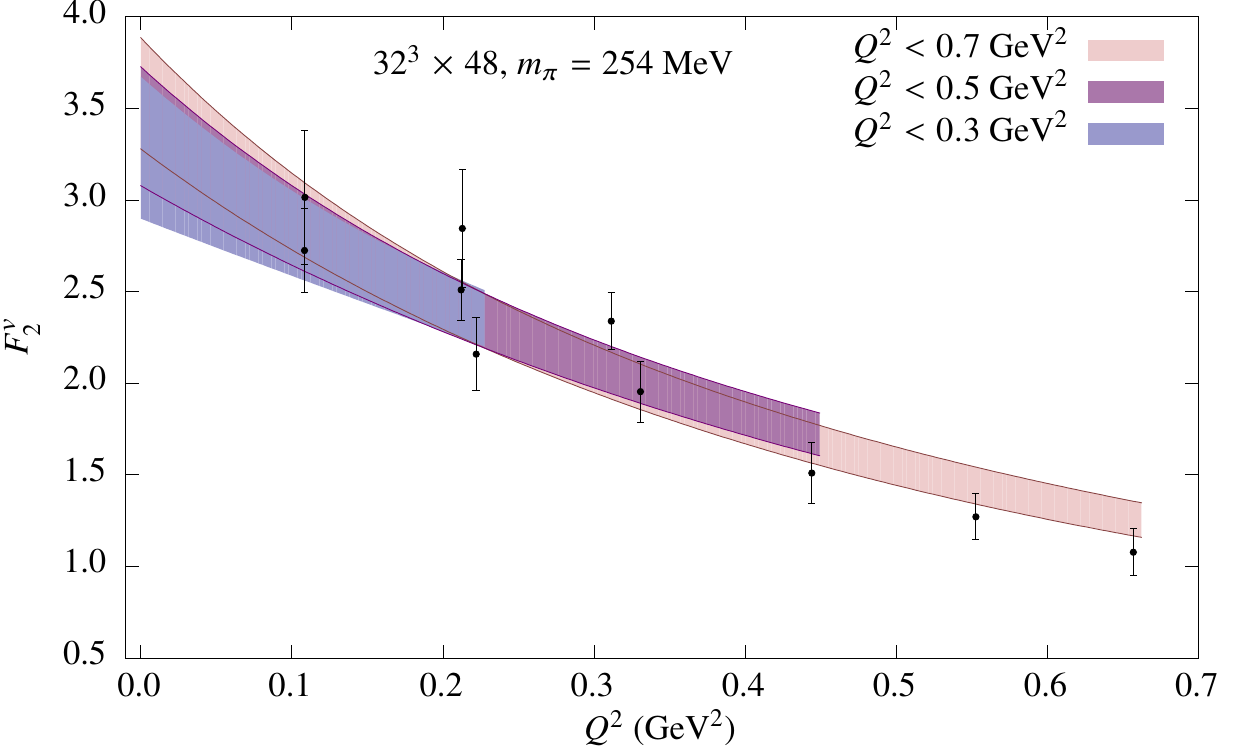}
  \includegraphics[width=0.49\textwidth]{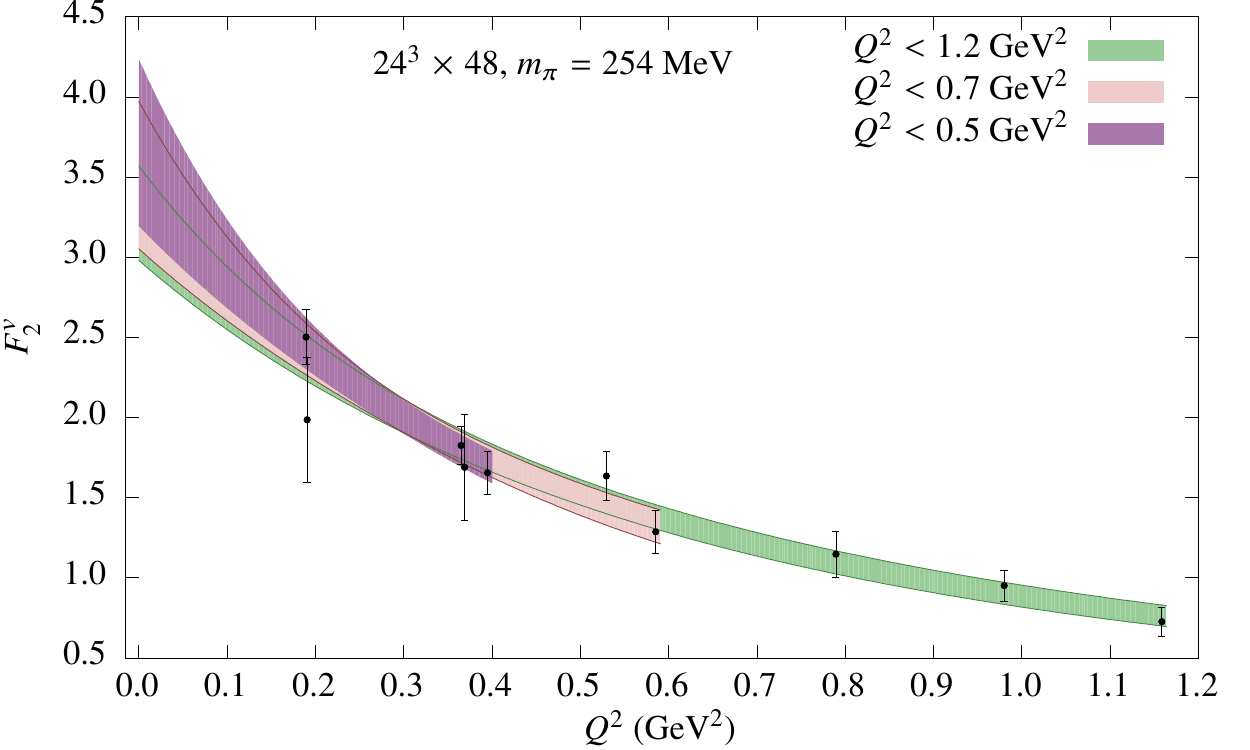}
\\
  \includegraphics[width=0.49\textwidth]{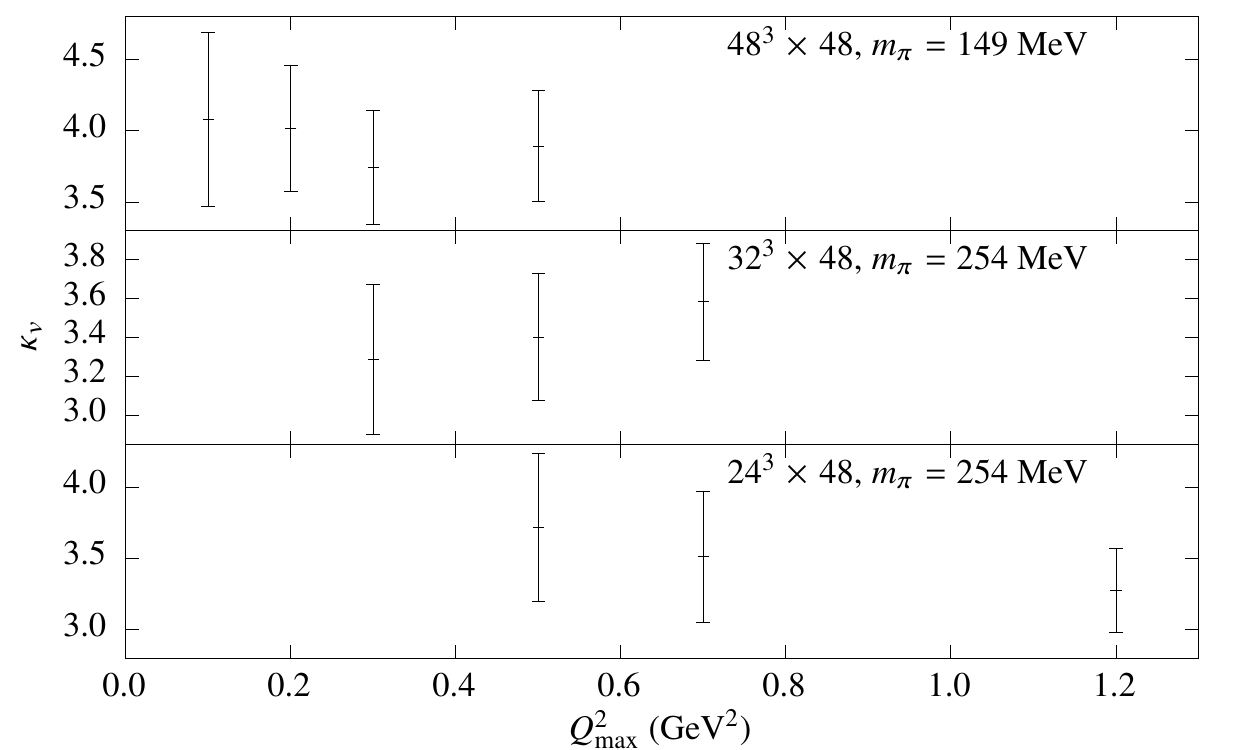}
  \includegraphics[width=0.49\textwidth]{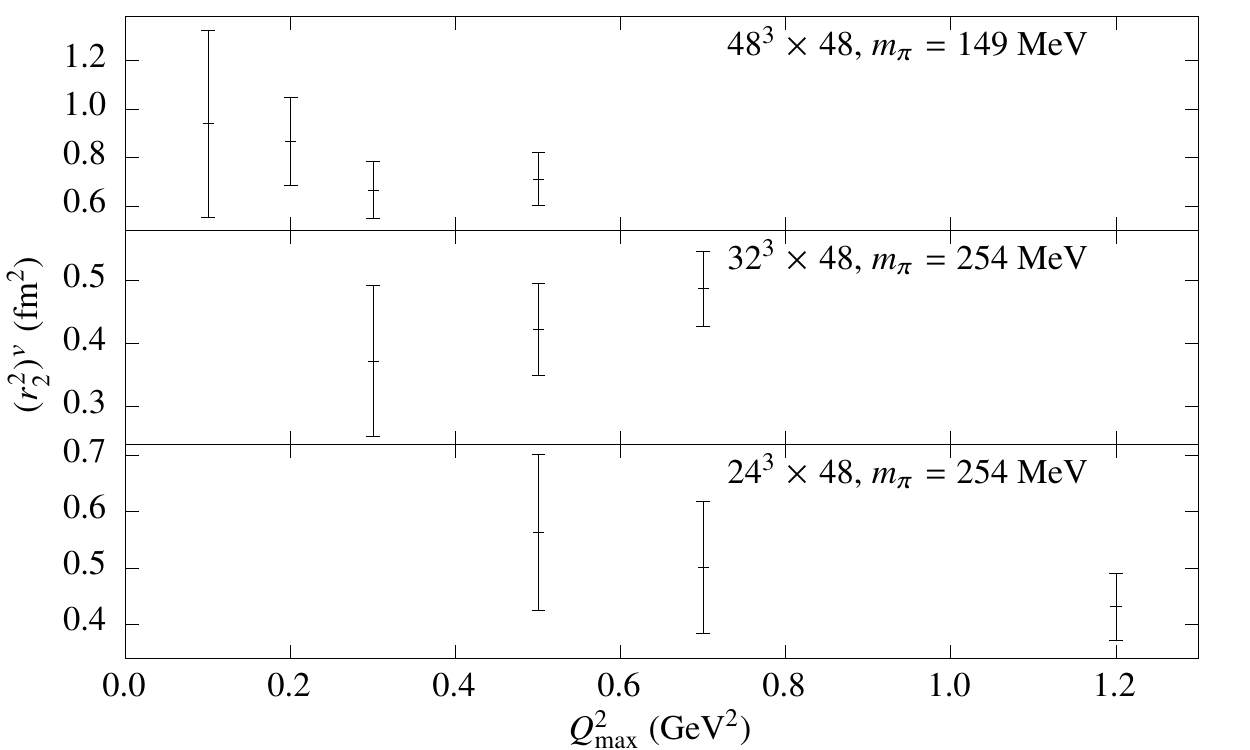}
  \caption{Dipole fits to $F_2^v(Q^2)$ with varying
    $Q^2_\text{max}$. The last two plots show the dependence on
    $Q^2_\text{max}$ of the fit parameters.}
  \label{fig:F2v_dipole}
\end{figure}

We again study dependence on the fit form by varying the maximum
momentum transfer included in the fit, $Q^2_\text{max}$, on three
ensembles, using form factors computed using the summation method;
these are shown in Fig.~\ref{fig:F2v_dipole}. Because of the need to
extrapolate to $Q^2=0$, the fit parameters have a greater variation
with $Q^2_\text{max}$ than occurred for the Dirac form factor;
although on the two shown $m_\pi=254$~MeV ensembles, this variation is
roughly within the statistical uncertainty of the fit done with our
choice of $Q^2_\text{max}=0.5\text{ GeV}^2$. On the 149~MeV ensemble,
this also holds true for $F_2^v(0)$, which varies between 3.74(40) and
4.08(61), and our chosen fit yields $F_2^v(0)=3.89(39)$; however,
$(r_2^2)^v\equiv\frac{12}{m_D^2}$ varies between $0.67(12)\text{
  fm}^2$ and $0.94(38)\text{ fm}^2$, and our chosen fit yields
$(r_2^2)^v=0.71(11)\text{ fm}^2$. Since the statistical uncertainty
increases significantly at small $Q^2_\text{max}$ and the results
remain consistent with our choice, we conclude that systematic errors
due to fitting are not large.

\subsubsection{Isovector Sachs form factors}

\begin{figure}
  \centering
  \includegraphics[width=0.49\textwidth]{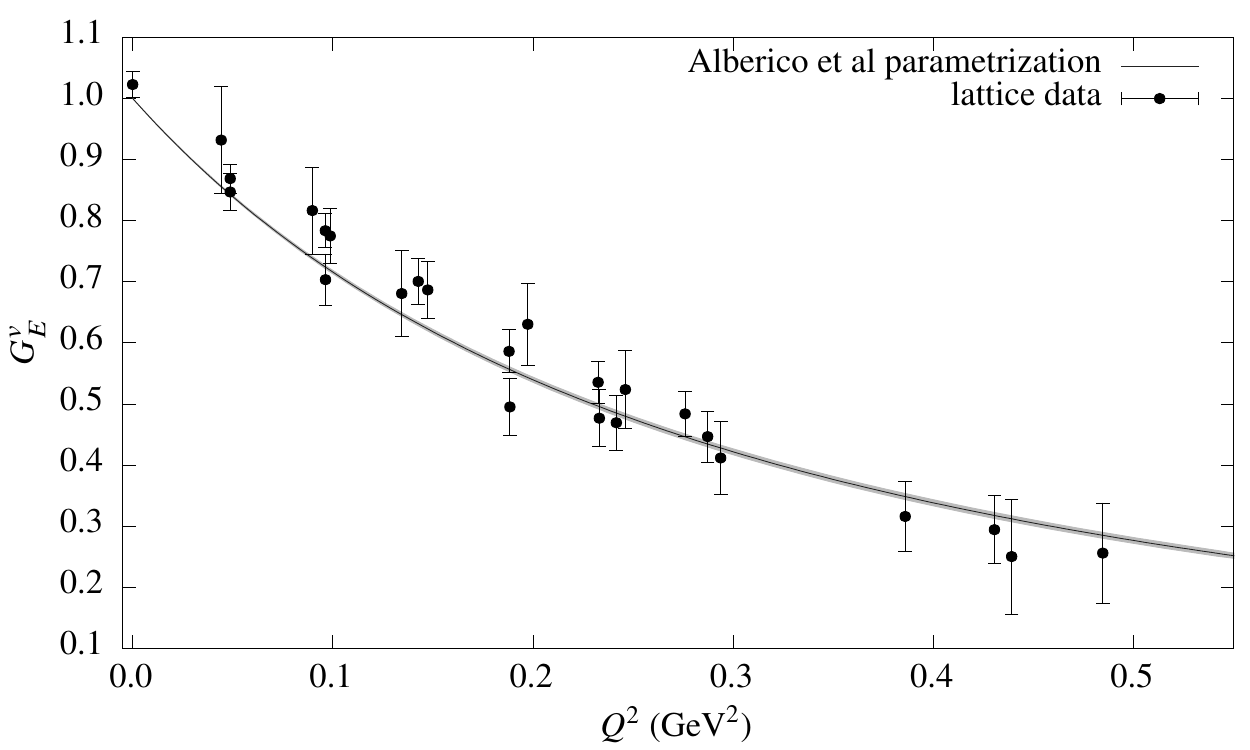}
  \includegraphics[width=0.49\textwidth]{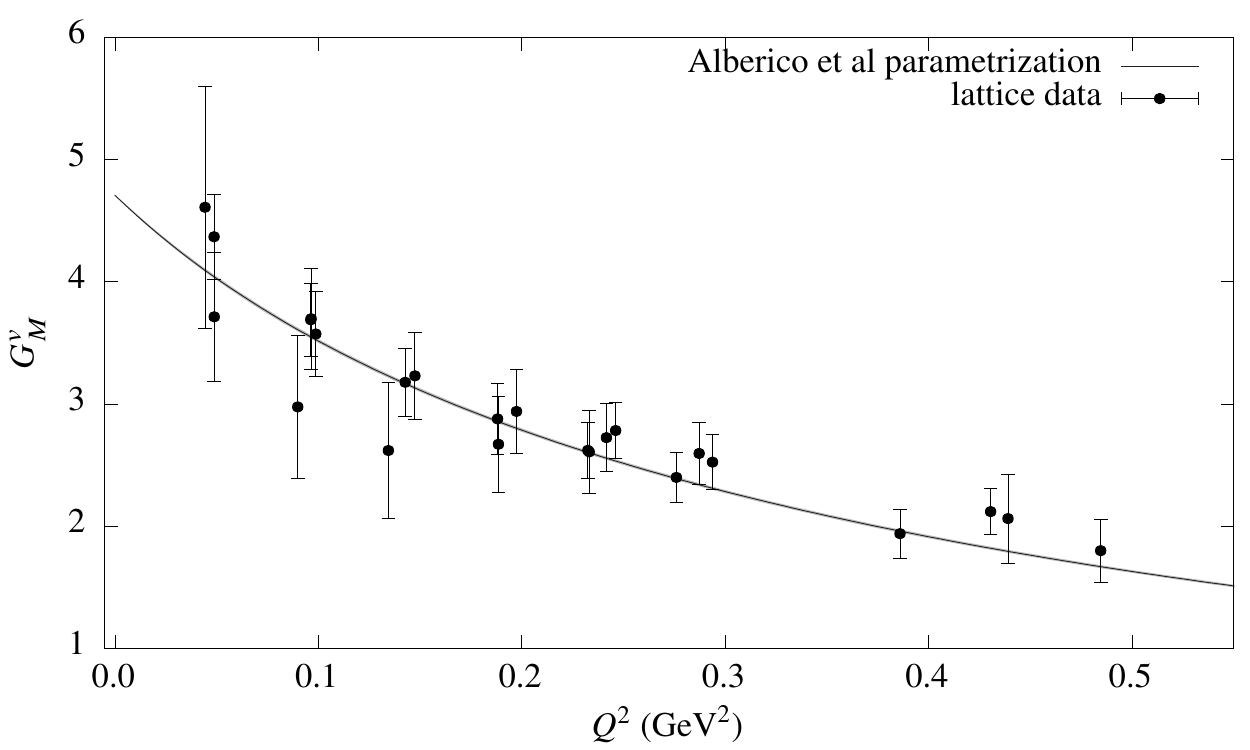}
  \caption{Isovector electric and magnetic form factors. Each plot
    contains the curve with error band from the fit to experiment in
    Ref.~\cite{Alberico:2008sz} and the summation data from the
    $m_\pi=149$~MeV ensemble.}
  \label{fig:isovector_expt}
\end{figure}

To avoid any model-dependence from fitting curves, we first compare
the lattice form factors themselves with experiment. In particular, we
use the experimentally-preferred electric and magnetic form factors,
$G_E$ and $G_M$, and make use of the phenomenological parameterization
of experimental data in Ref.~\cite{Alberico:2008sz}, for which
correlations between fit parameters have been made available, allowing
for the curves to be plotted with error bands. These are compared with
our summation data from the $m_\pi=149$~MeV ensemble in
Fig.~\ref{fig:isovector_expt}. Both of these form factors agree well
with experiment; a chi-squared comparison yields $p=0.64$ for $G_E$
and $p=0.81$ for $G_M$, a feat that only occurs when both the pion
mass is near-physical and excited-state contaminations are reasonably
controlled. Using the ratio method with the largest source-sink
separation or the GPoF method also produces reasonable agreement, with
$p>0.2$ in all cases.

\subsection{Isovector Radii and magnetic moment}
The isovector Dirac and Pauli radii, $(r_{1,2}^2)^v$, and the
isovector anomalous magnetic moment, $\kappa^v$, are defined from the
behavior of $F_{1,2}^v(Q^2)$ near $Q^2=0$:
\begin{align}
  F_1^v(Q^2) &= 1 - \frac{1}{6}(r_1^2)^vQ^2 + O(Q^4) \\
  F_2^v(Q^2) &= \kappa^v\left(1 - \frac{1}{6}(r_2^2)^vQ^2 + O(Q^4)\right).
\end{align}
On each ensemble, these quantities are determined from the dipole fits
to the form factor data described in the previous subsection.

In order to compare these results with experiment at the physical pion
mass, which is 134.8~MeV in the isospin limit~\cite{Colangelo:2010et},
we perform extrapolations employing physically well-motivated
functional forms taken from chiral perturbation theory (ChPT). Rather
than attempting a fully \emph{ab initio} prediction of nucleon
observables, we make use of ChPT with parameters input from
phenomenology; the compatibility of the lattice data with the
phenomenological fit forms corroborates the validity of the
extrapolations. To observe the congruence with ChPT, we include a
certain limited range of data, namely, the first four ensembles listed
in Tab.~\ref{tab:gauge_ens}, which are those with the smallest pion
masses and largest lattice volumes; by confining the fits to this
region, we concentrate on the regime where the predictions of ChPT are
most significant. Details of the extrapolations are given in
Appendix~\ref{app:chpt}. We note that more recent works in chiral
effective
theory~\cite{Tiburzi:2007ep,Greil:2011aa,Hall:2012pk,Hall:2012yx,Hall:2013oga}
have also included the infinite-volume extrapolation, however we do
not attempt to apply them here.

\subsubsection{Isovector Dirac radius \texorpdfstring{$(r_1^2)^v$}{(r₁²)v}}

\begin{figure}
  \centering
  \includegraphics[width=0.7\textwidth]{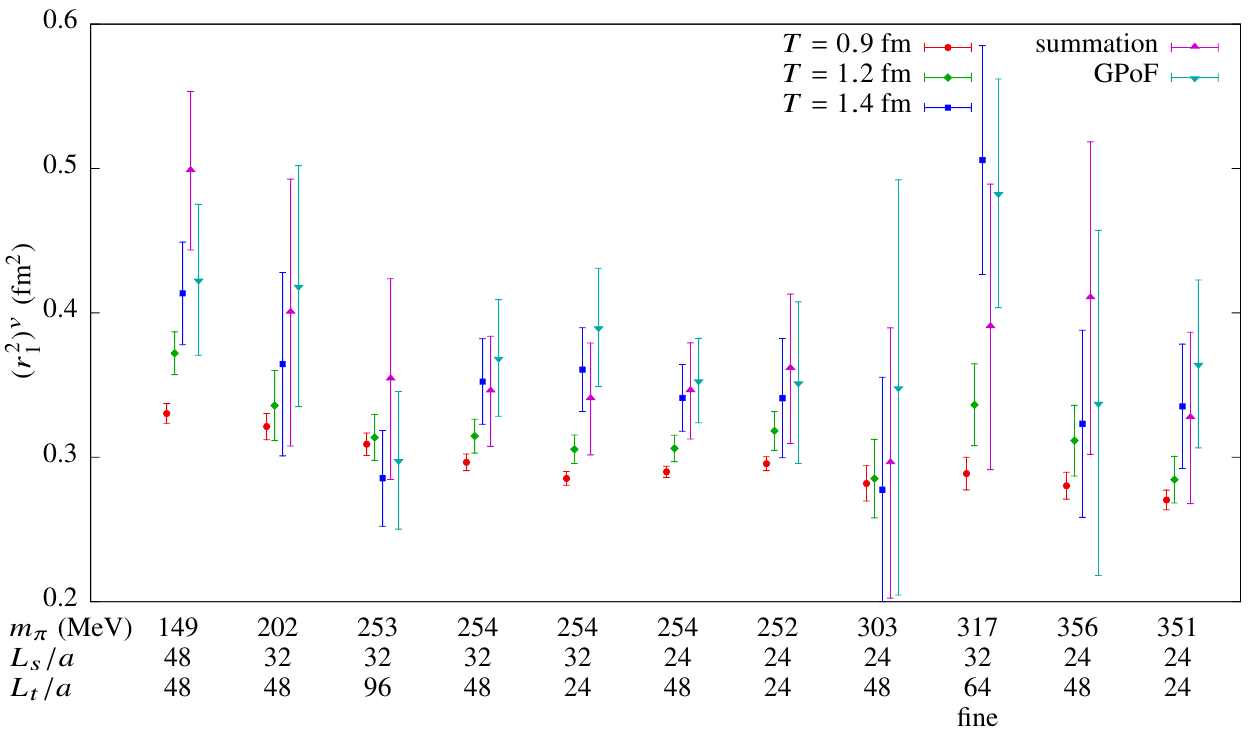}
  \caption{Isovector Dirac radius $(r_1^2)^v$, determined on each lattice
  ensemble using different analysis methods for computing form factors.}
  \label{fig:r1v2_ens}
\end{figure}

For each ensemble, the Dirac radius determined from a dipole fit to
$F_1^v(Q^2)$, determined using the ratio, summation, and GPoF methods,
is shown in Fig.~\ref{fig:r1v2_ens}. The ratio-method data show a
clear trend: the computed Dirac radius increases with the source-sink
separation. This indicates the presence of excited-state contamination
that is still poorly controlled when using the largest source-sink
separation. The dependence on source-sink separation is particularly
large on the $m_\pi=149$~MeV ensemble; on that ensemble, the summation
method yields an even larger Dirac radius.

\begin{figure}
  \centering
  \includegraphics[width=0.7\textwidth]{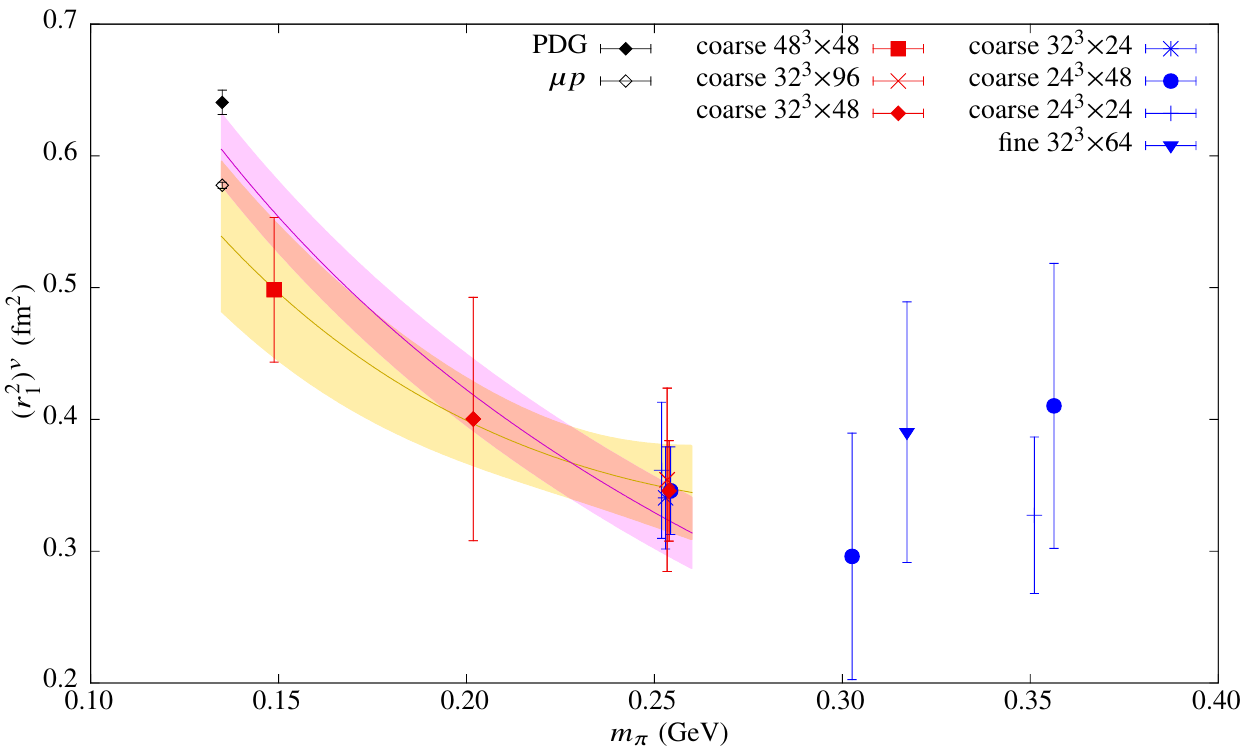}
  \caption{Chiral extrapolation of the isovector Dirac radius $(r_1^2)^v$,
  determined on each lattice ensemble using the summation method.
  Two experimental points are shown, where $(r_E^2)^p$ is taken from
  either the CODATA 2010 result~\cite{Mohr:2012tt} used by the
  PDG~\cite{Beringer:1900zz} or the measurement from spectroscopy of
  muonic hydrogen~\cite{Antognini:1900ns}. Both points use the PDG
  value for $(r_E^2)^n$. The magenta band results from fitting
  with the formula from Appendix~\ref{app:chpt}, whereas the orange
  band results from including an additional term proportional
  to $m_\pi^2$.}
  \label{fig:r1v2_extrap}
\end{figure}

The chiral fit form for the isovector Dirac radius has one free
parameter; the fit to the summation data is of good quality and is
shown in magenta in Fig.~\ref{fig:r1v2_extrap}. Extrapolation to the
physical pion mass produces good agreement with the experimental
data. Although this fit is entirely compatible with our lattice data,
its slope constrained by ChPT appears larger than the slope suggested
by the data alone. Therefore we also perform a fit with an additional
higher-order term proportional to $m_\pi^2$, which is shown in orange
in Fig.~\ref{fig:r1v2_extrap}. The resulting extrapolated value has a
considerably larger uncertainty, but is also consistent with both
experimental points, within 1--2$\sigma$.

\subsubsection{Isovector anomalous magnetic moment \texorpdfstring{$\kappa^v$}{κv}}

\begin{figure}
  \centering
  \includegraphics[width=0.7\textwidth]{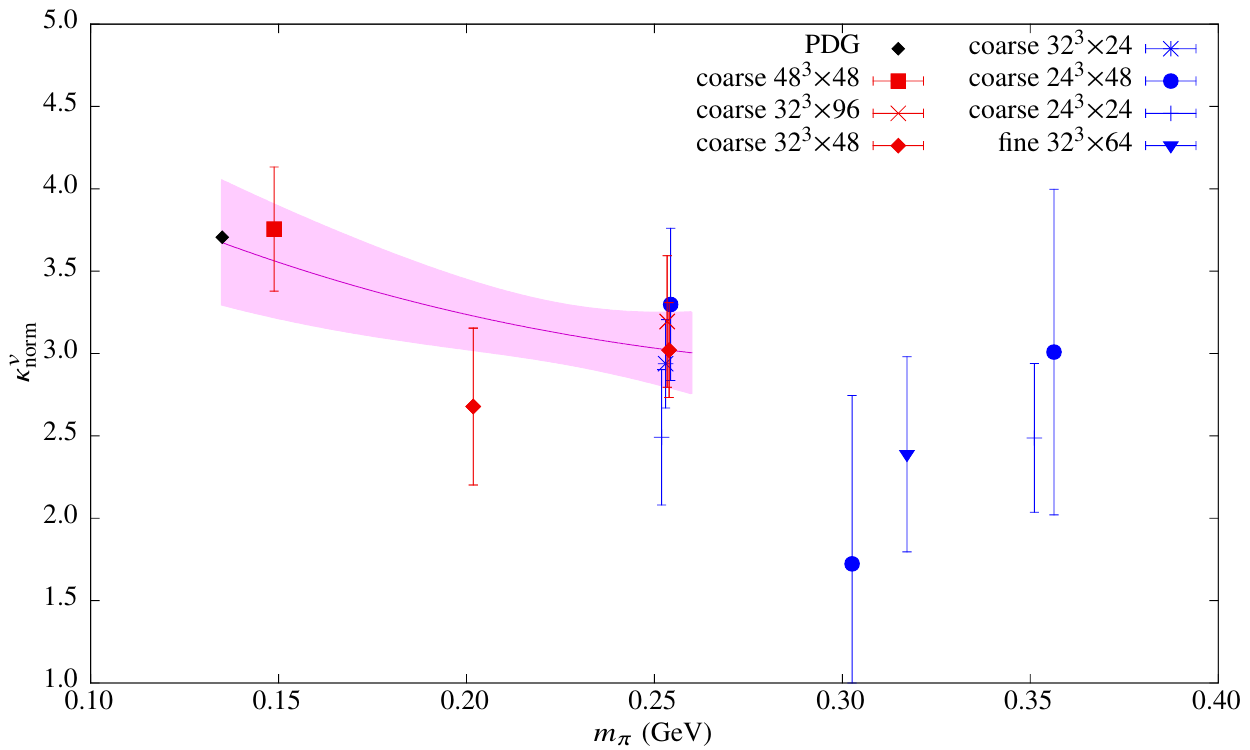}
  \caption{Chiral extrapolation of the isovector anomalous magnetic
    moment $\kappa^v_\text{norm}$, determined on each lattice ensemble
    using the summation method. The experimental point is from the
    PDG~\cite{Beringer:1900zz}}
  \label{fig:kvnorm_extrap}
\end{figure}

For comparing across different ensembles, we normalize the isovector
anomalous magnetic moment relative to the physical magneton, rather
than using the ensemble-dependent nucleon mass as in
Eq.~(\ref{eq:formfac_def}):
\begin{equation}\label{eq:kv_norm}
  \kappa^v_\text{norm} =
  \frac{m_N^\text{phys}}{m_N^\text{lat}}F_2^{v,\text{lat}}(0).
\end{equation}
As shown in Fig.~\ref{fig:kvnorm_extrap}, the summation method on the
$m_\pi=149$~MeV ensemble produces a value of $\kappa^v$ consistent
with experiment, as does the two-parameter chiral extrapolation to the
physical pion mass. We only find a clear sign of excited-state effects
on the $m_\pi=149$~MeV ensemble; see Appendix~\ref{app:exc_states}.

\subsubsection{Isovector Pauli radius \texorpdfstring{$(r_2^2)^v$}{(r₂²)v}}

\begin{figure}
  \centering
  \includegraphics[width=0.7\textwidth]{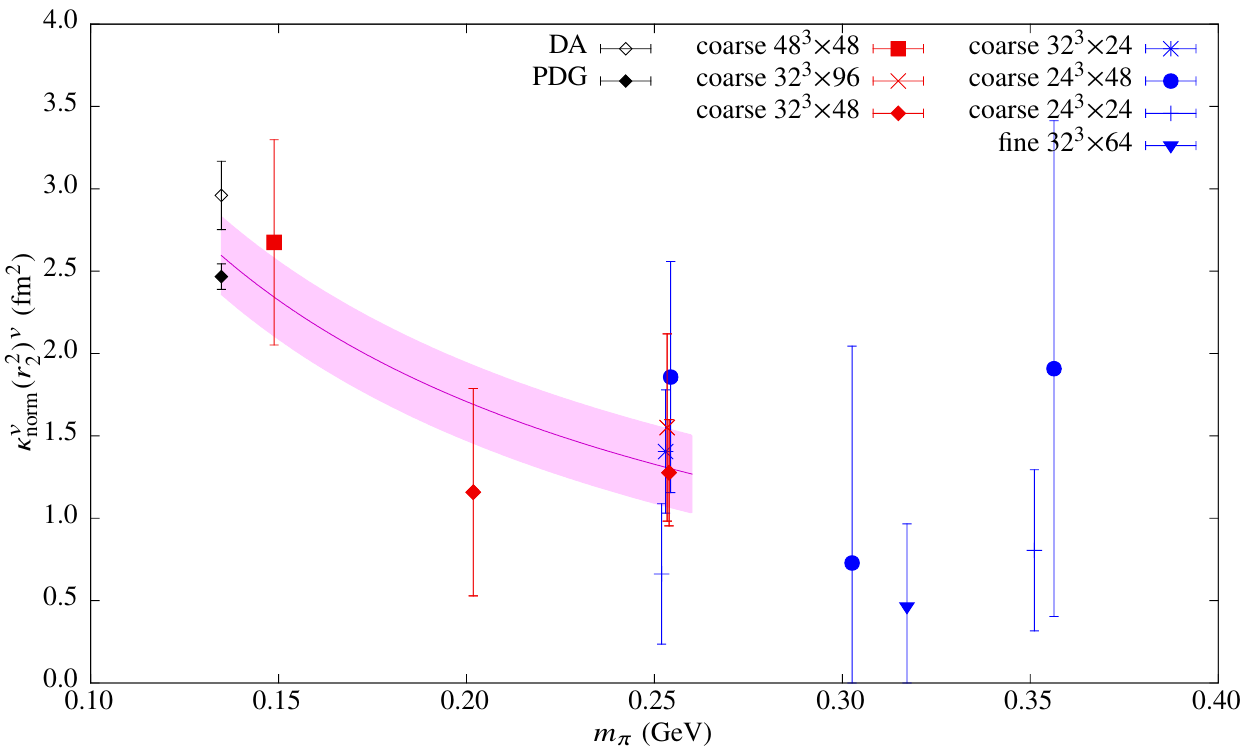}
  \caption{Chiral extrapolation of the product of the isovector
    anomalous magnetic moment and Pauli radius,
    $\kappa^v_\text{norm}(r_2^2)^v$, determined on each lattice
    ensemble using the summation method. We show two experimental
    values, where the radii are taken either from the 2012
    PDG~\cite{Beringer:1900zz} or from the dispersion analysis in
    Ref.~\cite{Lorenz:2012tm} (the difference mostly comes from
    different values for the proton magnetic radius).}
  \label{fig:r2v2_kvnorm_extrap}
\end{figure}

For chiral extrapolation, it is more natural to use the combination
$\kappa^v(r_2^2)^v$. As shown in Fig.~\ref{fig:r2v2_kvnorm_extrap},
when using the summation method, this quantity on the $m_\pi=149$~MeV
ensemble is consistent with the experimental points, as is the value
obtained using the one-parameter extrapolation to the physical pion
mass. Excited-state effects for the Pauli radius are similar to those
for the anomalous magnetic moment; see Appendix~\ref{app:exc_states}.

\section{\label{sec:isoscalar}Isoscalar form factors}
We also compute isoscalar form factors. Since we do not include the
contributions from disconnected quark contractions, these results
suffer from an uncontrolled systematic error. Despite this, these
results are still useful for illustrating qualitative features and the
effects of other systematic errors. They will also give some insight
into the size of disconnected contributions.

At relatively high pion masses, light quark disconnected contributions
have now been calculated directly using lattice QCD. In
Ref.~\cite{Abdel-Rehim:2013wlz}, disconnected contributions to $G_E^p$
and $G_M^p$ were found to be consistent with zero and at most 1\% when
using a pion mass of about 370~MeV. Preliminary results from a
high-statistics calculation at pion mass 317~MeV find nonzero values
for the disconnected contributions (positive for $G_E$ and negative
for $G_M$) that are also less than 1\% of the connected
contribution~\cite{Meinel_privcomm}.

At the physical point, total disconnected contributions have been
determined using form factors from experiment together with chiral
extrapolations of connected-contraction lattice data, sometimes
supplemented with both experimental and lattice data on octet
baryons. Using chiral perturbation theory, this is divided into
strange and light quark contributions, in order to obtain the strange
quark contribution alone, which is presented in
Refs.~\cite{Leinweber:2004tc,Leinweber:2006ug,Wang:1900ta,Shanahan:2014tja}. Most
precisely determined is the disconnected contribution to $G_M^p(0)$,
where experimental data on octet baryon magnetic moments were used;
taking the result from Ref.~\cite{Shanahan:2014tja} and undoing the
division into strange and light contributions yields a roughly $-3\%$
disconnected contribution, with a 20\% relative uncertainty on the
estimate. At $Q^2=0.26\text{ GeV}^2$, the disconnected contributions
to $G_M^p$ and $G_E^p$ are estimated to be $-3\%$ and $-1.6\%$,
respectively, albeit with roughly 100\% relative uncertainties in both
cases.

\subsection{Form factors}

Comparisons of the different methods for computing matrix elements,
applied to the isoscalar Dirac and Pauli form factor, are in
Appendix~\ref{app:exc_states}. The isoscalar Dirac form factor behaves
similarly to the isovector case, whereas the isoscalar Pauli form
factor is generally consistent with zero, without any clear trends
visible in the data.

\subsubsection{Isoscalar Dirac form factor \texorpdfstring{$F_1^s(Q^2)$}{F₁s(Q²)}}

As we did for the isovector Dirac form factor, we also perform dipole
fits to $F_1^s(Q^2)$ in the range $0\leq Q^2 < 0.5\text{ GeV}^2$. This
produces fits of generally good quality, except on some ensembles when
using the shortest source-sink separation, where the data have smaller
statistical uncertainties. On the $m_\pi=149$~MeV ensemble, these data
suffer from excited-state contamination, and the fit has
$\chi^2=44(13)$ for 23 degrees of freedom. As in the isovector case,
it is unclear whether this level of deviation from a dipole persists
when excited-state effects are reduced.

\begin{figure}
  \centering
  \includegraphics[width=0.49\textwidth]{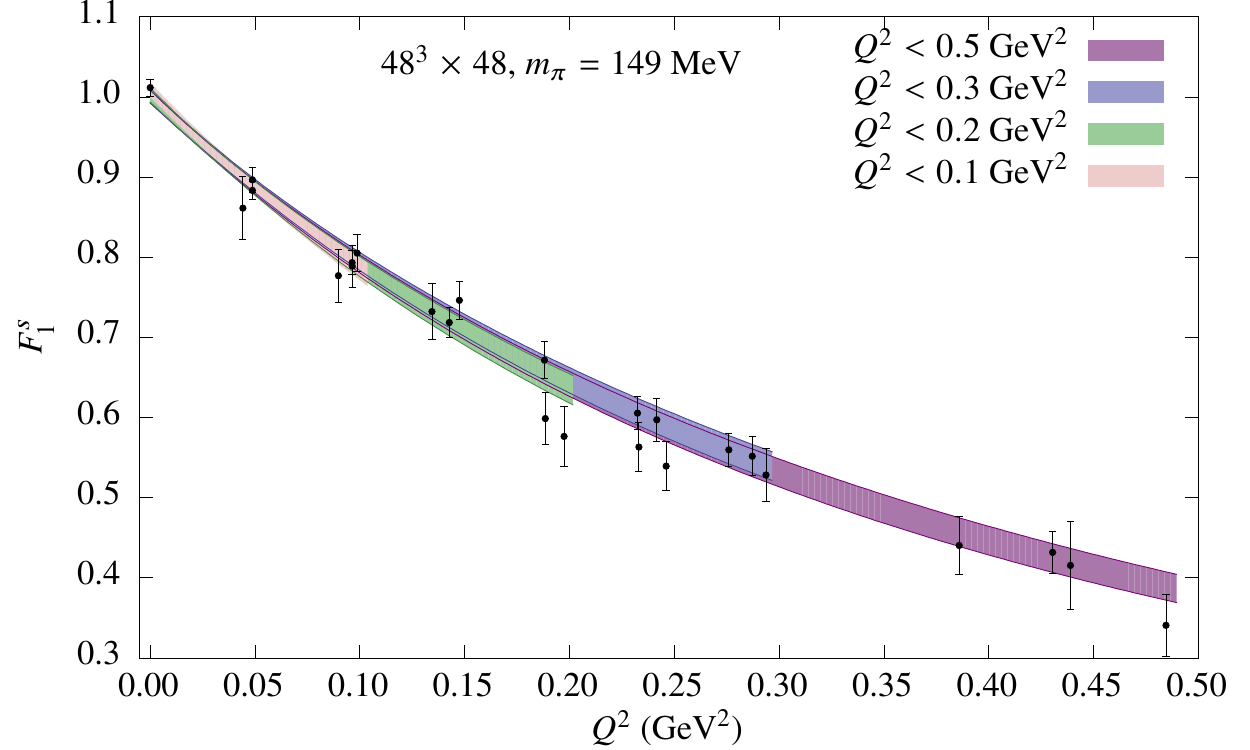}
  \includegraphics[width=0.49\textwidth]{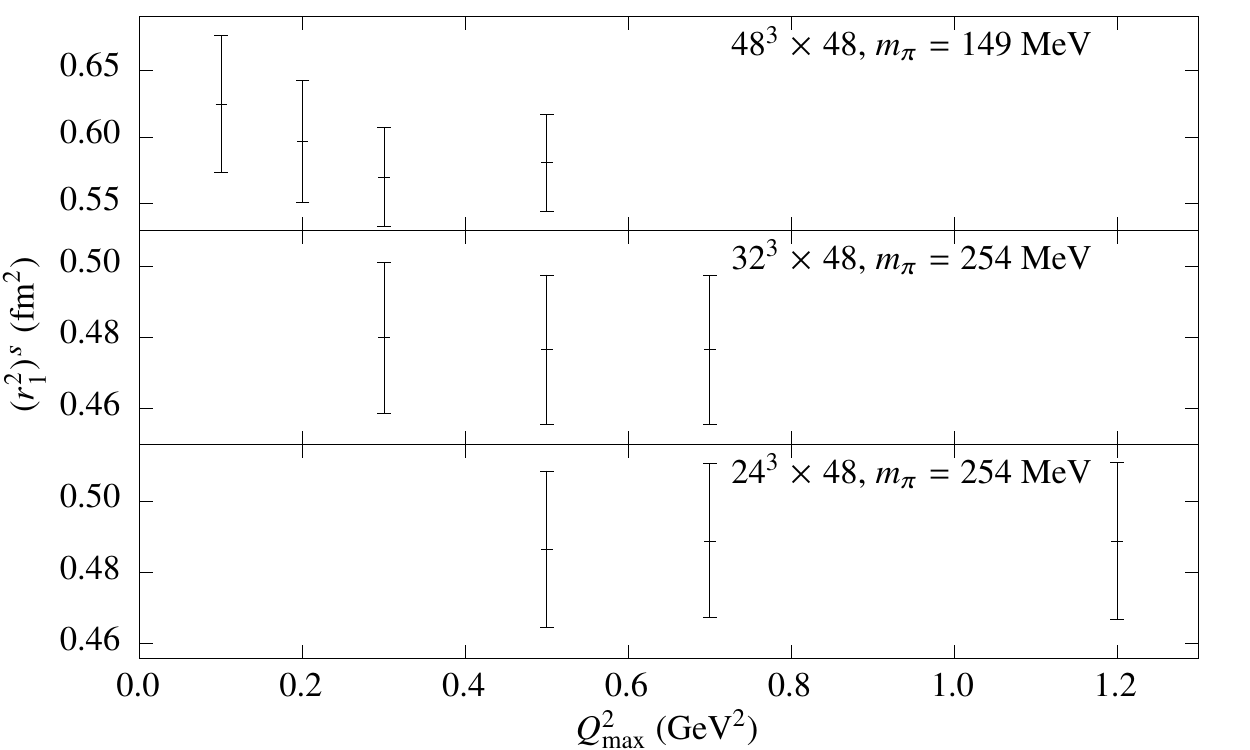}
\\
  \includegraphics[width=0.49\textwidth]{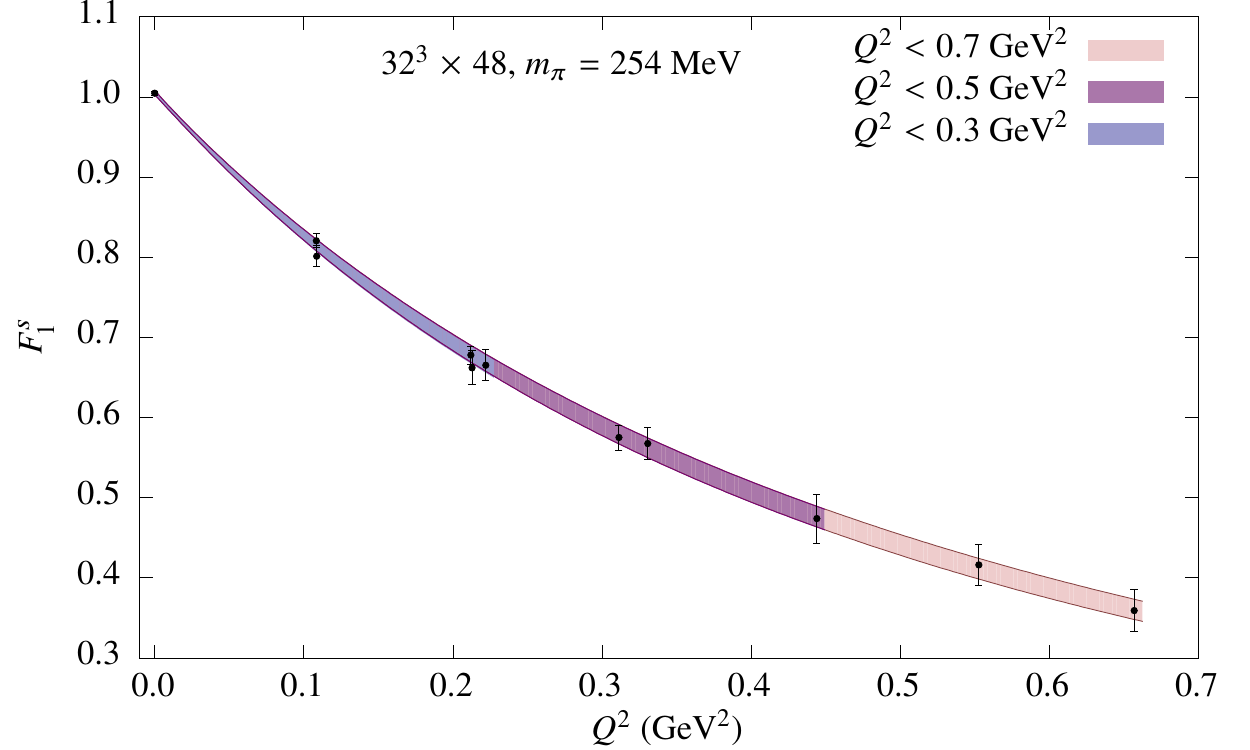}
  \includegraphics[width=0.49\textwidth]{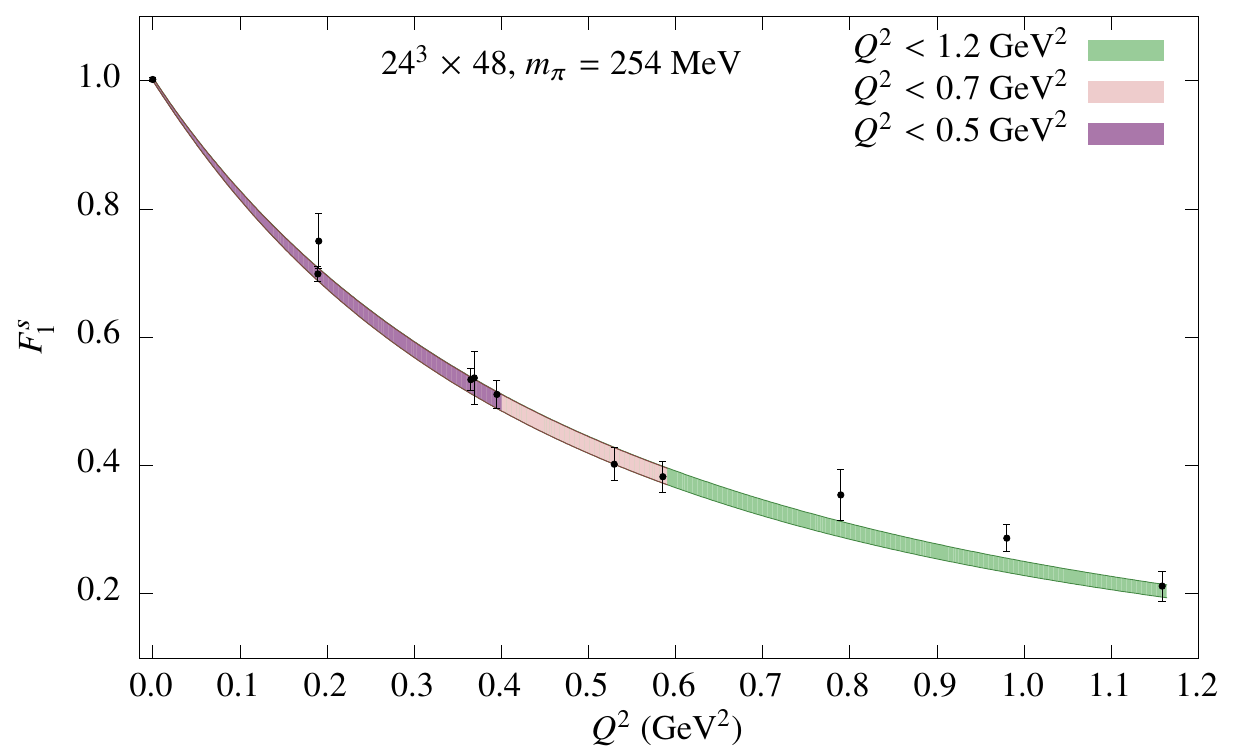}
  \caption{Dipole fits to $F_1^s(Q^2)$ with varying
    $Q^2_\text{max}$. The upper-right plot shows the dependence on
    $Q^2_\text{max}$ of the isoscalar Dirac radius derived from the
    fits.}
  \label{fig:F1s_dipole}
\end{figure}

To study dependence on the fit, we vary the upper bound of the range
in $Q^2$ on three ensembles; these are shown in
Fig.~\ref{fig:F1s_dipole}. We again find that the fit results vary by
less than the statistical uncertainty and we conclude that errors
caused by fitting are smaller than the statistical uncertainty.

\subsubsection{Isoscalar Pauli form factor \texorpdfstring{$F_2^s(Q^2)$}{F₂s(Q²)}}

As our isoscalar Pauli form factor data do not show a clear shape, we
fit them with a line,
\begin{equation}
  F_2^s(Q^2) = A + BQ^2,
\end{equation}
in our standard range $0<Q^2<0.5\text{ GeV}^2$. The fits are generally
of reasonable quality, except in some cases when using the shortest
source-sink separation, such as on the $m_\pi=149$~MeV ensemble, where
using the shortest source-sink separation yields $\chi^2=46(13)$ for
22 degrees of freedom.

\begin{figure}
  \centering
  \includegraphics[width=0.49\textwidth]{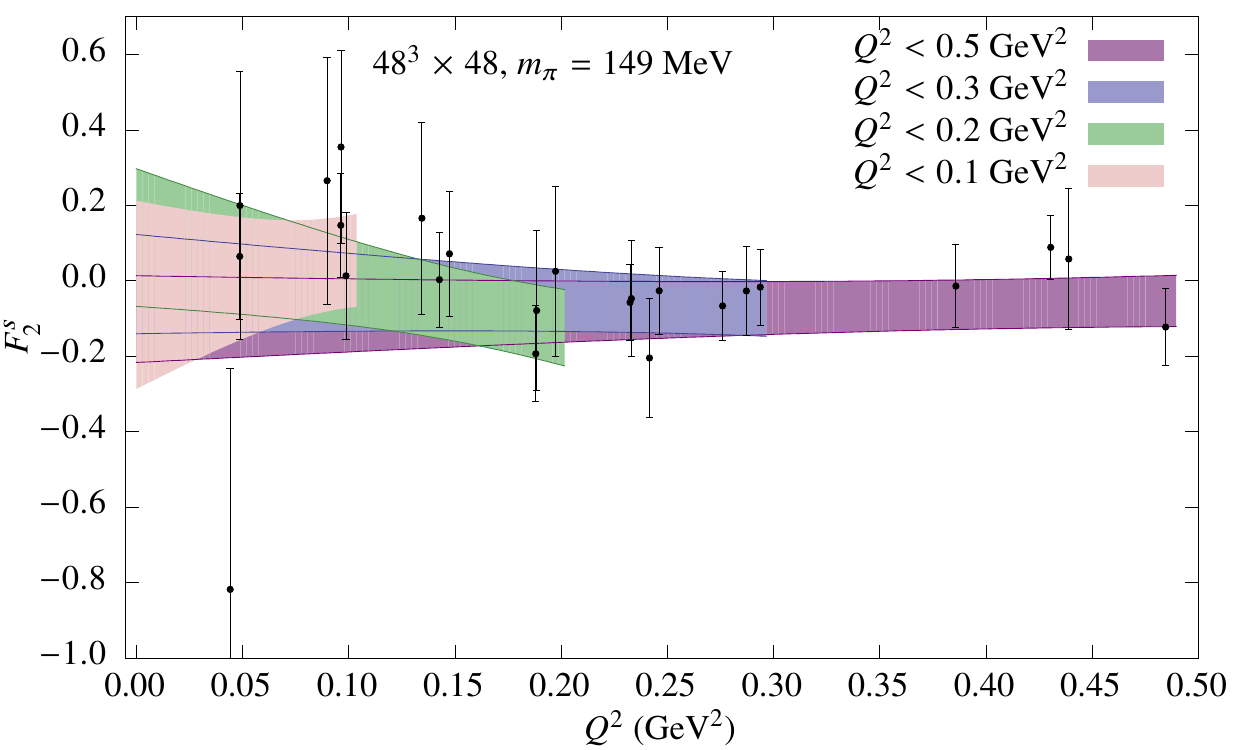}\\
  \includegraphics[width=0.49\textwidth]{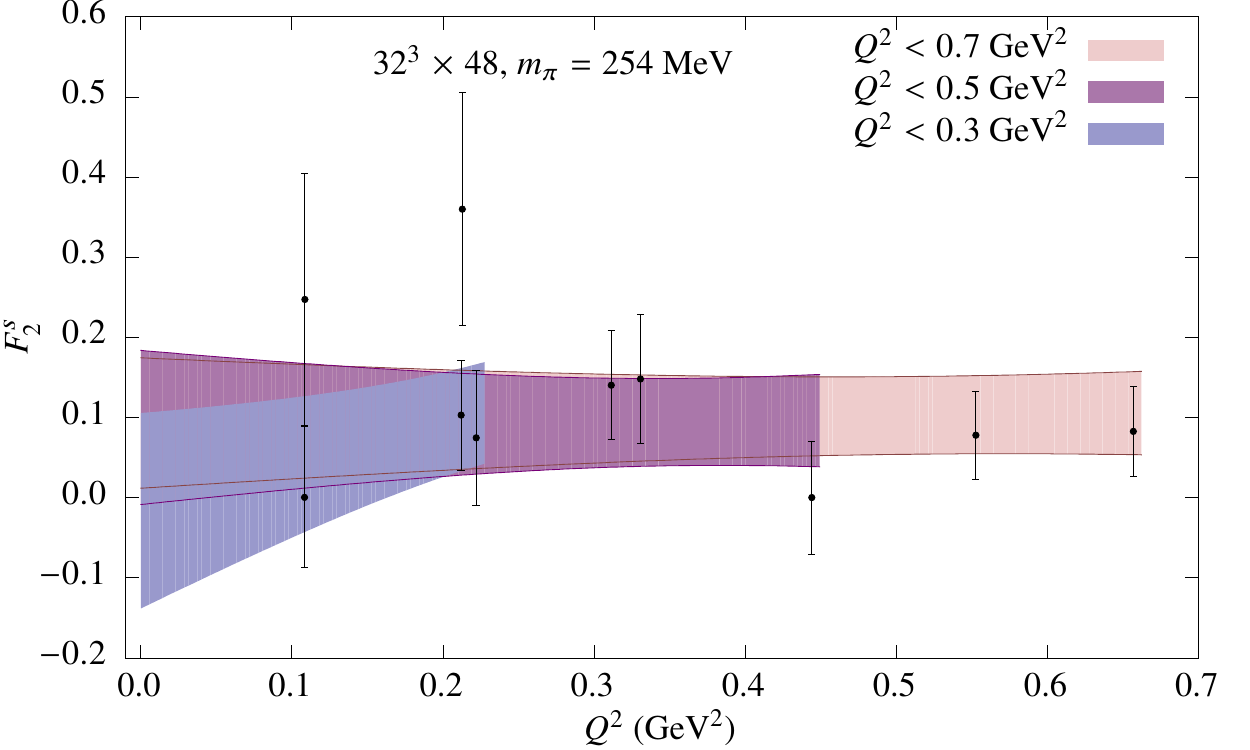}
  \includegraphics[width=0.49\textwidth]{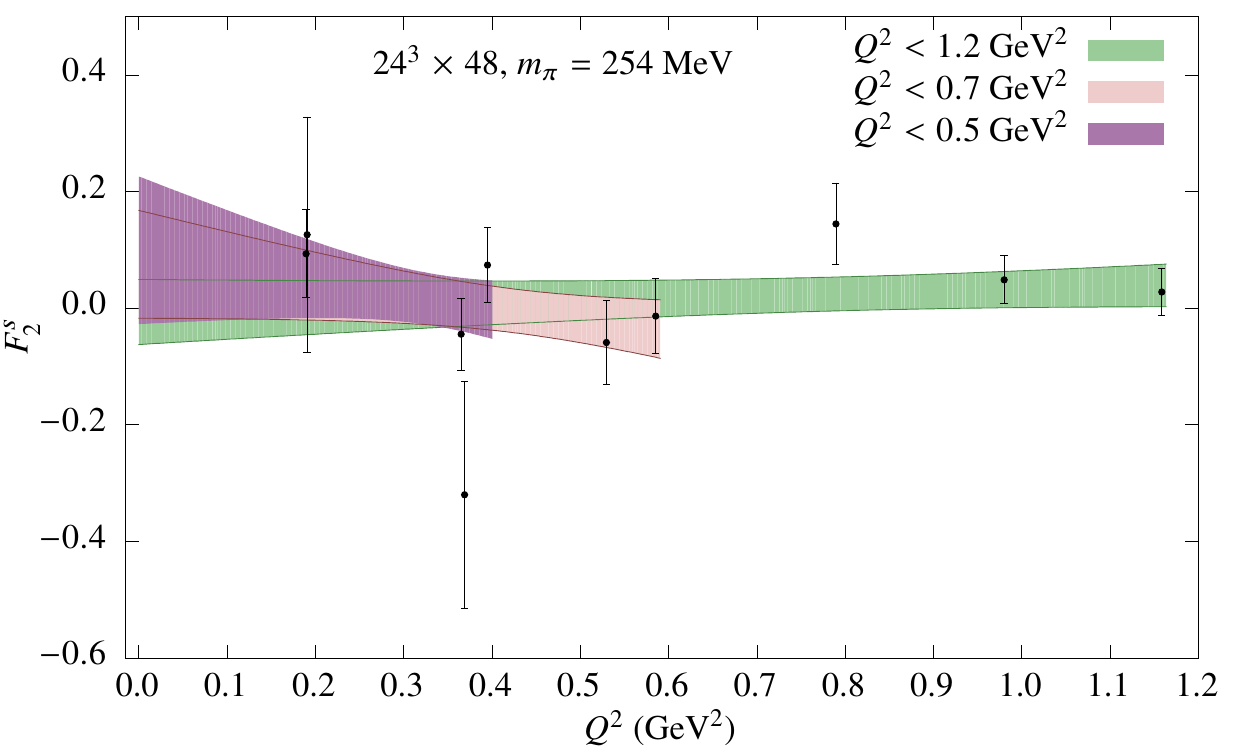}
\\
  \includegraphics[width=0.49\textwidth]{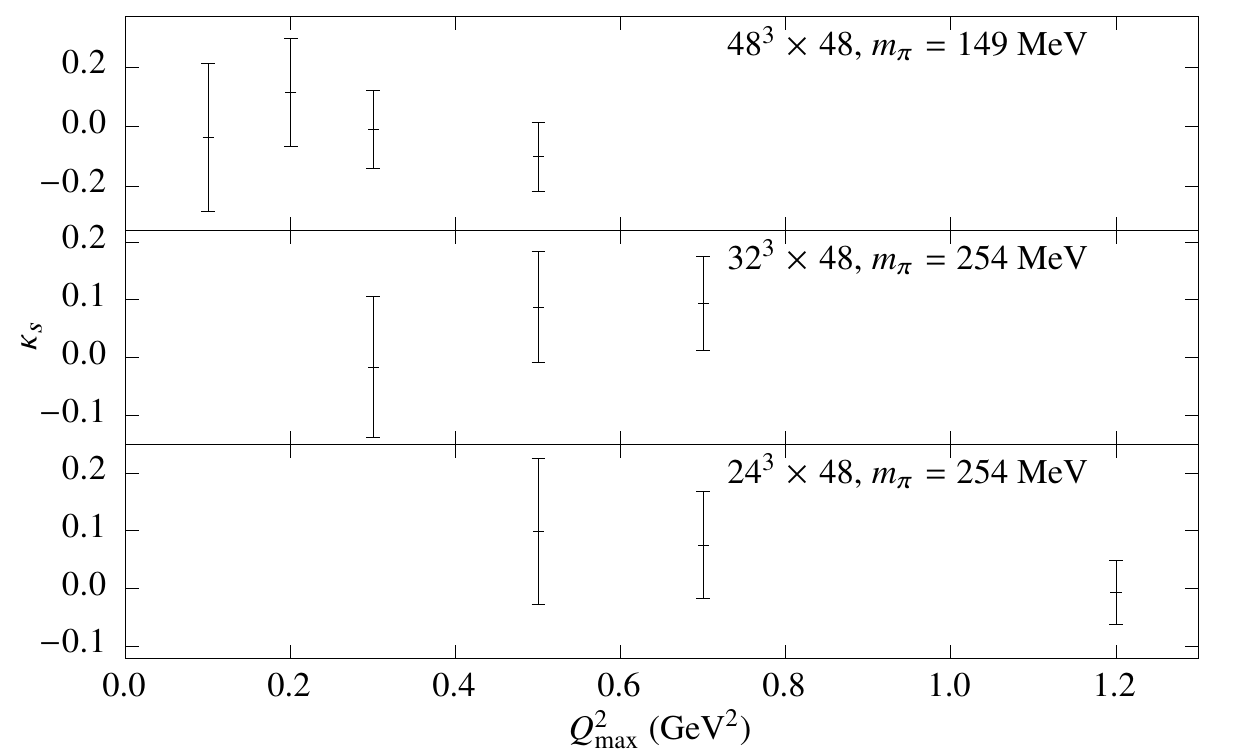}
  \includegraphics[width=0.49\textwidth]{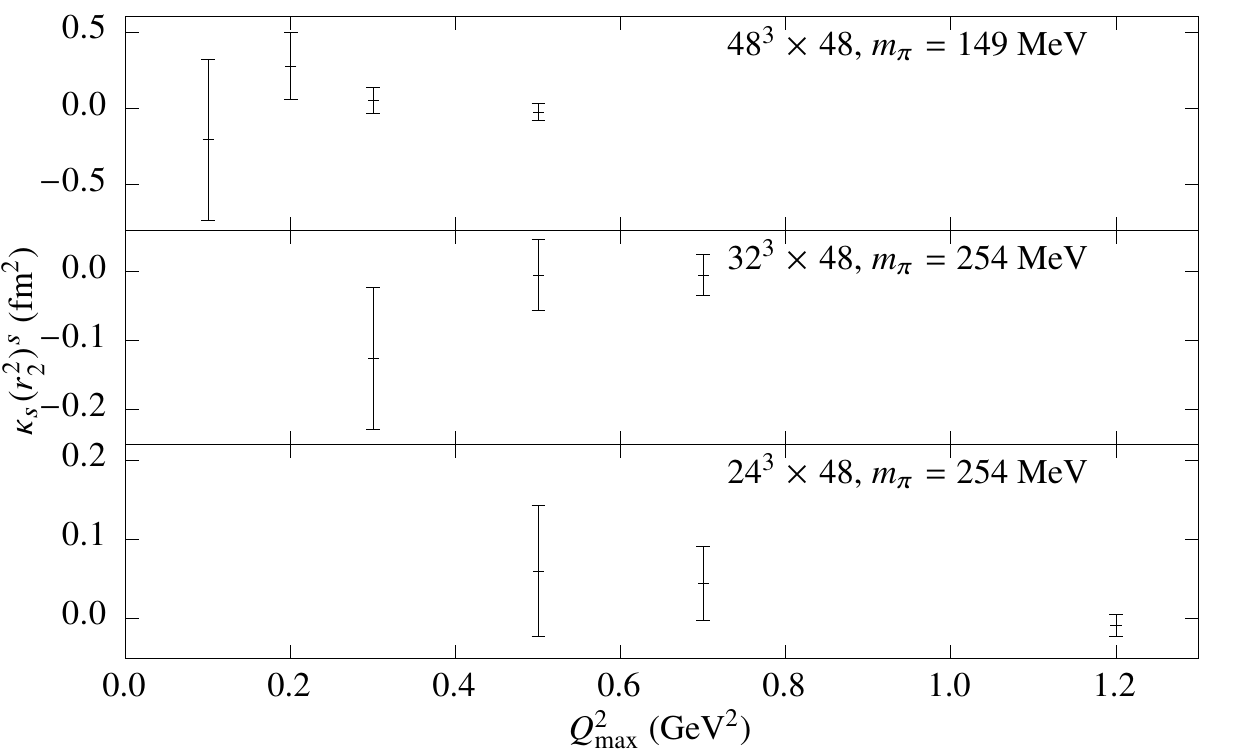}
  \caption{Line fits to $F_2^s(Q^2)$ with varying
    $Q^2_\text{max}$. The last two plots show the dependence on
    $Q^2_\text{max}$ of the fit parameters.}
  \label{fig:F2s_dipole}
\end{figure}

Varying, on three ensembles, the upper bound of the range of $Q^2$
included in the fit, yields the results shown in
Fig.~\ref{fig:F2s_dipole}. The intercept at $Q^2=0$ shows a small
variation with $Q^2_\text{max}$, with a moderate increase in its
statistical uncertainty as $Q^2_\text{max}$ is decreased. The slope at
$Q^2=0$, which is proportional to $\kappa^s(r^2_2)^s$, shows a strong
increase in its statistical uncertainty as $Q^2_\text{max}$ is
decreased. This is caused by the $F_2^s(Q^2)$ data being close to zero
over the sampled range of $Q^2$, which strongly constrains a line that
fits the data to have a small slope when the fitting range is
wider. Although the resulting slopes are statistically consistent with
the result from our choice of $Q^2_\text{max}=0.5\text{ GeV}^2$, it is
clear that data that were more precise and/or at smaller $Q^2$ could
yield significantly different values for the isoscalar Pauli radius.

\subsubsection{Isoscalar Sachs form factors}

\begin{figure}
  \centering
  \includegraphics[width=0.49\textwidth]{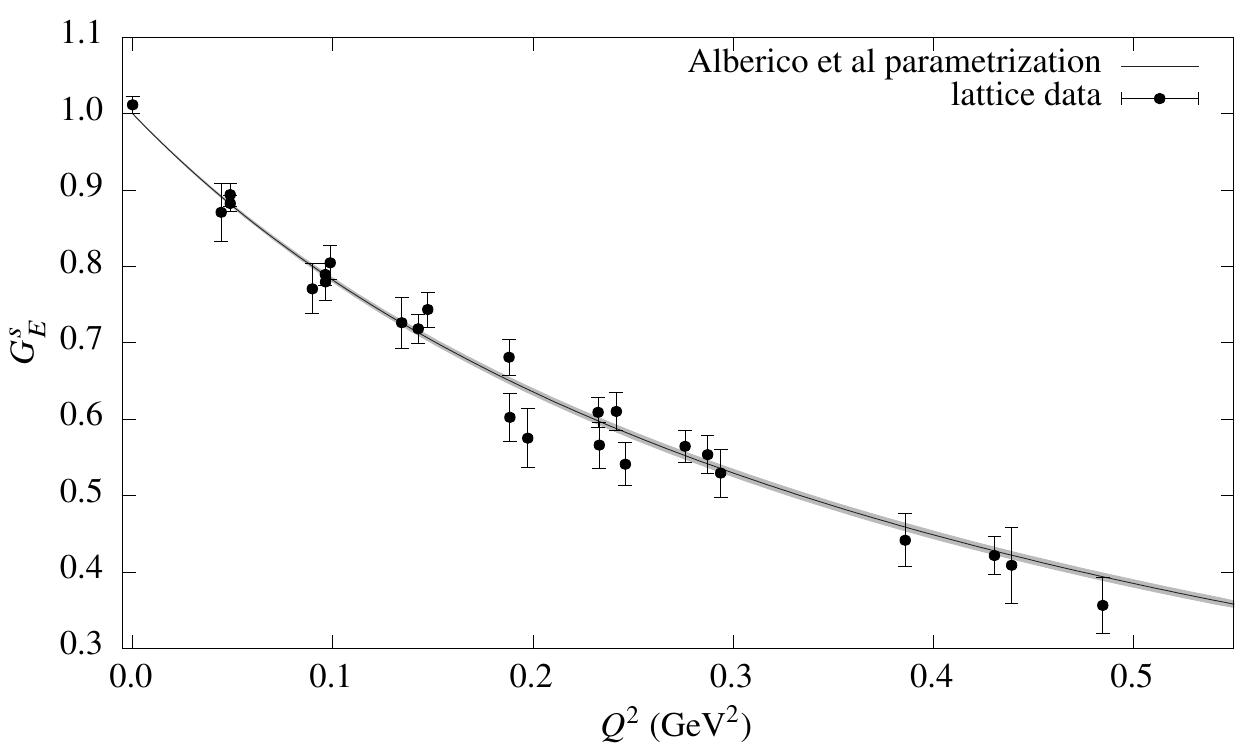}
  \includegraphics[width=0.49\textwidth]{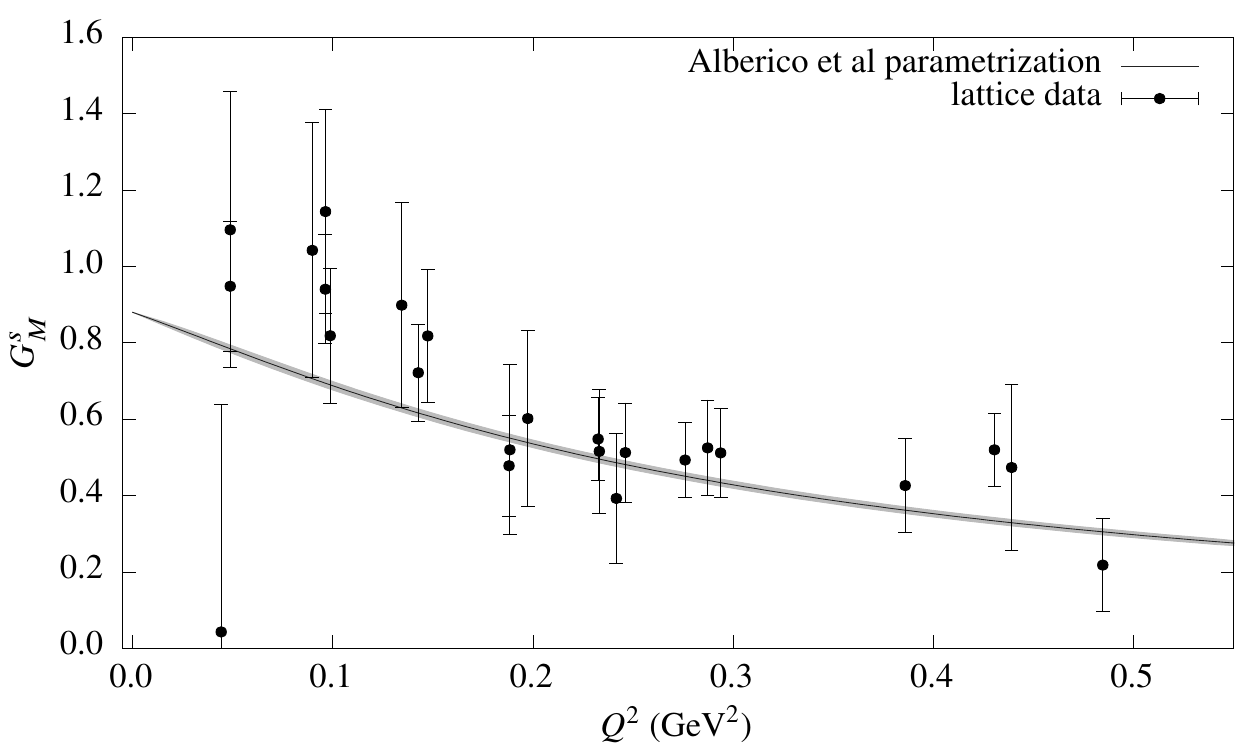}
  \caption{Isoscalar electric and magnetic form factors. Each plot
    contains the curve with error band from the fit to experiment in
    Ref.~\cite{Alberico:2008sz} and the summation data from the
    $m_\pi=149$~MeV ensemble.}
  \label{fig:isoscalar_expt}
\end{figure}

For comparison with experiment without using fits to the lattice data,
we again take the Sachs electric and magnetic form factors, $G_E$ and
$G_M$, on the $m_\pi=149$~MeV ensemble, and compare with the
parameterization of experimental data from
Ref.~\cite{Alberico:2008sz}. This is shown in
Fig.~\ref{fig:isoscalar_expt}. The $G_E$ data are in fairly good
agreement with the curve, whereas the $G_M$ data tend to lie somewhat
above the curve. It should be noted that, as computed at
$m_\pi=317$~MeV, disconnected $G_M$ is
negative~\cite{Meinel_privcomm}, so adding it would bring the data
closer to the curve, although the tendency for $G_M$ to be high could
be caused by other sources, including statistical noise.
Quantitatively, we find $p=0.25$ for $G_E$ and
$p=0.47$ for $G_M$, which are smaller than we found for the isovector
case. We should expect worse agreement with experiment due to the
absence of contributions from disconnected quark contractions, but the
fact that the data are still fairly close to experiment suggests that
the disconnected contributions are not large.

\subsection{Isoscalar Radii and magnetic moment}
The isoscalar Dirac and Pauli radii $(r_{1,2}^2)^s$, and the isoscalar
anomalous magnetic moment are related to the behavior of
$F_{1,2}^s(Q^2)$ near $Q^2=0$ in the same way as for the isovector
case:
\begin{align}
  F_1^s(Q^2) &= 1 - \frac{1}{6}(r_1^2)^sQ^2 + O(Q^4) \\
  F_2^s(Q^2) &= \kappa^s\left(1 - \frac{1}{6}(r_2^2)^sQ^2 + O(Q^4)\right).
\end{align}
We again determine these quantities from the fits described in the
previous section: dipole for $F_1^s(Q^2)$ and line for $F_2^s(Q^2)$.

The version of chiral perturbation theory that we used for isovector
observables is less useful for the isoscalar case, since, at the
presently-available one-loop order, it predicts $(r_1^2)^s$ and
$\kappa^s$ to be independent of $m_\pi$ and $(r_2^2)^s$ to be
zero. Furthermore, the isoscalar obervables are also missing
contributions from disconnected diagrams, so we will not perform a
careful extrapolation to the physical pion mass; instead, we will
simply plot the dependence of the observables on the pion mass and
compare the $m_\pi=149$~MeV ensemble with the experimental results.

\subsubsection{Isoscalar Dirac radius \texorpdfstring{$(r_1^2)^s$}{(r₁²)s}}

\begin{figure}
  \centering
  \includegraphics[width=0.7\textwidth]{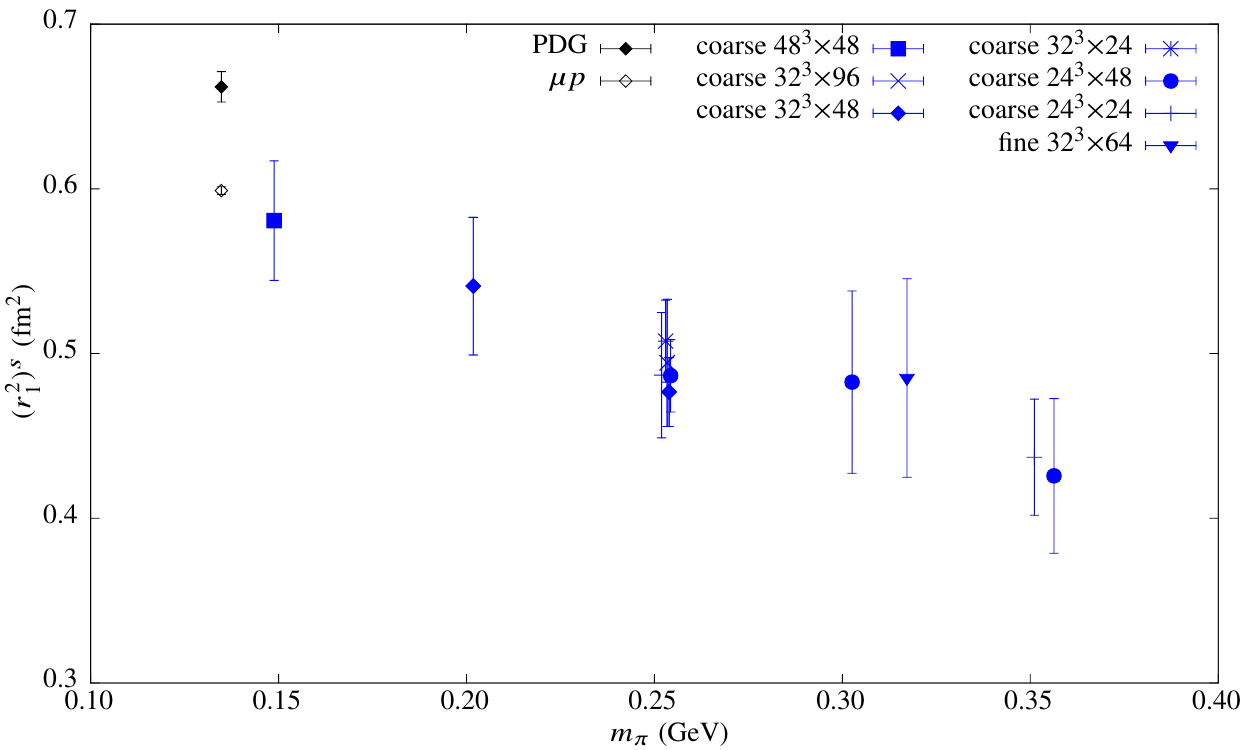}
  \caption{Isoscalar Dirac radius $(r_1^2)^s$, determined on each
    lattice ensemble using the summation method.  Two experimental
    points are shown, where $(r_E^2)^p$ is taken from either the
    CODATA 2010 result~\cite{Mohr:2012tt} used by the
    PDG~\cite{Beringer:1900zz} or the measurement from spectroscopy of
    muonic hydrogen~\cite{Antognini:1900ns}. Both points use the PDG
    value for $(r_E^2)^n$.}
  \label{fig:r1s2_vs_mpi}
\end{figure}

As in the isovector case, the isoscalar Dirac radius shows significant
excited-state effects, with a clear trend of increasing with the
source-sink separation; see Appendix~\ref{app:exc_states}.
The summation-method results, along with the experimental data, are
plotted versus the pion mass in Fig.~\ref{fig:r1s2_vs_mpi}. As the pion
mass decreases, the isoscalar Dirac radius increases, and the result
from the $m_\pi=149$~MeV ensemble is consistent with the lower
experimental point. The multiple ensembles at $m_\pi\approx 250$~MeV
with different volumes and temporal extents all agree well with one
another, indicating the absence of significant finite-volume effects.

\subsubsection{Isoscalar anomalous magnetic moment \texorpdfstring{$\kappa^s$}{κs}}

As in the isovector case [Eq.~(\ref{eq:kv_norm})], we normalize the
isoscalar anomalous magnetic moment to the physical magneton. The
results are shown in Fig.~\ref{fig:ksnorm_vs_mpi} and in
Appendix~\ref{app:exc_states}. There is no clear, consistent sign of
significant excited-state effects or a dependence on the pion
mass. The $m_\pi=149$~MeV ensemble is consistent with the experimental
measurement, albeit with a 100\% statistical uncertainty.

\begin{figure}
  \centering
  \includegraphics[width=0.7\textwidth]{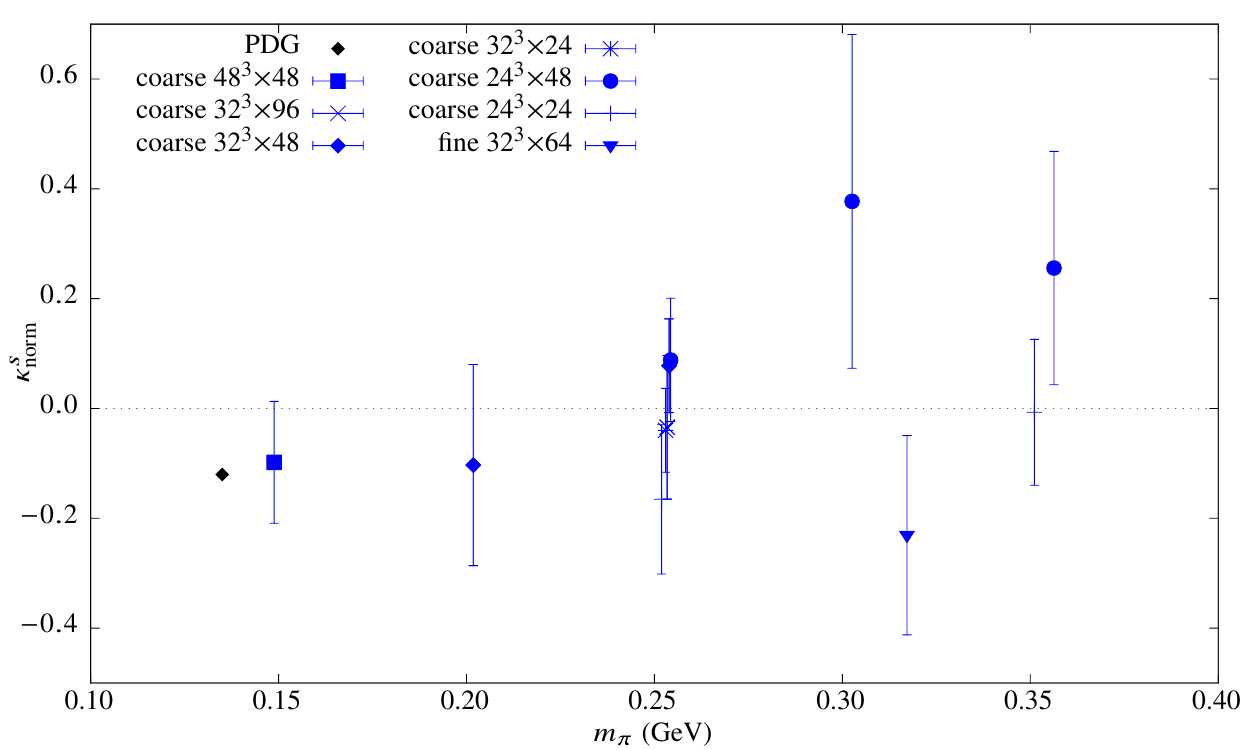}
  \caption{Isoscalar anomalous magnetic moment $\kappa^s_\text{norm}$,
  determined on each lattice ensemble using the summation method.}
  \label{fig:ksnorm_vs_mpi}
\end{figure}

\subsubsection{Isoscalar Pauli radius \texorpdfstring{$(r_2^2)^s$}{(r₂²)s}}

Because $\kappa^s$ is poorly determined from our fits to $F_2^s(Q^2)$,
the combination $\kappa^s(r_2^2)^s$, which is simply proportional to
the slope of $F_2^s$ at $Q^2=0$, is better to work with than the Pauli
radius by itself. We find no clear signal of excited-state effects, as
shown in Appendix~\ref{app:exc_states}. In
Fig.~\ref{fig:r2s2_ksnorm_vs_mpi}, we show the comparison with
experiment. The previously-discussed tendency of the fits to produce
small values for the slope of $F_2^s(Q^2)$ leads to values of
$\kappa^s(r_2^2)^s$ that are close to zero, which is consistent with
the result from dispersion-analysis fits to experimental data.

\begin{figure}
  \centering
  \includegraphics[width=0.7\textwidth]{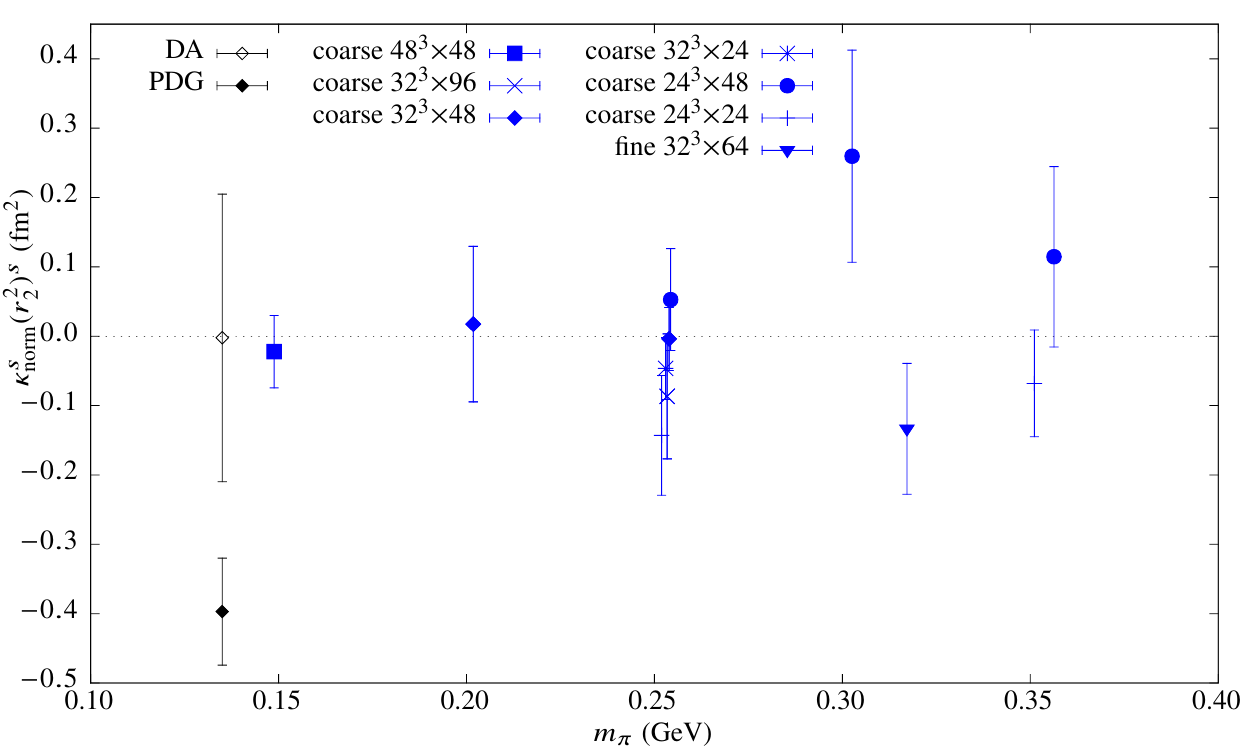}
  \caption{Product of the isoscalar anomalous magnetic moment and
    Pauli radius, $\kappa^s_\text{norm}(r_2^2)^s$, determined on each
    lattice ensemble using the summation method. We show two
    experimental values, where the radii are taken either from the
    2012 PDG~\cite{Beringer:1900zz} or from the dispersion analysis in
    Ref.~\cite{Lorenz:2012tm} (the difference mostly comes from
    different values for the proton magnetic radius).}
  \label{fig:r2s2_ksnorm_vs_mpi}
\end{figure}

\section{\label{sec:proton}Proton sachs form factors}
\begin{figure}
  \centering
  \includegraphics[width=0.49\textwidth]{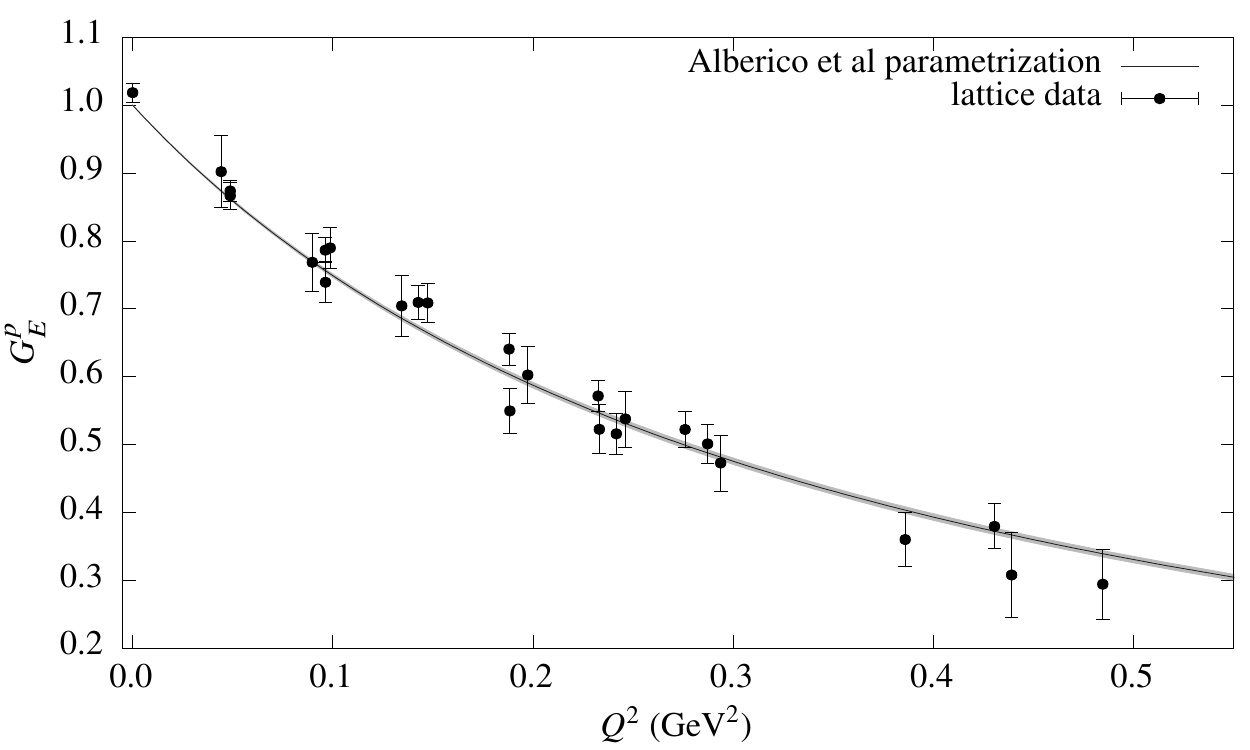}
  \includegraphics[width=0.49\textwidth]{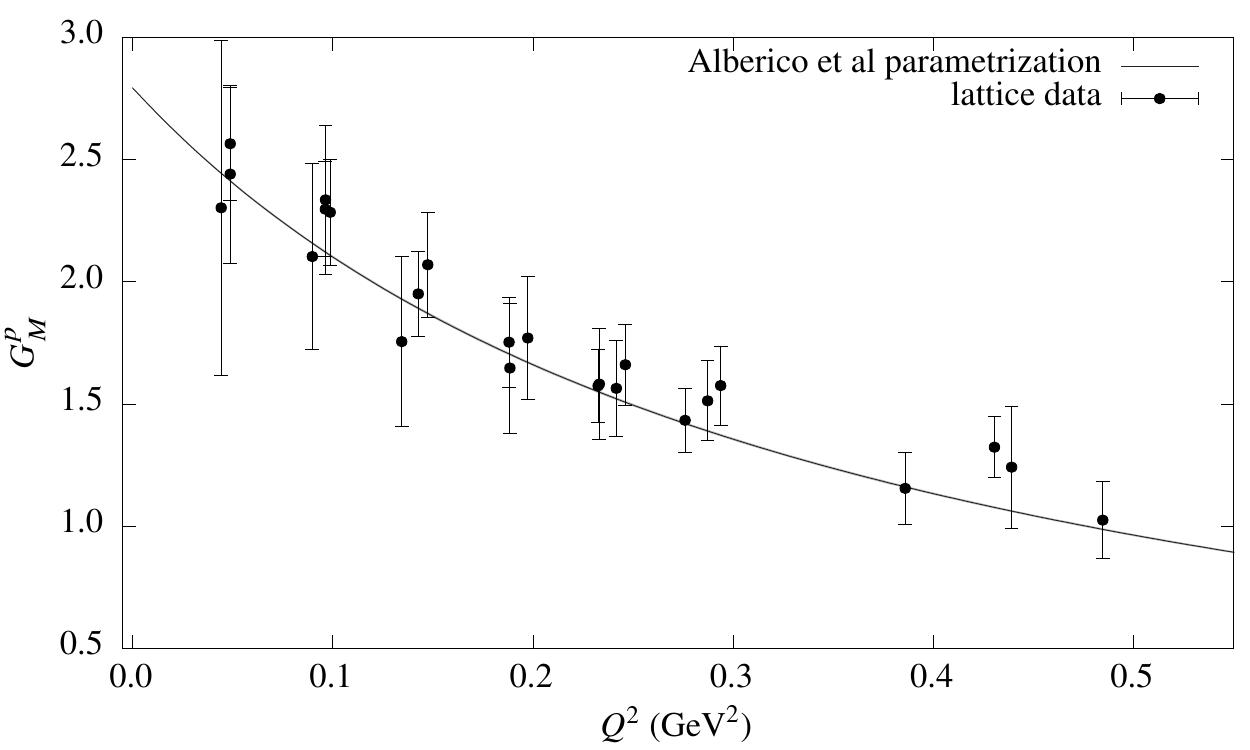}\\
  \includegraphics[width=0.49\textwidth]{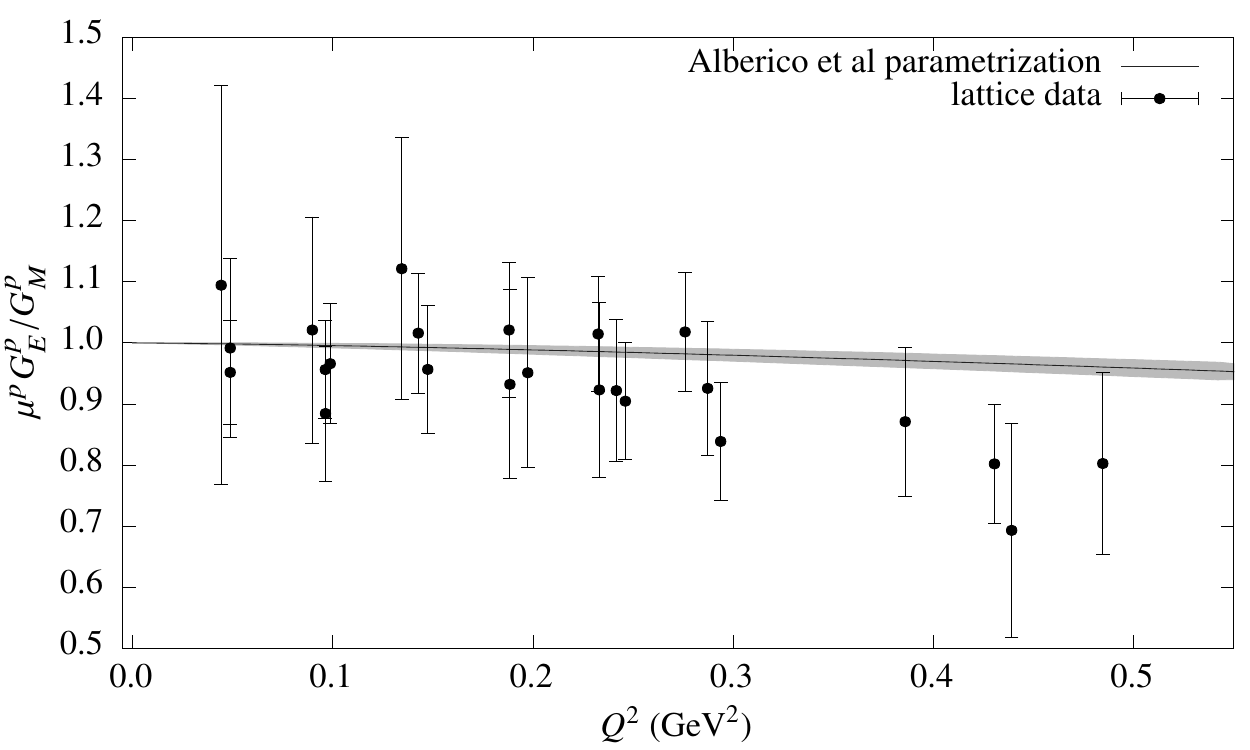}
  \caption{Proton electric and magnetic form factors, and their
    ratio. Each plot contains the curve with error band from the fit
    to experiment in Ref.~\cite{Alberico:2008sz} and the summation
    data from the $m_\pi=149$~MeV ensemble. For the third plot, the
    lattice values for $\mu^pG_E^p/G_M^p$ are scaled using the proton
    magnetic moment from experiment, and not from the fits to lattice
    data.}
  \label{fig:proton_expt}
\end{figure}

For a final comparison with experiment, we consider the proton electric
and magnetic form factors. As in the isoscalar case, the lattice data
are missing the contributions from quark-disconnected diagrams,
although their magnitude here is halved. Furthermore, the magnitude of
the proton magnetic form factor is more than double that of the
isoscalar magnetic form factor, so the relative size of disconnected
contributions is even smaller.

We show the proton $G_E$ and $G_M$ in Fig.~\ref{fig:proton_expt}, for
the summation method on the $m_\pi=149$~MeV ensemble. Unsurprisingly,
given what we saw for the isovector and isoscalar cases in
Figs.~\ref{fig:isovector_expt} and \ref{fig:isoscalar_expt}, there is
again good agreement between the lattice data and the parameterization
of experimental data. Finally, the figure also shows the ratio $\mu
G_E/G_M$, which is often used to probe the discrepancy between
scattering experiments using Rosenbluth separation and those using
polarization transfer. Although the lattice data hint at a decline at
the highest $Q^2$ probed on this ensemble, as seen in the polarization
transfer experiments, much higher values of $Q^2$ are needed to settle
the issue.

\section{\label{sec:conclusions}Conclusions}
\begin{table}
  \caption{\label{tab:results}Comparison of isovector and isoscalar radii
   and magnetic moments with experiment. For all observables, the
   lattice result from the summation method on the $m_\pi=149$~MeV
   ensemble is shown, and for isovector observables the extrapolated
   value and the goodness-of-fit are also shown. The first set of
   experimental values are derived using inputs from the
   PDG~\cite{Beringer:1900zz}, while for the second values for the
   Dirac radii, the proton charge radius was taken from muonic
   hydrogen spectroscopy~\cite{Antognini:1900ns}, and for the second
   values for the Pauli radii, the proton and neutron radii were taken
   from the dispersion-analysis fits in Ref.~\cite{Lorenz:2012tm}.}
  \begin{tabular}{l|D{.}{.}{8}D{.}{.}{7}c|D{.}{.}{6}cD{.}{.}{6}}
    \hline\hline
    $X$ & \multicolumn{1}{c}{$X^\text{lat}_{m_\pi=149\text{ MeV}}$} &
     \multicolumn{1}{c}{$X^\text{lat}_\text{extrap}$} & $\chi^2/\text{dof}$ &
     \multicolumn{3}{c}{$X^\text{exp}$} \\
    \hline
    $(r_1^2)^v$ ($\text{fm}^2$) & 0.498(55) &
    0.605(27)\footnote{Including an additional term proportional to $m_\pi^2$ yields an extrapolated $(r_1^2)^v=0.539(57)\text{ fm}^2$ with $\chi^2/\text{dof}=0.01/2$.} & 1.7/3 &
    0.640(9) & or & 0.578(2) \\
    $\kappa^v$ & 3.76(38) &
    3.68(38) & 1.8/2 &
    3.706 & & \\
    $\kappa^v(r_2^2)^v$ ($\text{fm}^2$) & 2.68(62) &
    2.59(24) & 1.2/3 &
    2.47(8) & or & 2.96(21) \\
    \hline
    $(r_1^2)^s$ ($\text{fm}^2$) & 0.581(36) &
    & &
    0.662(9) & or & 0.599(2) \\
    $\kappa^s$ & -0.10(11) &
    & &
    -0.120 & & \\
    $\kappa^s(r_2^2)^s$ ($\text{fm}^2$) & -0.02(5) &
    & &
    -0.40(8) & or & 0.00(21) \\
    \hline\hline 
  \end{tabular}
\end{table}

The essential result of this work is that we have achieved excellent
agreement with experiment for the Sachs form factors, shown in
Figs.~\ref{fig:isovector_expt}, \ref{fig:isoscalar_expt}, and
\ref{fig:proton_expt}, and the Dirac radius, Pauli radius, and
magnetic moment, as summarized in Tab.~\ref{tab:results}. This was
achieved by using the near-physical pion mass of 149~MeV and reducing
the amount of contamination from excited states. For the Dirac radius,
we found a strong signal of significant excited-state effects across
all lattice ensembles, whereas for other observables these effects
were most clearly seen in the $m_\pi=149$~MeV ensemble.

Because of the importance of controlling the systematic error due to
contamination from excited states, we have studied the three methods,
ratio, summation, and GPoF, and provided the most comprehensive
comparison to date. We used the summation method, which is robust and
widely used by the community, for our primary analysis and showed that
within the present statistics, the results of all three are
consistent.

The multiple ensembles with the same pion mass $m_\pi\approx 250$~MeV
and varying spatial and temporal extents $L_s$ and $L_t$ allow for
studying finite-volume and finite-temperature effects; we find
excellent agreement for the Dirac radius between these ensembles and
also good agreement for the other observables. This was reported in
more detail in a separate study~\cite{Green:2013hja}. We also used one
ensemble with a finer lattice spacing and find no sign of large
discretization effects.

For the isoscalar form factors, we found similar results as in the
isovector case, except that the current level of precision is
insufficient for the isoscalar Pauli form factor to clearly differ
from zero. Their consistency with experiment, as again summarized in
Tab.~\ref{tab:results}, in the absence of contributions from
disconnected diagrams suggests that the latter are small. This is
consistent with the size of disconnected contributions from studies
with pion masses between 300 and 400~MeV and with indirect
determinations at the physical point, although these also need to be
calculated directly using lattice QCD close to the physical pion mass.

An important goal is an \emph{ab initio} calculation of the proton
charge radius and form factors at very low momentum transfer to help
understand the origin of the apparently inconsistent experimental
results. Although finite-volume and discretization effects appear to
be small, confirmation at the physical pion mass is required in order
to have fully-controlled systematic errors. Better control over
excited-state effects is needed, ideally using several source-sink
separations and very high statistics to confirm that different
analysis methods converge to the same ground-state matrix
elements. Finally, the determination of the derivative of $F_1$ at
$Q^2=0$ needs to be better-controlled; this will be helped by the use
of larger volumes which give access to $F_1$ at smaller values of
$Q^2$, or by the exploration of alternative techniques such as the one
proposed in Ref.~\cite{deDivitiis:2012vs} for directly computing
momentum-derivatives of matrix elements.

\begin{acknowledgments}
We thank Zoltan Fodor for useful discussions and the
Budapest-Marseille-Wuppertal collaboration for making some of their
configurations available to us. This research used resources of the
Argonne Leadership Computing Facility at Argonne National Laboratory,
which is supported by the Office of Science of the U.S.\ Department of
Energy under contract \#DE--AC02--06CH11357, and resources at
Forschungszentrum Jülich.

During this research JRG, SK, JWN, AVP and SNS were supported in part
by the U.S.\ Department of Energy Office of Nuclear Physics under
grant \#DE--FG02--94ER40818, ME was supported in part by DOE grant
\#DE--FG02--96ER40965, SNS was supported by Office of Nuclear Physics
in the US Department of Energy's Office of Science under Contract
\#DE--AC02--05CH11231, SK was supported in part by Deutsche
Forschungsgemeinschaft through grant SFB--TRR~55, and JRG was
supported in part by the PRISMA Cluster of Excellence at the
University of Mainz.

Calculations for this project were done using the Qlua software
suite~\cite{Qlua}.
\end{acknowledgments}

\appendix

\section{\label{app:chpt}Chiral extrapolation}
We will largely use the same methods and phenomenological inputs for
chiral perturbation theory as
Refs.~\cite{Green:2012ud,Bratt:2010jn}. In particular, we use the
following values in the chiral limit: the pion decay constant,
\begin{equation}
  F_\pi=86.2\text{ MeV},
\end{equation}
the delta-nucleon mass splitting,
\begin{equation}
  \Delta = 293\text{ MeV},
\end{equation}
and the nucleon axial charge,
\begin{equation}
  g_A = 1.26.
\end{equation}

The nucleon isovector Dirac and Pauli form factors are given in heavy
baryon ChPT including the delta baryon, to order $\epsilon^3$ in the
small-scale expansion ($\epsilon\in\{p,m_\pi,\Delta\}$) in
Ref.~\cite{Bernard:1998gv}. This gives an expression for the Dirac
radius~\cite{Syritsyn:2009mx},
\begin{equation}
\begin{aligned}\label{eq:r1v2_chpt}
  (r_1^v)^2 &= - \frac{1}{(4\pi F_\pi)^2}\left[1 + 7g_A^2 + (2+10g_A^2)\log\left(\frac{m_\pi}{\lambda}\right)\right] - \frac{12B_{10}^r(\lambda)}{(4\pi F_\pi)^2}\\
  &\quad + \frac{c_A^2}{54\pi^2 F_\pi^2}\left[26 + 30\log\left(\frac{m_\pi}{\lambda}\right) + 30\frac{\Delta}{\sqrt{\Delta^2-m_\pi^2}}\log\left(\frac{\Delta}{m_\pi} + \sqrt{\frac{\Delta^2}{m_\pi^2}-1}\right)\right], \\
\end{aligned}
\end{equation}
where $c_A$ is the leading-order pion-nucleon-delta coupling in the
chiral limit, which we set to 1.5~\cite{Syritsyn:2009mx}, and
$B_{10}^r(\lambda)$ is a counterterm and the single free parameter.

For the anomalous magnetic moment, we include the modification from
Ref.\ \cite{Hemmert:2002uh}:
\begin{equation}
\begin{aligned}\label{eq:kv_chpt}
\kappa^v &= \kappa^v_0 - \frac{g_A^2 m_\pi m_N}{4\pi F_\pi^2} 
+ \frac{2c_A^2\Delta m_N}{9\pi^2 F_\pi^2}\left[\sqrt{1-\frac{m_\pi^2}{\Delta^2}}\log\left(\frac{\Delta}{m_\pi}+\sqrt{\frac{\Delta^2}{m_\pi^2}-1}\right) + \log\frac{m_\pi}{2\Delta}\right] \\
&\quad - 8E_1^r(\lambda) m_N m_\pi^2 + \frac{4c_Ac_Vg_Am_Nm_\pi^2}{27\pi^2 F_\pi^2\Delta}\left( 3\Delta\log\frac{2\Delta}{\lambda} + \pi m_\pi\right)\\
&\quad - \frac{8c_Ac_Vg_A\Delta^2 m_N}{27\pi^2 F_\pi^2}\left[\left(1\!-\!\frac{m_\pi^2}{\Delta^2}\right)^{3/2}\log\left(\frac{\Delta}{m_\pi}+\sqrt{\frac{\Delta^2}{m_\pi^2}\!-\!1}\right) + \left(1\!-\!\frac{3m_\pi^2}{2\Delta^2}\right)\log\frac{m_\pi}{2\Delta}\right],
\end{aligned}
\end{equation}
where $c_V$ is the leading photon-nucleon-delta coupling in the chiral
limit, which we set to $-2.5\text{ GeV}^{-1}$, and we use the physical
nucleon mass $m_N=939$~MeV. The two free parameters are $\kappa^v_0$
and the counterterm $E_1^r(\lambda)$.

The combination $\kappa^v(r_2^2)^v$ is more natural in ChPT than the
Pauli radius alone; we include the $O(m_\pi^0)$ ``core'' contribution
from Ref.~\cite{Gockeler:2003ay} in the expression for it:
\begin{equation}\label{eq:r2v2_chpt}
\kappa^v(r_2^v)^2 = \frac{g_A^2 m_N}{8\pi F_\pi^2 m_\pi} + \frac{c_A^2 m_N}{9\pi^2 F_\pi^2 \sqrt{\Delta^2 - m_\pi^2}}\log\left(\frac{\Delta}{m_\pi} + \sqrt{\frac{\Delta^2}{m_\pi^2}-1}\right) + 24m_N\mathcal{C},
\end{equation}
where $\mathcal{C}$ is the single free parameter.

\section{\label{app:exc_states}Additional comparisons of methods for computing matrix elements}
In this appendix, we show comparisons of the ratio, summation, and
GPoF methods discussed in Sec.~\ref{sec:ff_extract} to compute matrix
elements, for observables where these details were omitted in the main
text and for additional ensembles.

\subsection{Form factors}

\begin{figure}
  \centering
  \begin{minipage}{.5\textwidth}
    \centering
    \includegraphics[width=\textwidth]{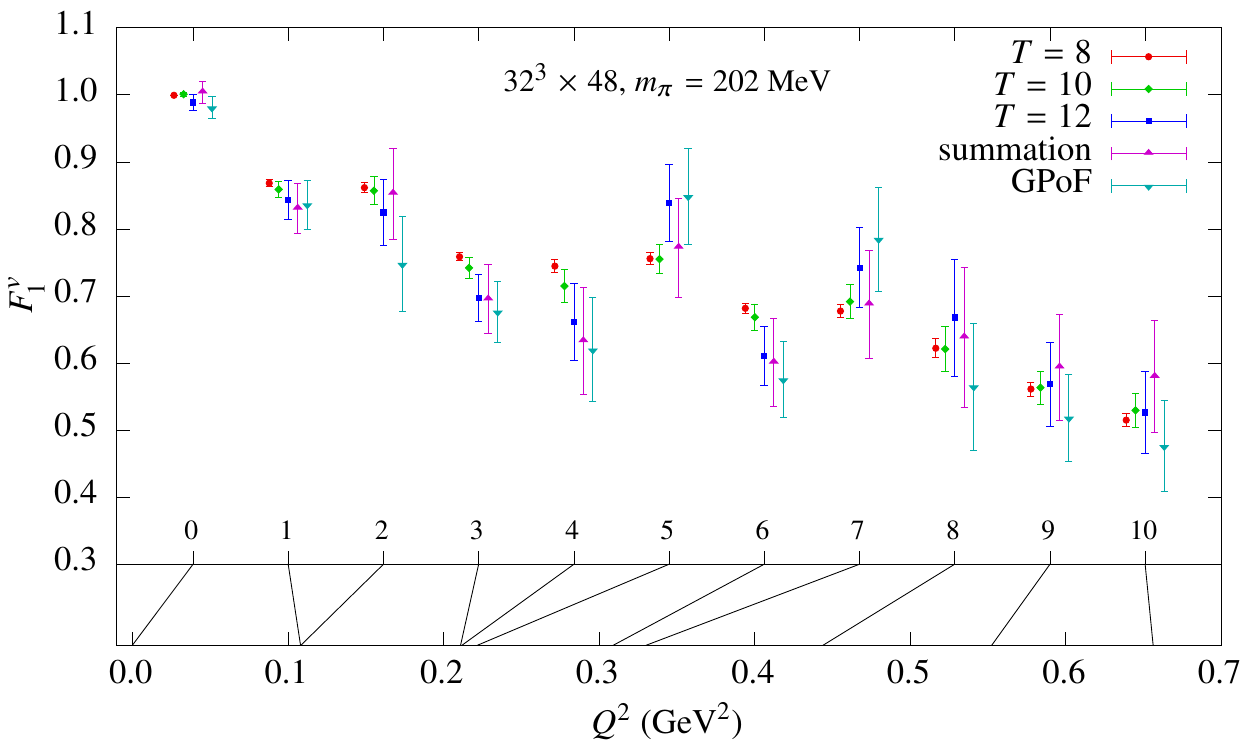}\\
    \includegraphics[width=\textwidth]{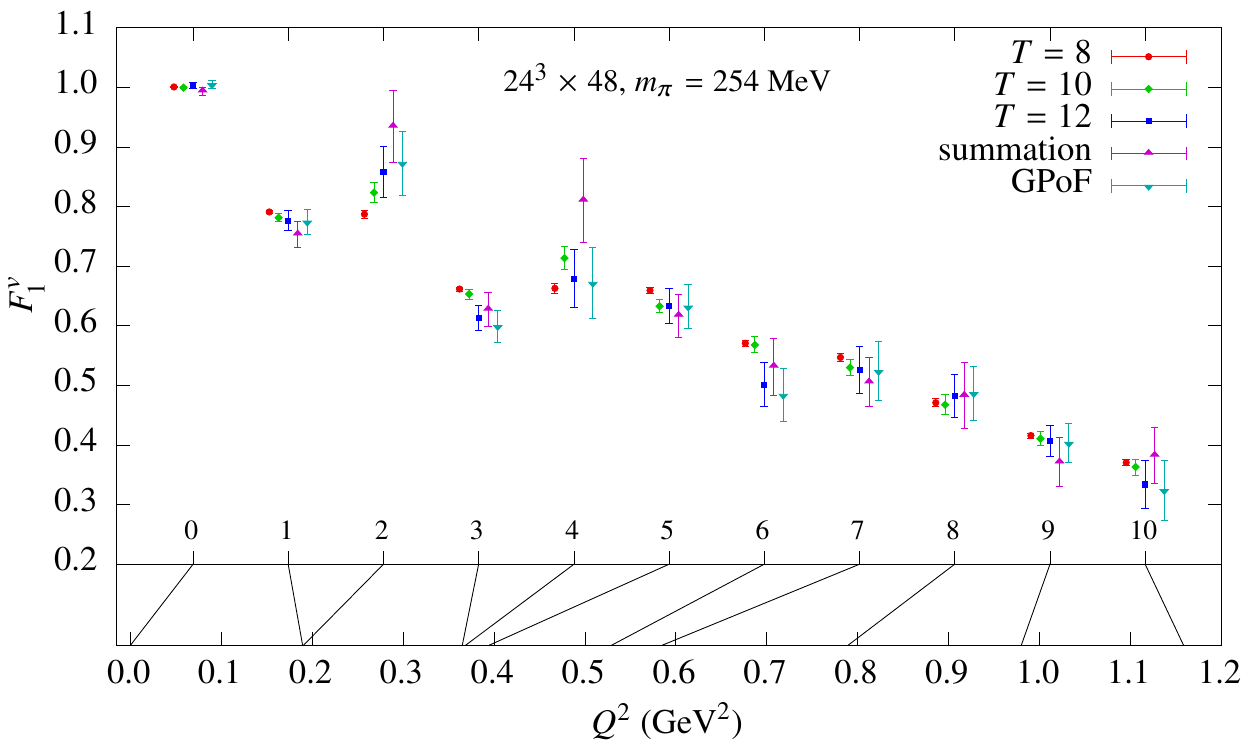}
  \end{minipage}~
  \begin{minipage}{.5\textwidth}
    \centering
    \includegraphics[width=\textwidth]{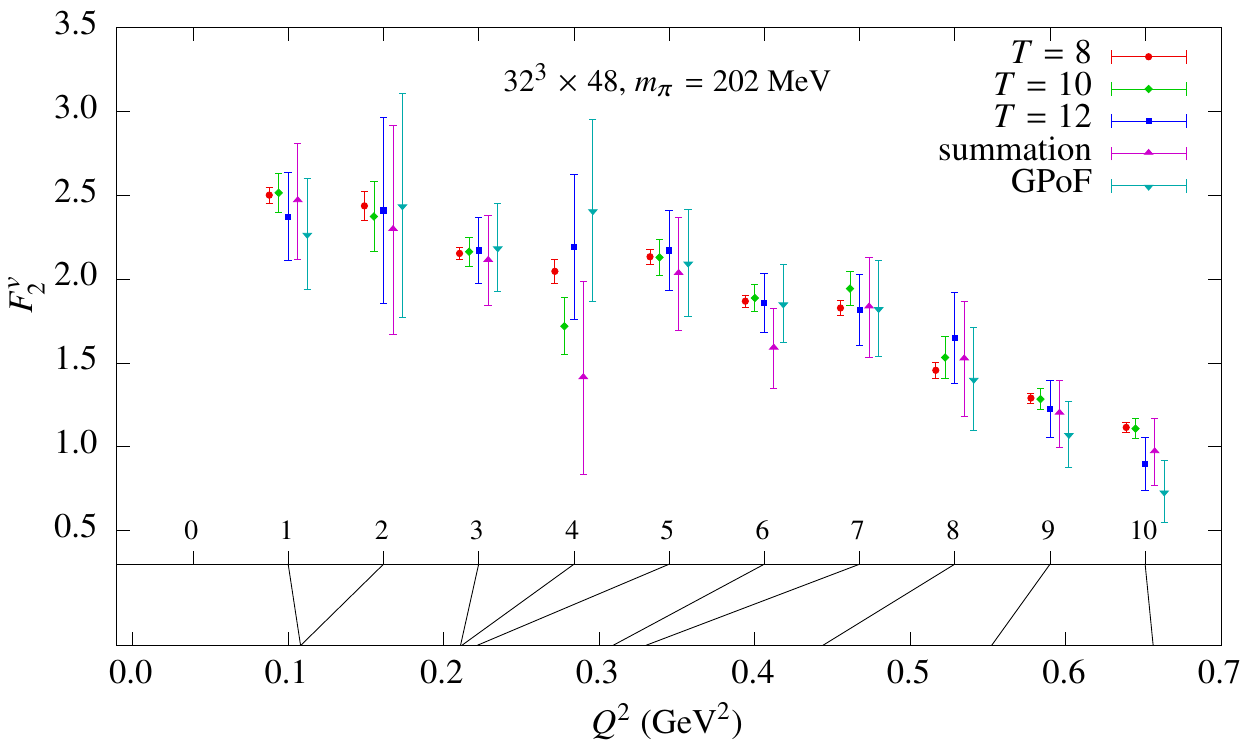}\\
    \includegraphics[width=\textwidth]{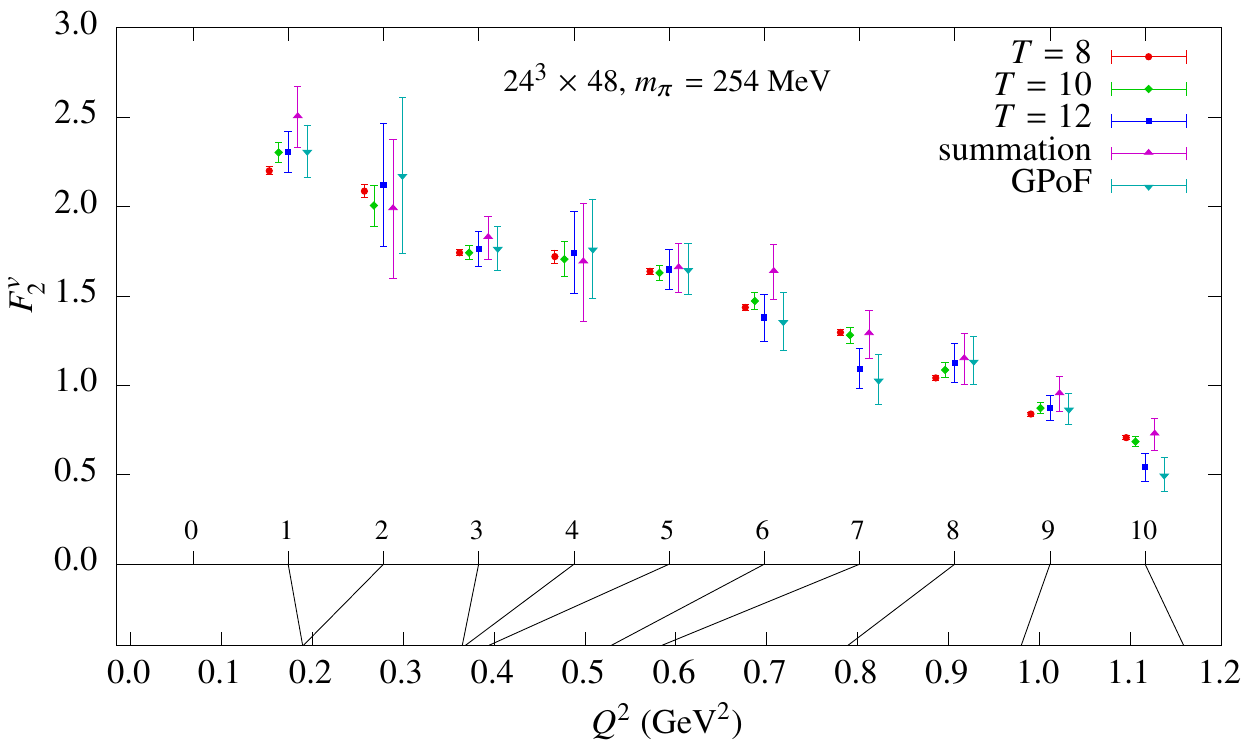}
  \end{minipage}
  \caption{\label{fig:ff_extract_comparison_2}Comparison
    of different methods to extract the ground state isovector form
    factors $F_1^v(Q^2)$ and $F_2^v(Q^2)$. The upper plots show
    the $m_\pi=202$~MeV ensemble and the lower plots show the
    $m_\pi=254$~MeV, $24^3\times 48$ ensemble.}
\end{figure}

In Fig.~\ref{fig:ff_extract_comparison_2}, we show the isovector Dirac
and Pauli form factors for two additional ensembles, cf.\
Fig.~\ref{fig:ff_extract_comparison}. Signs of excited-state effects
are much less clear and consistent here than they were for the two
previously-shown ensembles.

\begin{figure}
  \centering
  \begin{minipage}{.5\textwidth}
    \centering
    \includegraphics[width=\textwidth]{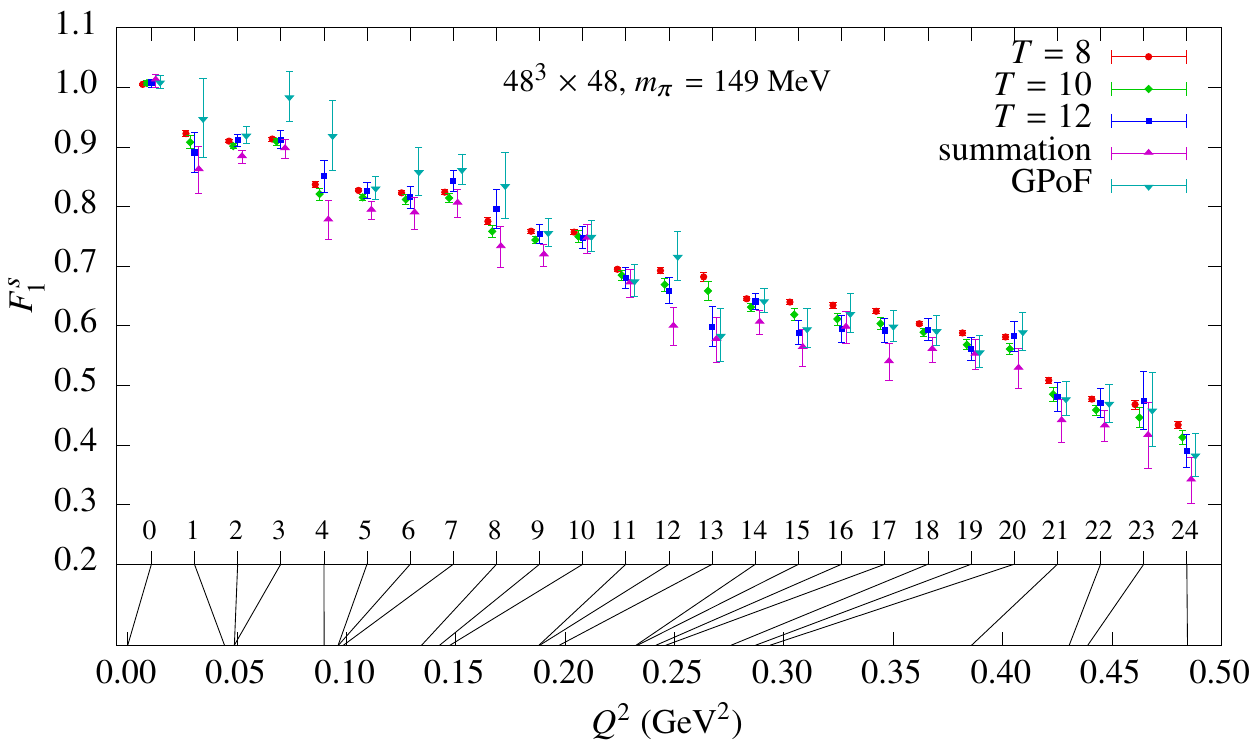}\\
    \includegraphics[width=\textwidth]{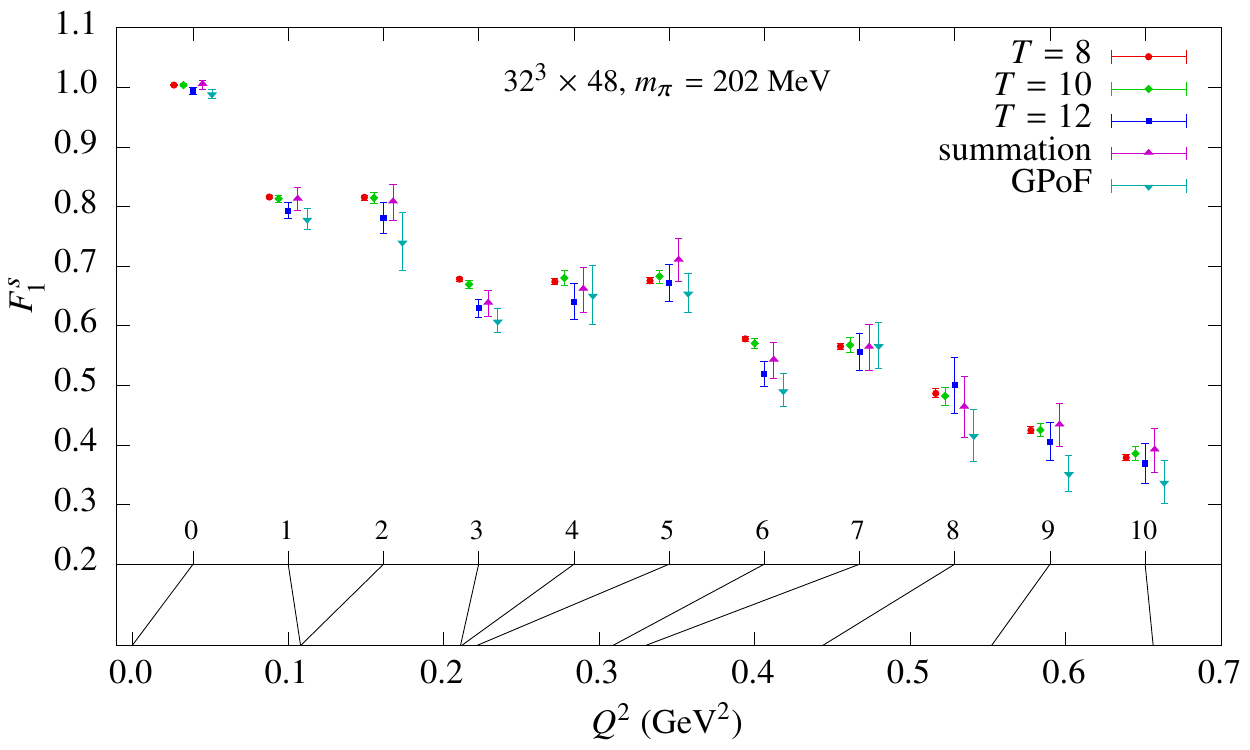}\\
    \includegraphics[width=\textwidth]{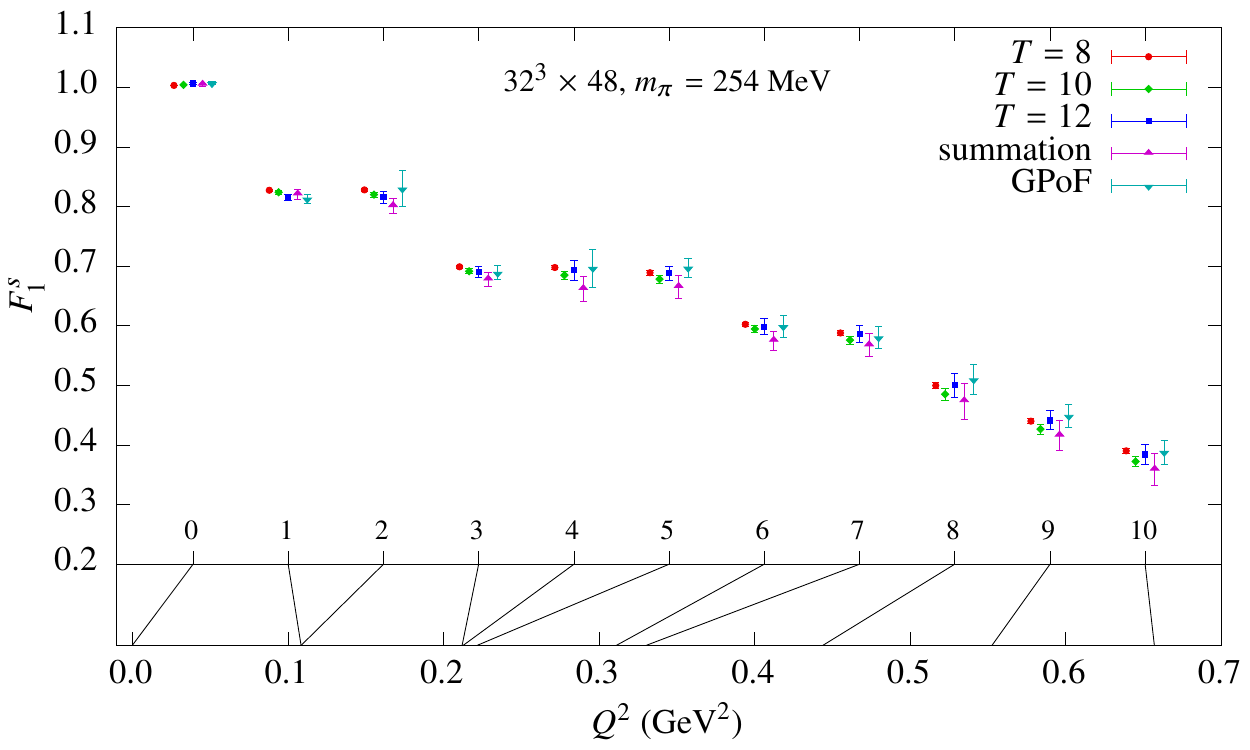}\\
    \includegraphics[width=\textwidth]{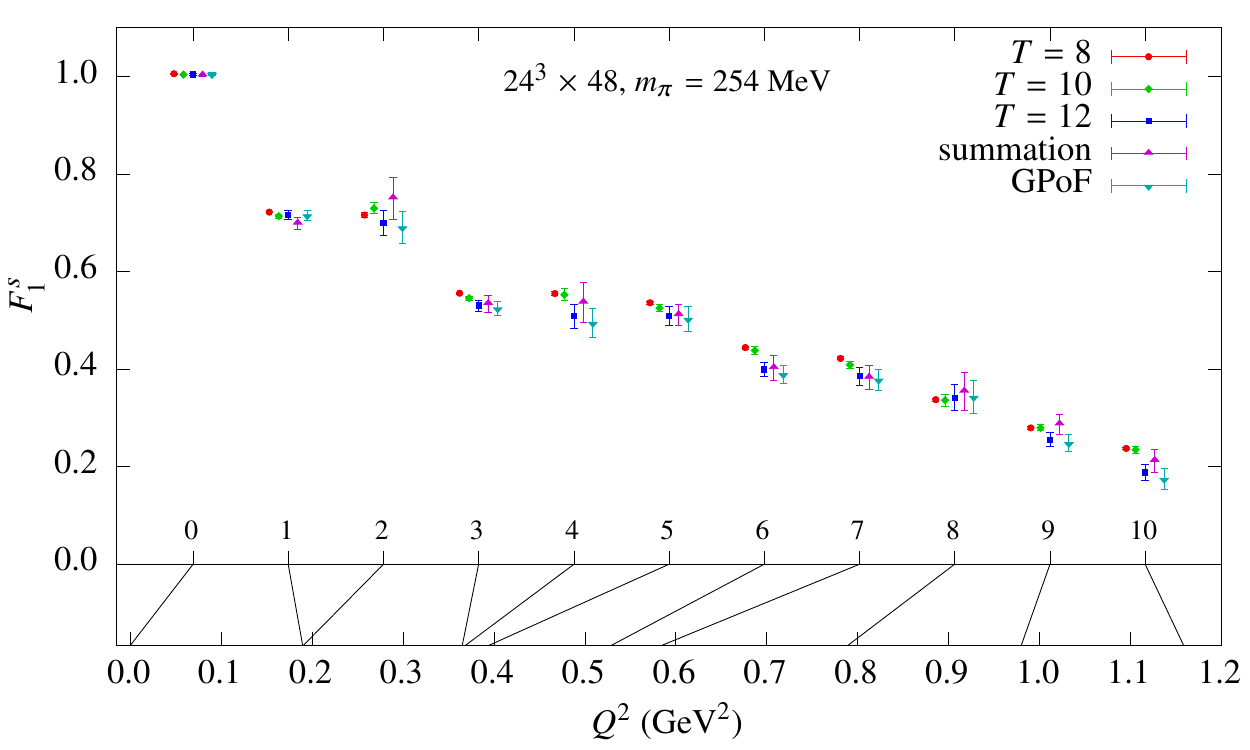}
  \end{minipage}~
  \begin{minipage}{.5\textwidth}
    \centering
    \includegraphics[width=\textwidth]{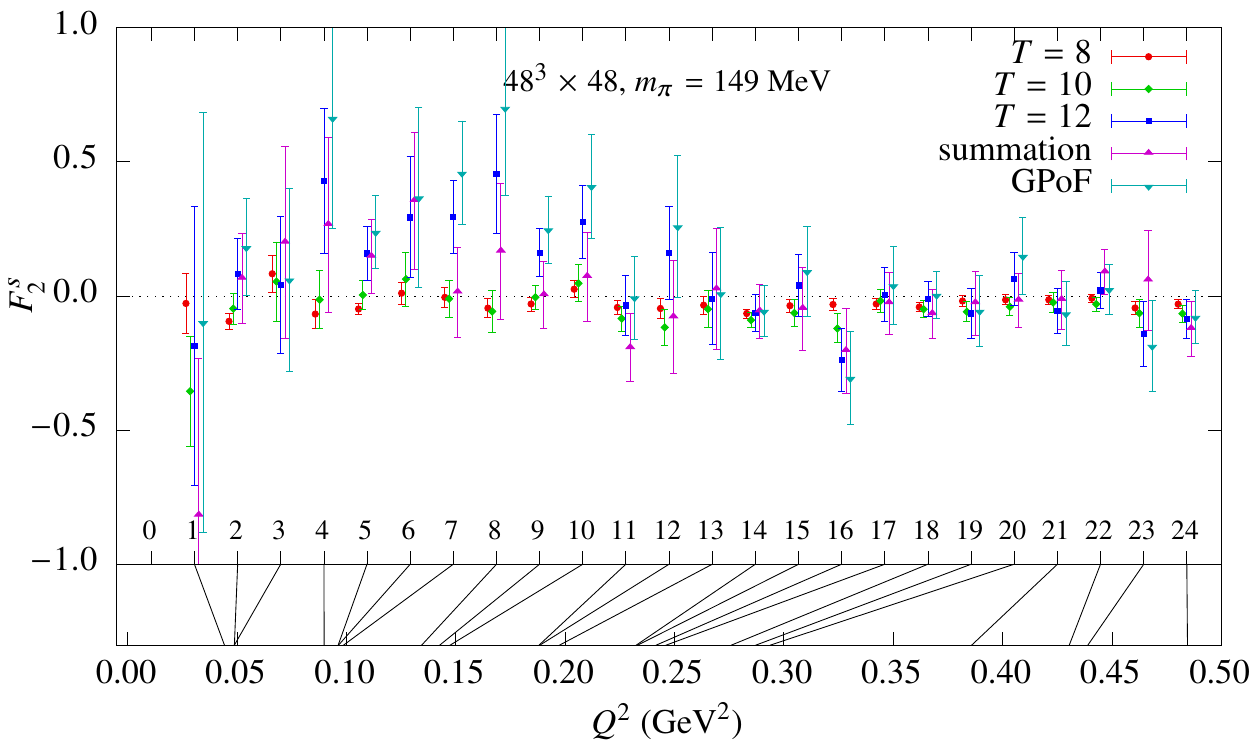}\\
    \includegraphics[width=\textwidth]{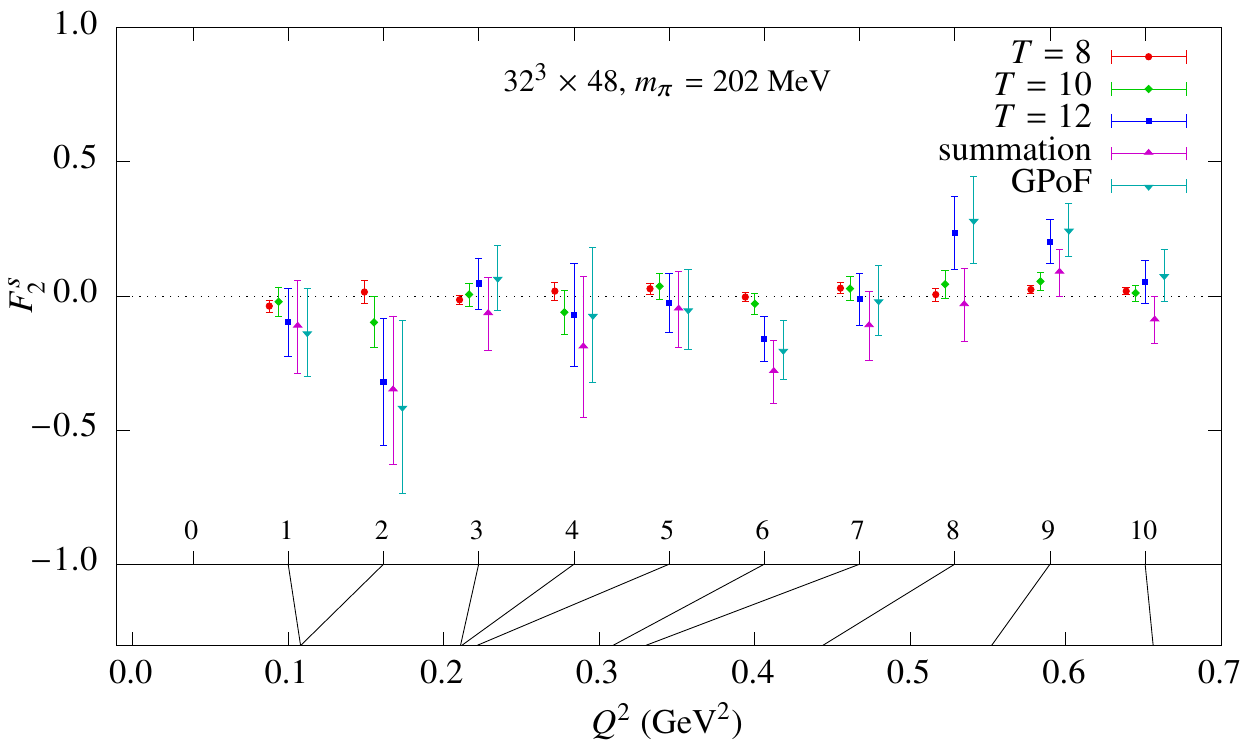}\\
    \includegraphics[width=\textwidth]{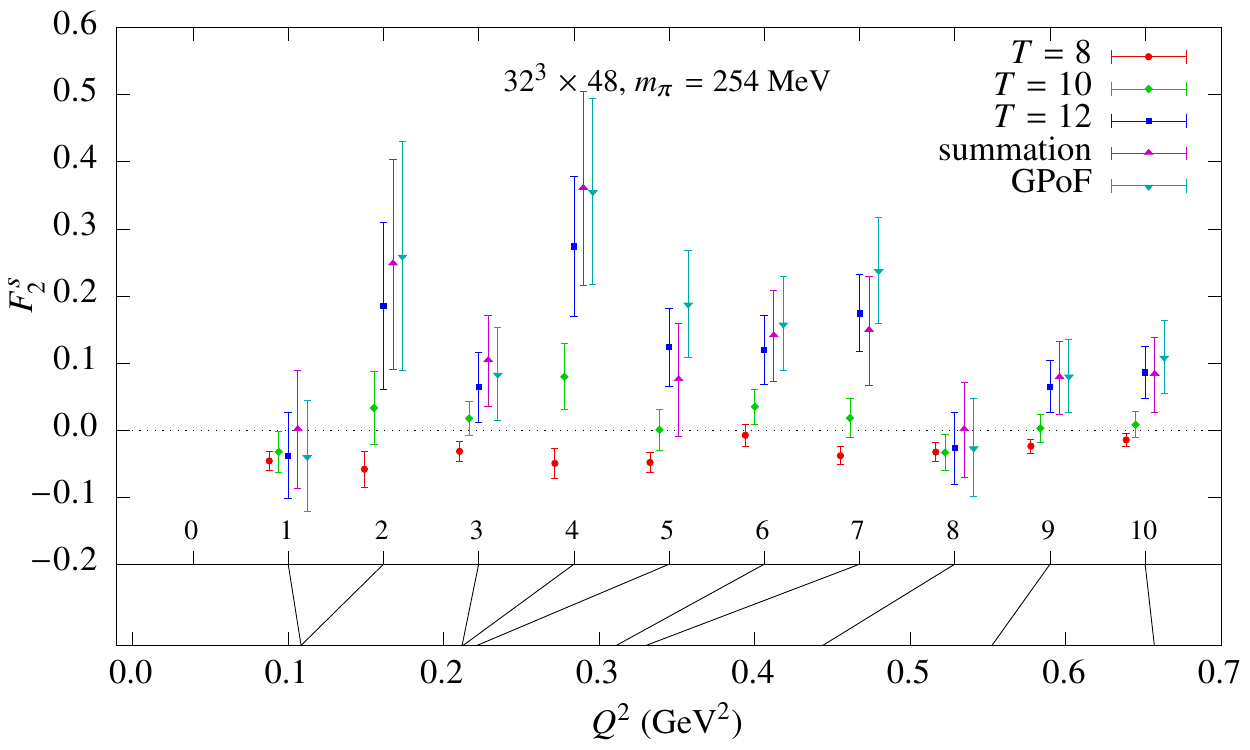}\\
    \includegraphics[width=\textwidth]{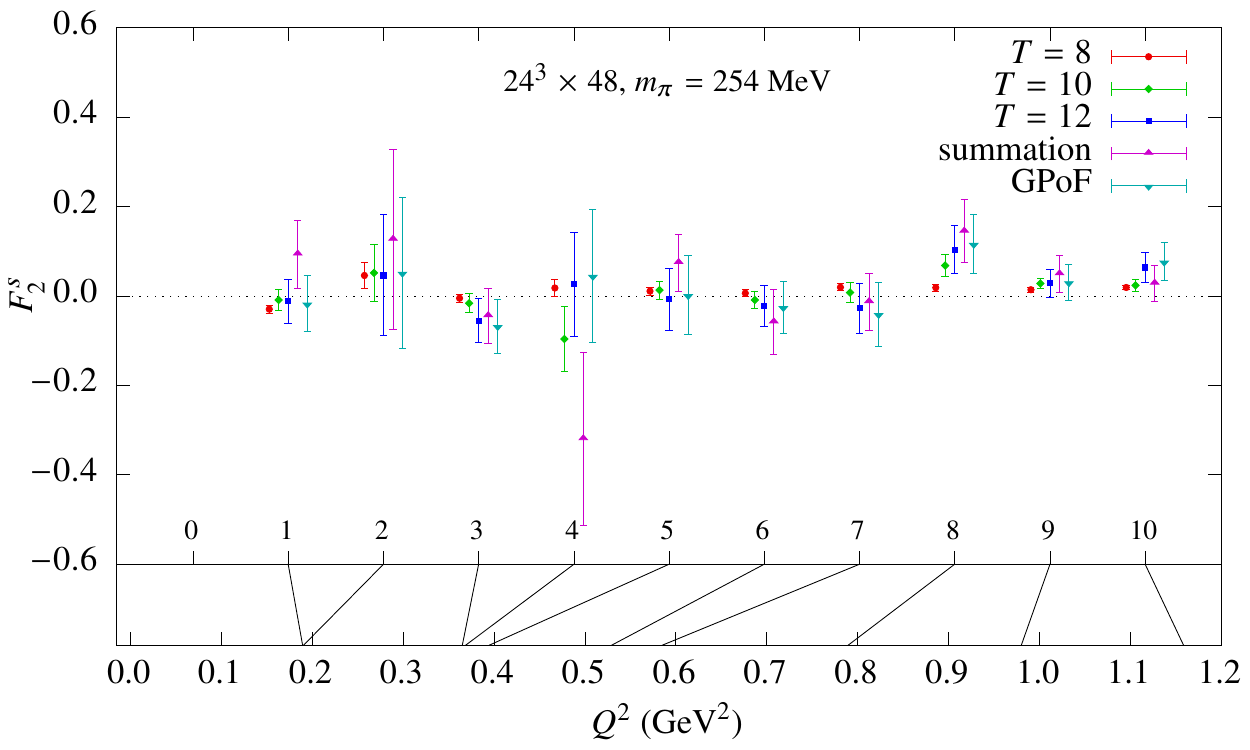}
  \end{minipage}
  \caption{\label{fig:isoscalar_ff_extract_comparison}Comparison
    of different methods to extract the connected-quark-contraction
    contribution to the ground state isoscalar form
    factors $F_1^s(Q^2)$ and $F_2^s(Q^2)$.  From top to bottom, data
    from the $m_\pi=149$~MeV, 202~MeV, 254~MeV ($32^3\times 48$), and
    254~MeV ($24^3\times 48$) lattice ensembles are shown.}
\end{figure}

Isoscalar Dirac and Pauli form factors on four ensembles are shown in
Fig.~\ref{fig:isoscalar_ff_extract_comparison}. Increasing the
source-sink separation from $8a$ to $10a$ tends to cause $F_1^s$ to
decrease, whereas for $F_2^s$, the trend is unclear.  In general, the
ratio method with $T=12a$ tends to agree with the summation and GPoF
methods, except for the Dirac form factor on the $m_\pi=149$~MeV
ensemble, where the summation method produces results that generally
lie below the others. This suggests that, as for the isovector form
factors, excited-state effects are small except at the lightest pion
mass.

\subsection{Radii and magnetic moments}

\begin{figure}
  \centering
  \includegraphics[width=0.7\textwidth]{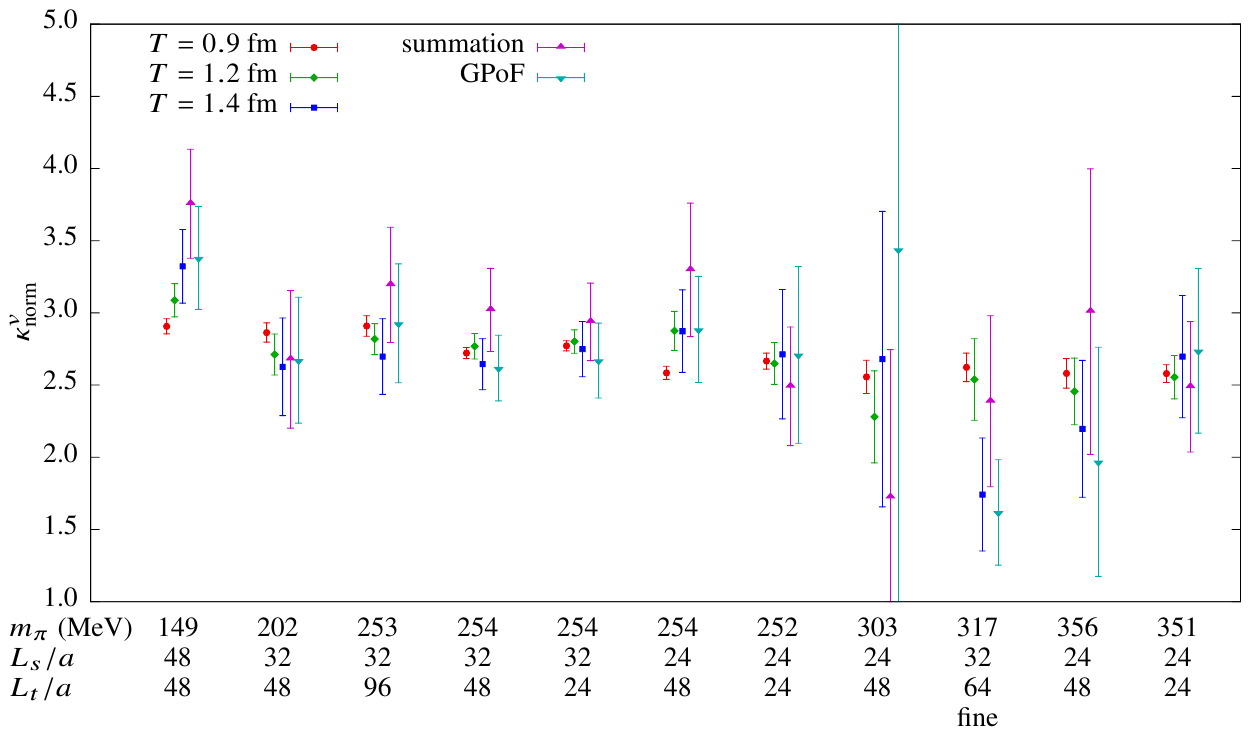}
  \caption{Isovector anomalous magnetic moment $\kappa^v_\text{norm}$,
  determined on each lattice ensemble using different analysis methods
  for computing form factors.}
  \label{fig:kvnorm_ens}
\end{figure}

The isovector anomalous magnetic moment, $\kappa^v_\text{norm}$, is
shown in Fig.~\ref{fig:kvnorm_ens}. There is no broad trend of
dependence on source-sink separation, except at the lightest pion
mass, where the extracted magnetic moment increases with the
source-sink separation, and the summation method produces a
still-higher value.

\begin{figure}
  \centering
  \includegraphics[width=0.7\textwidth]{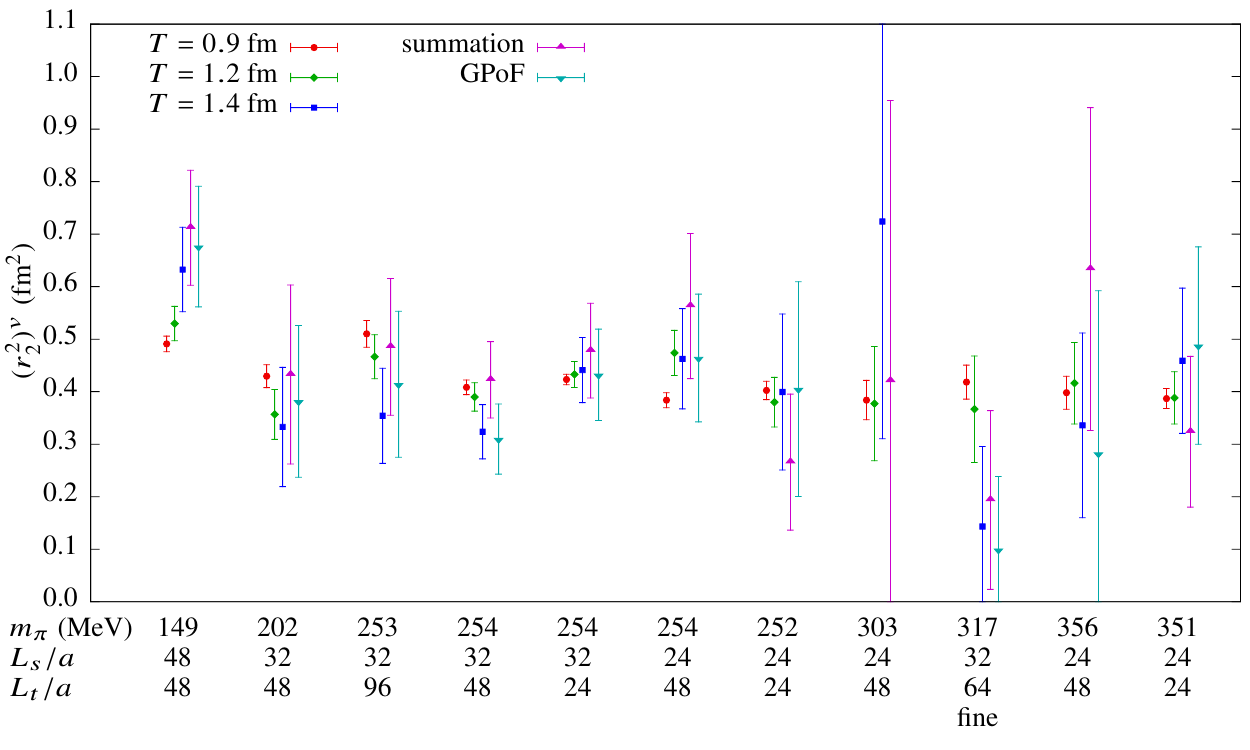}
  \caption{Isovector Pauli radius $(r_2^2)^v$, determined on each lattice
  ensemble using different analysis methods for computing form factors.}
  \label{fig:r2v2_ens}
\end{figure}

For the isovector Pauli radius, we show the dependence on the method
used for computing matrix elements in Fig.~\ref{fig:r2v2_ens}. The
result is very similar to $\kappa^v$: there is no broad trend of
dependence on source-sink separation, but $(r_2^2)^v$ does appear to
increase with source-sink separation on the $m_\pi=149$~MeV ensemble,
and the summation method produces a still-higher value.

\begin{figure}
  \centering
  \includegraphics[width=0.7\textwidth]{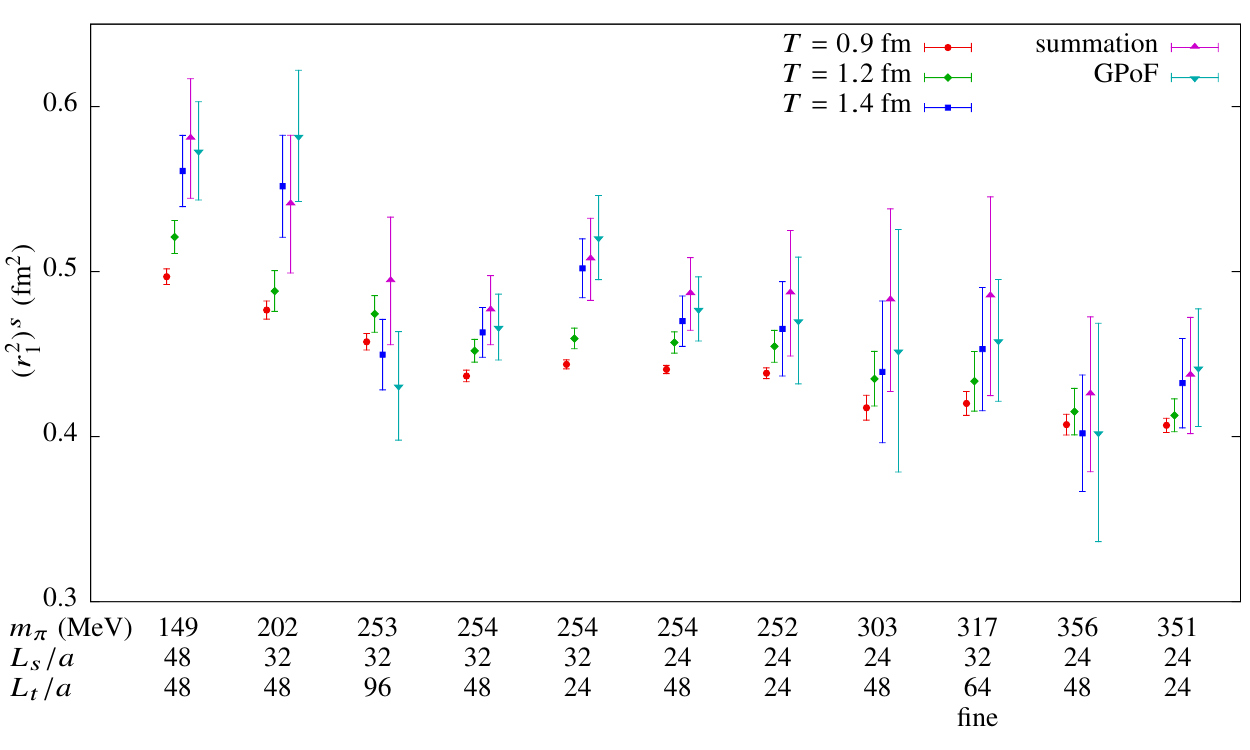}
  \caption{Isoscalar Dirac radius $(r_1^2)^s$, determined on each lattice
  ensemble using different analysis methods for computing form factors.}
  \label{fig:r1s2_ens}
\end{figure}

\begin{figure}
  \centering
  \includegraphics[width=0.7\textwidth]{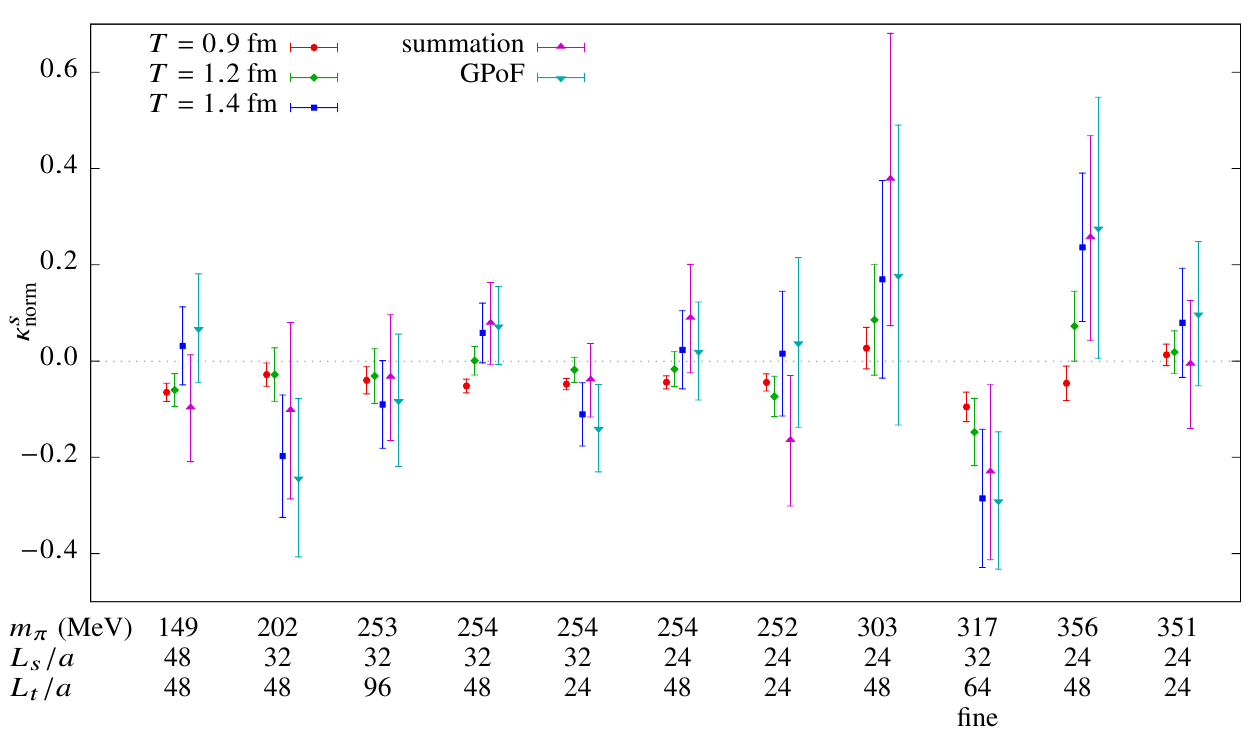}
  \caption{Isoscalar anomalous magnetic moment $\kappa^s_\text{norm}$,
  determined on each lattice ensemble using different analysis methods
  for computing form factors.}
  \label{fig:ksnorm_ens}
\end{figure}

\begin{figure}
  \centering
  \includegraphics[width=0.7\textwidth]{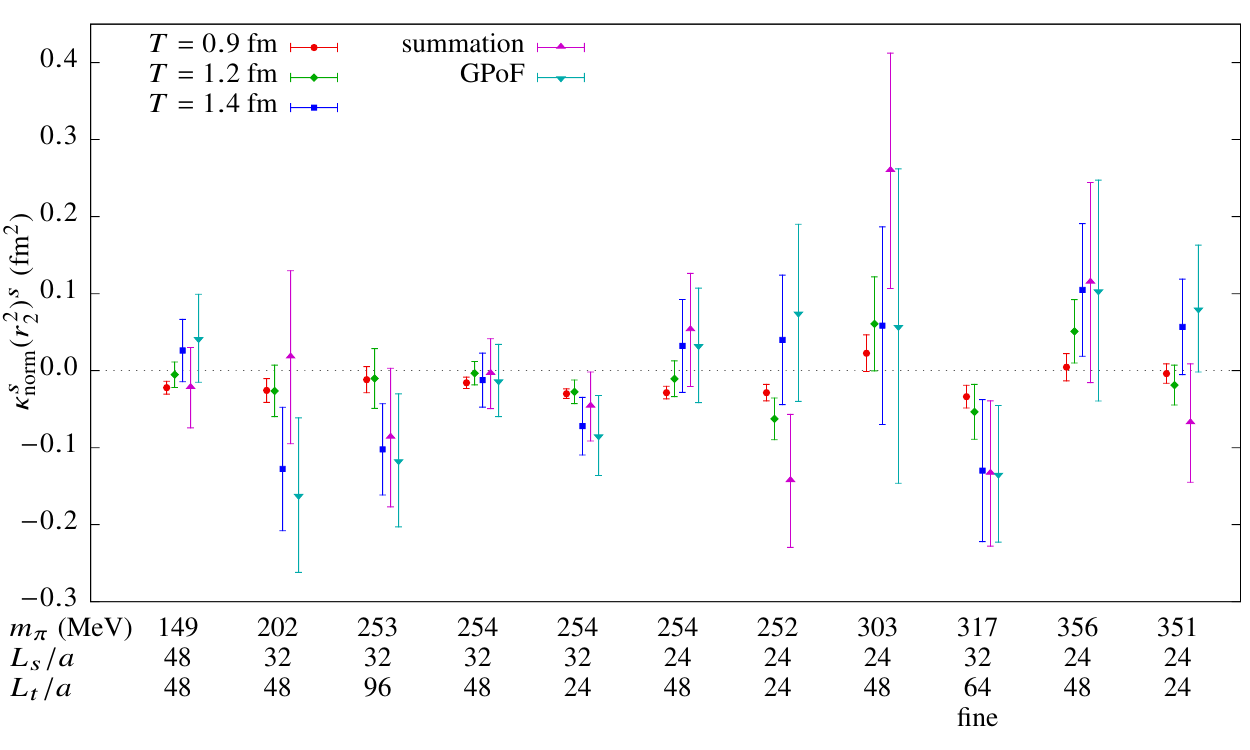}
  \caption{Product of the isoscalar anomalous magnetic moment and Pauli
  radius, $\kappa^s_\text{norm}(r_2^2)^s$, determined on each lattice
  ensemble using different analysis methods for computing form factors.}
  \label{fig:r2s2_ksnorm_ens}
\end{figure}

In Figs.~\ref{fig:r1s2_ens}--\ref{fig:r2s2_ksnorm_ens}, we show the
isoscalar radii and anomalous magnetic moment. These behave similarly to the
isovector case: we find large excited-state effects for the Dirac
radius but not for observables related to the Pauli form factor at
$Q^2=0$.

\section{\label{app:table}Tables of results}
We list isovector and isoscalar Dirac and Pauli form factors for four
ensembles, computed using the summation method, in
Tabs.~\ref{tab:149}--\ref{tab:254_24c48}.

\newcommand{\mm}{\hphantom{-}}
\begin{table}
  \caption{\label{tab:149}Electromagnetic form factors from the $m_\pi=149$~MeV
    ensemble, computed using the summation method. The first column lists
    representative source and sink momenta ($\vec p = \frac{2\pi}{L_s}\vec n$
    and $\vec p\,'=\frac{2\pi}{L_s}\vec n'$, respectively) for each momentum
    transfer $Q^2$.}
  \begin{tabular}{D{|}{|\:|}{7.7}|c|D{.}{.}{1.7}D{.}{.}{1.6}|D{.}{.}{1.7}D{.}{.}{2.6}}
    \hline\hline
    & & \multicolumn{2}{c|}{Isovector} & \multicolumn{2}{c}{Isoscalar} \\
    \la \vec n'|\vec n \ra & $Q^2 (\text{GeV}^2)$
    & \multicolumn{1}{c}{$F_1(Q^2)$} & \multicolumn{1}{c|}{$F_2(Q^2)$}
    & \multicolumn{1}{c}{$F_1(Q^2)$} & \multicolumn{1}{c}{$F_2(Q^2)$} \\
    \hline
\la  0,0,0|\mm 0,0,0\ra & 0     & 1.023(21) &          & 1.012(11) &          \\
\la -1,0,0|   -2,0,0\ra & 0.044 & 0.974(87) & 3.63(98) & 0.861(39) & -0.82(59)\\
\la  0,0,0|\mm 1,0,0\ra & 0.049 & 0.913(24) & 3.45(34) & 0.883(11) &  0.06(17)\\
\la -1,0,0|   -1,1,0\ra & 0.049 & 0.883(31) & 2.83(52) & 0.897(16) &  0.20(36)\\
\la -1,0,0|   -2,1,0\ra & 0.090 & 0.866(72) & 2.11(57) & 0.777(33) &  0.27(33)\\
\la  0,0,0|\mm 1,1,0\ra & 0.096 & 0.856(30) & 2.83(28) & 0.793(15) &  0.15(14)\\
\la -1,0,0|   -1,1,1\ra & 0.096 & 0.778(45) & 2.92(40) & 0.789(26) &  0.35(26)\\
\la -1,0,0|\mm 0,1,0\ra & 0.099 & 0.846(44) & 2.73(34) & 0.805(23) &  0.01(17)\\
\la -1,0,0|   -2,1,1\ra & 0.134 & 0.747(73) & 1.87(53) & 0.732(35) &  0.17(25)\\
\la  0,0,0|\mm 1,1,1\ra & 0.143 & 0.791(38) & 2.39(27) & 0.719(18) &  0.00(13)\\
\la -1,0,0|\mm 0,1,1\ra & 0.147 & 0.782(47) & 2.45(35) & 0.746(24) &  0.07(16)\\
\la  0,0,0|\mm 2,0,0\ra & 0.188 & 0.695(38) & 2.18(27) & 0.671(23) & -0.19(13)\\
\la -1,0,0|   -1,2,0\ra & 0.188 & 0.598(51) & 2.07(37) & 0.599(32) & -0.08(21)\\
\la -1,0,0|\mm 1,0,0\ra & 0.197 & 0.745(66) & 2.19(34) & 0.576(38) &  0.03(23)\\
\la  0,0,0|\mm 2,1,0\ra & 0.232 & 0.656(38) & 1.96(21) & 0.605(20) & -0.06(10)\\
\la -1,0,0|   -1,2,1\ra & 0.233 & 0.600(48) & 2.01(32) & 0.563(31) & -0.05(15)\\
\la -1,0,0|\mm 0,2,0\ra & 0.242 & 0.605(47) & 2.12(26) & 0.597(27) & -0.20(16)\\
\la -1,0,0|\mm 1,1,0\ra & 0.246 & 0.661(64) & 2.12(21) & 0.539(30) & -0.03(12)\\
\la  0,0,0|\mm 2,1,1\ra & 0.276 & 0.614(40) & 1.78(19) & 0.560(20) & -0.07(9) \\
\la -1,0,0|\mm 0,2,1\ra & 0.287 & 0.598(44) & 2.00(23) & 0.552(25) & -0.03(12)\\
\la -1,0,0|\mm 1,1,1\ra & 0.294 & 0.564(59) & 1.96(21) & 0.528(33) & -0.02(10)\\
\la -1,0,0|\mm 1,2,0\ra & 0.386 & 0.466(58) & 1.47(18) & 0.440(36) & -0.01(11)\\
\la -1,0,0|\mm 1,2,1\ra & 0.430 & 0.481(53) & 1.64(17) & 0.432(26) &  0.09(8) \\
\la -1,0,0|\mm 2,0,0\ra & 0.439 & 0.439(96) & 1.62(33) & 0.416(55) &  0.06(19)\\
\la -1,0,0|\mm 2,1,0\ra & 0.485 & 0.431(87) & 1.37(21) & 0.341(39) & -0.12(10)\\
    \hline\hline
  \end{tabular}
\end{table}

\begin{table}
  \caption{\label{tab:202}Electromagnetic form factors from the $m_\pi=202$~MeV 
    ensemble, computed using the summation method. The first column lists
    representative source and sink momenta ($\vec p = \frac{2\pi}{L_s}\vec n$
    and $\vec p\,'=\frac{2\pi}{L_s}\vec n'$, respectively) for each momentum
    transfer $Q^2$.}
  \begin{tabular}{D{|}{|\:|}{7.7}|c|D{.}{.}{1.8}D{.}{.}{1.6}|D{.}{.}{1.7}D{.}{.}{2.6}}
    \hline\hline
    & & \multicolumn{2}{c|}{Isovector} & \multicolumn{2}{c}{Isoscalar} \\
    \la \vec n'|\vec n \ra & $Q^2 (\text{GeV}^2)$
    & \multicolumn{1}{c}{$F_1(Q^2)$} & \multicolumn{1}{c|}{$F_2(Q^2)$}
    & \multicolumn{1}{c}{$F_1(Q^2)$} & \multicolumn{1}{c}{$F_2(Q^2)$} \\
    \hline
\la  0,0,0|\mm 0,0,0\ra & 0     & 1.004(16) &          & 1.004(8) &          \\
\la  0,0,0|\mm 1,0,0\ra & 0.108 & 0.831(37) & 2.47(35) & 0.812(19) & -0.11(17)\\
\la -1,0,0|   -1,1,0\ra & 0.108 & 0.853(68) & 2.29(62) & 0.807(30) & -0.35(28)\\
\la  0,0,0|\mm 1,1,0\ra & 0.211 & 0.695(51) & 2.11(27) & 0.637(22) & -0.07(14)\\
\la -1,0,0|   -1,1,1\ra & 0.212 & 0.633(80) & 1.41(57) & 0.661(38) & -0.19(26)\\
\la -1,0,0|\mm 0,1,0\ra & 0.222 & 0.772(74) & 2.03(34) & 0.710(36) & -0.05(14)\\
\la  0,0,0|\mm 1,1,1\ra & 0.309 & 0.601(65) & 1.59(24) & 0.542(29) & -0.28(12)\\
\la -1,0,0|\mm 0,1,1\ra & 0.330 & 0.688(81) & 1.83(30) & 0.564(39) & -0.11(13)\\
\la -1,0,0|\mm 1,0,0\ra & 0.444 & 0.638(104)& 1.52(34) & 0.463(51) & -0.03(14)\\
\la -1,0,0|\mm 1,1,0\ra & 0.552 & 0.594(79) & 1.20(20) & 0.433(36) &  0.09(9) \\
\la -1,0,0|\mm 1,1,1\ra & 0.656 & 0.580(84) & 0.97(20) & 0.391(37) & -0.09(9) \\
    \hline\hline
  \end{tabular}
\end{table}

\begin{table}
  \caption{\label{tab:254_32c48}Electromagnetic form factors from the $m_\pi=254$~MeV, $32^3\times 48$ ensemble, computed using the summation method. Source and sink momenta are the same as given in Tab.~\ref{tab:202}.}
  \begin{tabular}{c|D{.}{.}{1.7}D{.}{.}{1.6}|D{.}{.}{1.7}D{.}{.}{2.6}}
    \hline\hline
    & \multicolumn{2}{c|}{Isovector} & \multicolumn{2}{c}{Isoscalar} \\
    $Q^2 (\text{GeV}^2)$
    & \multicolumn{1}{c}{$F_1(Q^2)$} & \multicolumn{1}{c|}{$F_2(Q^2)$}
    & \multicolumn{1}{c}{$F_1(Q^2)$} & \multicolumn{1}{c}{$F_2(Q^2)$} \\
    \hline
    0     & 1.007(6)  &          & 1.005(3)  & \\
    0.108 & 0.849(18) & 2.72(23) & 0.821(9)  &  0.00(9) \\
    0.109 & 0.893(30) & 3.01(36) & 0.801(13) &  0.25(16) \\
    0.212 & 0.751(24) & 2.51(17) & 0.678(11) &  0.10(7) \\
    0.213 & 0.800(44) & 2.84(32) & 0.662(21) &  0.36(15) \\
    0.222 & 0.711(38) & 2.16(20) & 0.665(19) &  0.07(8) \\
    0.311 & 0.686(33) & 2.34(16) & 0.575(16) &  0.14(7) \\
    0.331 & 0.666(39) & 1.95(17) & 0.567(20) &  0.15(8) \\
    0.444 & 0.478(57) & 1.51(17) & 0.474(30) &  0.00(7) \\
    0.553 & 0.464(43) & 1.27(13) & 0.416(26) &  0.08(5) \\
    0.657 & 0.409(44) & 1.08(13) & 0.359(27) &  0.08(6) \\
    \hline\hline
  \end{tabular}
\end{table}

\begin{table}
  \caption{\label{tab:254_24c48}Electromagnetic form factors from the $m_\pi=254$~MeV, $24^3\times 48$ ensemble, computed using the summation method. Source and sink momenta are the same as given in Tab.~\ref{tab:202}, scaled by 4/3 due to the smaller box size.}
  \begin{tabular}{c|D{.}{.}{1.7}D{.}{.}{1.6}|D{.}{.}{1.7}D{.}{.}{2.6}}
    \hline\hline
    & \multicolumn{2}{c|}{Isovector} & \multicolumn{2}{c}{Isoscalar} \\
    $Q^2 (\text{GeV}^2)$
    & \multicolumn{1}{c}{$F_1(Q^2)$} & \multicolumn{1}{c|}{$F_2(Q^2)$}
    & \multicolumn{1}{c}{$F_1(Q^2)$} & \multicolumn{1}{c}{$F_2(Q^2)$} \\
    \hline
    0     & 0.993(7)  &          & 1.002(3)  & \\
    0.189 & 0.753(21) & 2.50(17) & 0.699(12) &  0.09(8) \\
    0.190 & 0.934(61) & 1.99(39) & 0.750(43) &  0.13(20) \\
    0.365 & 0.627(29) & 1.83(12) & 0.534(17) & -0.04(6) \\
    0.369 & 0.810(71) & 1.69(33) & 0.537(41) & -0.32(19) \\
    0.395 & 0.617(36) & 1.66(14) & 0.511(22) &  0.07(6) \\
    0.529 & 0.532(48) & 1.63(15) & 0.402(25) & -0.06(7) \\
    0.585 & 0.505(41) & 1.29(13) & 0.383(24) & -0.01(6) \\
    0.790 & 0.483(55) & 1.15(14) & 0.354(39) &  0.14(7) \\
    0.980 & 0.371(41) & 0.95(10) & 0.287(21) &  0.05(4) \\
    1.158 & 0.382(47) & 0.73(9)  & 0.212(23) &  0.03(4) \\
    \hline\hline
  \end{tabular}
\end{table}

\bibliography{paper}

\printtables

\printfigures

\end{document}